\documentclass[%
 aip,
 amsmath,amssymb,
 reprint,%
]{revtex4-1}

\usepackage{graphicx}

\usepackage{dcolumn}
\usepackage{bm}

\usepackage[utf8]{inputenc}
\usepackage[T1]{fontenc}
\usepackage{mathptmx}
\usepackage{etoolbox}
\usepackage{accents}
\usepackage{stmaryrd}

\usepackage{color}

\newcommand{\ojoo}[1]{\textcolor{black}{#1}}

\makeatletter
\def\@email#1#2{%
 \endgroup
 \patchcmd{\titleblock@produce}
  {\frontmatter@RRAPformat}
  {\frontmatter@RRAPformat{\produce@RRAP{*#1\href{mailto:#2}{#2}}}\frontmatter@RRAPformat}
  {}{}
}%
\makeatother
\begin{document}

\preprint{AIP/123-QED}

\title[Random Close Packing of Semi-Flexible Polymers in Two Dimensions: Emergence of Local and Global Order]{Random Close Packing of Semi-Flexible Polymers in Two Dimensions: Emergence of Local and Global Order}
\author{Daniel Martínez-Fernández}
\author{Clara Pedrosa}%
\author{Miguel Herranz}%
\author{Katerina Foteinopoulou}%
\author{Nikos Ch. Karayiannis$^*$}%
 \email{n.karayiannis@upm.es}
\author{Manuel Laso}
\affiliation{ 
Institute for Optoelectronic Systems and Microtechnology (ISOM) and Escuela Técnica Superior de Ingenieros Industriales (ETSII), Universidad Politécnica de Madrid (UPM) C/ Jose Gutierrez Abascal 2, 28006 Madrid (Spain) 
}%

\date{\today}

\begin{abstract}
Through extensive Monte Carlo simulations, we systematically study the effect of chain stiffness on the packing ability of linear polymers composed of hard spheres in extremely confined monolayers, corresponding effectively to 2D films. First, we explore the limit of random close packing as a function of the equilibrium bending angle and then quantify the local and global order by the degree of crystallinity and the nematic or tetratic orientational order parameter, respectively. A multi-scale wealth of structural behavior is observed, which is inherently absent in the case of athermal individual monomers and is surprisingly richer than its 3D counterpart under bulk conditions. As a general trend an isotropic to nematic transition is observed at sufficiently high surfaces coverages which is followed by the establishment of the tetratic state, which in turn marks the onset of the random close packing. For chains with right-angle bonds, the incompatibility of the imposed bending angle with the neighbor geometry of the triangular crystal leads to a singular intra- and inter-polymer tiling pattern made of squares and triangles with optimal local filling at high surface concentrations. The present study could serve as a first step towards the design of hard colloidal polymers with tunable structural behavior for 2D applications.  
\end{abstract}

\maketitle

\section{Introduction}

\par The packing ability and phase behavior of atoms, particles and objects are intimately related to the macroscopic properties of the corresponding physical systems. Packing is thus a key factor in many important physical processes and diverse technological and engineering applications involving for example proteins in living cells, atomic motifs in amorphous and crystal solids, optical photonic and functional materials based on colloidal crystals, stacked piles of cannonballs, oranges in baskets and grains in silos \cite{RN2024,RN2003,RN1907,RN1831,RN2025,RN317,RN153,RN2028,RN2031}.
\par The packing of hard spheres and disks of uniform size in three and two dimensions, respectively, has been extensively studied, advancing our fundamental understanding of the salient features of optimal packing, jamming, crystallization and melting \cite{RN2002,RN1995,RN1994,RN1910,RN1899,RN1898,RN1887,RN1886,RN1884,RN1807,RN1782,RN1769,RN1740,RN1735,RN128,RN163,RN543,RN199,RN470,RN442,RN168,RN164,RN55,RN147,RN202,RN107,RN2027,RN106,RN2029,RN2043}.  
In three dimensions, it has been formally proven \cite{RN1881} that the maximum packing density of non-overlapping spheres, $\varphi^{max}_{3D}$, can be achieved in face centered cubic (FCC) or hexagonal close packed (HCP) crystals and corresponds to $\varphi^{max}_{3D} = \varphi^{HCP} = \varphi^{FCC} = \ojoo{\sqrt{2}\pi/6}$. 
\par In practice, in three dimensions, starting from dilute conditions and through compression, the system, be an experimental realization inside a container or a simulation cell subjected to periodic boundary conditions, will reach a jammed state of predominantly amorphous character, where a further increase in concentration is not achievable within the observation time. This state is known as random close packing (RCP) and occurs in a density range that is approximately 10\% lower than the maximum achievable one, i.e. $\varphi^{RCP}_{3D} \approx 0.64$. The maximally random jammed (MRJ) state, as introduced in \cite{RN202}, diminishes problems related to the robust definition of the RCP state, as it introduces a more rigorous concept based on volume fraction and the incipient level of order (or equivalently of disorder). 
\par In two dimensions the maximum packing density corresponds to the triangular (TRI) crystal of the $p6m$ wallpaper group, with a corresponding value of surface coverage, $\varphi^{*,max}_{2D} = \ojoo{\pi/2\sqrt{3}}$. If we consider a triangular monolayer of maximally packed hard spheres, the analogous packing density is approximately $\varphi^{max}_{2D} \approx 0.605$. Estimates for the RCP state for disks in two dimensions have been provided by recent theoretical studies \cite{RN1886,RN1884,RN1887}, according to which $\varphi^{*,RCP}_{2D}$ ranges between 0.853 and 0.887.
\par Independent of whether in experiments or simulations, important challenges must be addressed, first related to the generation of loose, dense, jammed, and maximal hard-body packings and then to the characterization of the resulting systems. The former is a particularly challenging task because of the very high volume fraction associated with jamming and crystallization \cite{RN364,RN55,RN1782,RN1900,RN1730}, especially when non-trivial hard shapes are considered \cite{RN434,RN508,RN226,RN227,RN371,RN368,RN116,RN627}. The latter is essential to quantify the degree of disorder among the possible jammed configurations so as to identify the MRJ state \cite{RN202,RN99}, or the degree of order to gauge crystallization and the corresponding crystal morphologies, or a melting transition \cite{RN1997,RN2029,RN1998,RN1999,RN2000}. Efficient simulation algorithms have been developed over the last decades and employed to both create \cite{RN552,RN181,RN159,RN43,RN31,RN45} and characterize \cite{RN52,RN1769,RN1730,RN128,RN71,RN147,RN146,RN69,RN1766,RN65} hard-body packings under a wide variety of conditions.

\par The complexity in studying packing problems is further augmented when polymer systems are considered, where hard monomers are bonded to form long linear or non-linear sequences. This is due to numerous reasons: first, in contrast to many-particle regular objects like polygons \cite{RN627,RN226,RN2044,RN2045,RN2046,RN2047} or polyhedra \cite{RN1803,RN1714,RN225,RN629,RN626,RN224,RN327,RN368}, the contour of polymer chains constantly changes so that their shape and size require a statistical description \cite{RN1477,RN291,RN293,RN376}; second, additional constraints and parameters in the form of bond lengths, and bending and torsion angles dictate their behavior including  the effect of chain length and stiffness, among others; third, due to the wide spectrum of characteristic time and lengths scales \cite{RN293,RN376}, special techniques \cite{RN1724,RN1732}, such as advanced Monte Carlo (MC) algorithms \cite{RN9,RN2033,RN1712,RN502}, are required not only to generate the corresponding polymer packings but to correctly sample the local and global structure of the chains. 

\par In the past we have studied through extensive MC simulations the effect of chain connectivity on the structure \cite{RN17,RN27,RN13}, packing ability \cite{RN17,RN12}, phase behavior \cite{RN25,RN23,RN497} and jamming \cite{RN17} of athermal systems made of polymers in the bulk and under various conditions of confinement \cite{RN1678,RN1892}. In three dimensions it has been demonstrated that linear, freely-jointed chains of tangent hard spheres i) reach the same RCP limit as monomers \cite{RN17,RN12}; ii) crystallize once a critical concentration range is reached, with the melting point being higher than the corresponding one of monomers and strongly dependent on the gaps between bonds \cite{RN25,RN497}; iii) form crystal polymorphs of mixed HCP and FCC character and eventually reach perfection in the form of a pure, defect-free FCC cystal \cite{RN1894}, which is has also been shown to be the thermodynamically stable phase \cite{RN2013} as in the case of monomers \cite{RN132,RN215,RN1736}. Through the use of the $Simu$-$D$ simulator-descriptor software \cite{RN1824}, our studies have been further extended to treat athermal packings of semi-flexible chains so as to identify the short- and long-range order, corresponding to close-packed crystals (at the level of atoms) and nematic phases of mesogens (at the level of chains), respectively \cite{RN2010}. The formation of nematic mesophases of prolate or oblate mesogens can precede or be synchronous to crystallization, depending on the equilibrium bending angle \cite{RN2010}.

\par The long-range orientational order of polymer chains is traditionally quantified by \ojoo{the nematic director, $\boldsymbol{n}$}, that defines the preferred direction of alignment of the molecules \cite{RN94}.  In three dimensions long-range orientation is quantified by the orientational (nematic) order parameter which gauges the degree of inter-chain alignment along the preferred director \ojoo{$\boldsymbol{n}$}. The factors that affect the isotropic to nematic transition for semi-flexible polymers in the bulk, solutions and under confinement have been studied in \cite{RN1929,RN1952,RN1958,RN2081,RN2082}.

\par In two dimensions, the reduction of dimensionality leads to a more complex behavior of anisotropic hard bodies where long-range orientational order can be absent or unstable and/or higher-order symmetries prevail like in the form of tetratic, sexatic or octatic phases \cite{RN2070,RN2060,RN2063,RN2071,RN2077,RN2066}. Theoretical predictions on the singular long-range phase behavior in two dimensions have been confirmed and expanded by simulation and experimental findings, further exploring systematically the effect of aspect ratio and local geometry of the particles and coverage and curvature of the surface \cite{RN2056,RN2061,RN1999,RN2072,RN2073,RN627,RN2065,RN2074,RN2075,RN2064,RN2076,RN2078} and studying hard-body realizations that include, among others, zig-zag \cite{RN2058}, rod \cite{RN2057,RN2061}, general-shape \cite{RN2059,RN2060,RN2080} objects, bend-core trimers \cite{RN1905} and semi-flexible polymers \cite{RN2062}. 

\par In three dimensions, studies on the crystallization of fully flexible athermal polymers have demonstrated that while monomers are orderly positioned in sites corresponding to close packed crystals \cite{RN25,RN23,RN497}, the corresponding chains behave as ideal random walks, with their statistics being practically unaltered between the initial amorphous state, the intermediate crystal polymorphs and the final stable FCC crystal \cite{RN1894,RN2013}. 

\par In extremely confined monolayers whose thickness approaches the monomer diameter, practically corresponding to 2D systems, freely-jointed chains of tangent hard spheres form a crystal of predominantly triangular (TRI) character whose surface coverage, $\varphi^{*,RCP}_{2D}(FJ) = 0.895$, is very close to the maximum achievable one $\varphi^{*,max}_{2D} \approx 0.908$ as demonstrated in \cite{RN2034}. In parallel, simple geometric arguments suggest that packings approaching $\varphi^{*,max}_{2D} \approx 0.908$ ever more closely can be achieved by properly tuning the number of monomers in the simulation cell to comply with the geometric condition of the height-to-width ratio of the perfect hexagon \cite{RN2034}. Effectively, the simulated systems correspond to experimental realizations of spheres being confined in two dimensions, in the form of nanoparticle monolayers \cite{RN2030}, as for example microgels between two glass coverslips \cite{RN2029},  hard colloidal spherical particles in a cell whose thickness is comparable to the particle size \cite{RN2026}, rod-like hard bodies in aluminum/steel cylindrical or cubic containers loaded on magnet shakers \cite{RN2064,RN2065}, or superparamagnetic particles in a quartz cuvette \cite{RN2032}. Extremely confined polymer thin films have drawn scientific and industrial attention \cite{RN2038,RN2039,RN1472,RN579,RN2040,RN529,RN2041,RN2042,RN1895,RN1473} due to their numerous applications especially as high-performance electronics and organic transistors \cite{RN2037}. \ojoo{Experimental realizations of ultra thin films made of polymers have been reported in the literature corresponding to molecularly flat films, 2D platelets, colloidal 2D crystals, and general self-assembled monolayers \cite{RN2114,RN2111,RN2112,RN2113,RN2117,RN2116,RN2115}}.

\par In the present contribution we analyze the effect of chain stiffness on the ability of polymers to pack in two dimensions, quantifying the established short- and long-range order and comparing the results with those obtained for freely-jointed chains (having unconstrained bend and torsion angles) and monomers (free of any constraints imposed by chain connectivity). Towards this,  we generate various polymer systems with different equilibrium bending angles under dilute conditions,  which are then compressed to the maximum achievable packing density by means of long constant-volume simulations. In a last step, quantitative descriptors are employed to gauge the local and global structure of the computer-generated, single-layer polymer packings.

\section{Model and Simulation Methodology}

\par In the present work we adopt the same polymer model as used in Ref. \cite{RN2010} in three dimensions. Linear chains consist of hard spheres of uniform size, with the sphere diameter, $\sigma$, being the characteristic length of the system. Successive monomers are tangent within a numerical tolerance of $dl = 6.5 \times 10^{-4}$, the ensuing bond gaps being too small to affect the phase behavior of the chains, as analyzed in \cite{RN497}. Bending angles are controlled by a harmonic potential of the form:
\begin{align}
{U_{bend}(\theta)}=k_{\theta}(\theta - \theta_{0})^2
\label{Bending_Eq}
\end{align}
where $U_{bend}$ is the energy, $\theta$ is the bending angle formed by a triplet of successive spheres along the chain backbone (see for example the sketch of Fig. 1 in \cite{RN2010}), $\theta_{0}$ is the equilibrium bending angle and $k_{\theta}$ is the harmonic constant. As in the case of semi-flexible chains in three dimensions \cite{RN2010} we set $k_{\theta} / k_{B}T= 9$ rad$^{-2}$. The following values have been used for $\theta_{0}$: 0, 60, 90, and $120^{\circ}$, the first and the last corresponding to fully extended (rod) and locally compact configurations, respectively. The $\theta_{0}=$ 0, 60, and $120^{\circ}$ equilibrium bending angles are compatible with the geometry imposed by the connectivity of the sites in the TRI crystal while the $\theta_{0} =$ $90^{\circ}$ is not. Results are compared against the ones of freely-jointed (FJ) chains ($k_{\theta} = 0$) and of monomers (free of any constraints imposed by chain connectivity) under the same simulation conditions \cite{RN2034}. Torsion angles are allowed to fluctuate freely in the case of three dimensions and are latter forced to adopt co-planar configurations once a monolayer is formed. 

\par All simulated systems consist of 100 chains ($n_{ch}=100$) with an average length of $N_{av} = 12$, \ojoo{measured in number of spheres}, for a total of $n_{at} = 1200$ interacting hard spheres. Due to the presence of chain-connectivity-altering moves \cite{RN240}, chain lengths of individual molecules fluctuate uniformly in the interval $N \in [6,18]$. Packing density, $\varphi$, is defined as the volume occupied by all hard spheres, divided by the total volume of simulation cell. In two dimensions, surface coverage, $\varphi^{*}$, is defined as the surface occupied by the parallel projection of the spheres on a large face, divided by the area of each of the two large faces. For monolayers, where film thickness is approximately equal to monomer diameter: $3\varphi=2\varphi^{*}$.

\par We use the $Simu$-$D$ software suite \cite{RN1824} for the generation, equilibration and characterization of the system configurations. The simulation protocol is a combination of the ones adopted in past works on semi-flexible athermal polymer packings in three dimensions \cite{RN2010} and fully-flexible ones in two dimensions \cite{RN2034}. First, initial configurations of freely-jointed chains are generated at very dilute conditions ($\varphi = 0.05$) in three dimensions under periodic boundary conditions (PBCs). Then, the bending potential with the appropirate $\theta_{0}$ is activated, while one cell dimension is confined by the introduction of parallel flat and impenetrable walls. Extended constant-volume MC simulations are conducted on the resulting polymer configurations.  In the continuation, strong attraction, in the form of a square-well potential, is enforced between one confining wall and all monomers, while repulsion of equal strength is applied on the opposite wall and the monomers. Eventually, this leads to the formation of a monolayer whose thickness is equal to the monomer diameter within a numerical tolerance of $10^{-5}$. Once the monolayer is formed the attractive potential between the wall and the chain monomers is deactivated.
\par Starting from this single-layer configuration the long dimensions are then isotropically compressed until a surface coverage of $\varphi^{*} = 0.70$ is reached. Then,  the process is continued with anisotropic shrinkage of the two large edges of the cell until no further compression can be achieved. This stage corresponds to the random close packed limit in two dimensions for a given value of the equilibrium bending angle, $\varphi_{2D}^{RCP}(\theta_{0})$. Constant-volume MC simulations are conducted at the maximum achieved packing density and at specific values of surface coverage, which are all common between the different systems: $\varphi^{*} = 0.50$, 0.60, 0.70 and $\varphi_{2D}^{*,RCP}(\theta_{0})=(3/2)\varphi_{2D}^{RCP}(\theta_{0})$. 

\par The duration of the MC simulations ranges between 8 and 20 $\times 10^{10}$, depending on the packing density and the equilibrium bending angle. Frames (system configurations), along with statistics, are recorded every $10^{7}$ MC steps. 
The constant-volume, production simulation corresponds to the following MC scheme \cite{RN1824,RN1252,RN16}: i) reptation (10\%), ii) end-mer rotation (10\%), iii) flip (34.8\%), iv) intermolecular reptation (25\%), v) simplified end-bridging, sEB (0.1\%) and vi) simplified intramolecular end-bridging (0.1\%); where numbers in parentheses denote attempt percentages. As described extensively in \cite{RN16,RN1824}, all local MC moves (types i) through iv)) are executed in a configurational bias pattern with the number of trial configurations, $n_{trials}$ depending on the packing density. As a general rule $n_{trials}$ is set to $50$ once surface coverage exceeds the threshold $\varphi^{*}=0.70$. Neither the attempt percentage nor the number of trial configurations per local MC move change with the value of the equilibrium bending angle.

\section{Descriptors of Local and Global Structure}

The main objective of the present work is to study the effect of chain stiffness, by varying the equilibrium bending angle, on the ability of chains to pack as densely as possible in extremely confined thin films, practically corresponding to monolayers. It is thus essential to quantify the degree of local (at the level of atoms) and global (at the level of chains) order of the corresponding packings. The definition of the MRJ state is, for example, intimately related to the incipient degree of structural disorder \cite{RN202}, which can be quantified indirectly by measuring its antithesis, order, manifested through the presence of sites with crystal similarity. Furthermore, the degree of order depends strongly on the protocol used for the generation of the systems in the vicinity of the RCP limit. Thus, refined metrics are required to gauge the local and global structure of the system configurations as the monolayers, made of athermal polymers, generated here.

\subsection{Local Order: Characteristic Crystallographic Element Norm}

\par A critical component of the analysis is to calculate, as precisely as possible, the degree of order (or, equivalently, disorder)  in the computer-generated packings in two dimensions. Towards this, the similarity of the local structure with respect to specific crystal templates is quantified through the descriptor part of the $Simu$-$D$ software \cite{RN1824}, which is based on the Characteristic Crystallographic Element (CCE) norm \cite{RN10,RN1542}. The CCE norm descriptor is built around the fundamental concept that a unique, and thus distinguishing, set of crystallographic elements and actions corresponds to each crystal in two and three dimensions \cite{RN243}. Practically the computer-generated local environment of the closest neighbors around an atom or particle, as identified by a Voronoi tessellation, is mapped onto the ideal one corresponding to a perfect reference crystal, $X$, and depending on the corresponding geometric actions a norm, $\epsilon^X_i$, is calculated (see for example Eq. 2 in \cite{RN1542} and related discussion).

Here, the honeycomb (HON), square (SQU), triangular (TRI) crystals and the pentagonal (PEN) local symmetry are used as the reference ideal structures in two dimensions. Their corresponding characteristic actions and elements together with the exact details on the quantification through the CCE norm, as well as its practical implementation, can be found in \cite{RN1542}. By further employing a threshold value, $\epsilon^{thres}$, if $\epsilon^X_i < \epsilon^{thres} $ then the site $i$ is labelled as $X$-type. Based on this, and given the highly discriminating nature of the CCE norm descriptor \cite{RN10,RN1542}, an order parameter can be assigned for each reference crystal $X$, $S^X$ ($S^X \in [0,1]$), which practically corresponds to the fraction of sites with $X$ similarity divided by $n_{at}$. The degree of (crystalline) order, $\tau^c$ and of (amorphous) disorder, $S^{AMO}$ can be calculated as:
\begin{align}
\tau^c = \sum S^X, \:\:\:\:\:\:  S^{AMO}=1-\tau^c
\label{order_parameter}
\end{align}
where index $X$ above runs over all reference crystals in two dimensions (TRI, SQU and HON). The salient features and the exact algorithmic implementation of the CCE norm descriptor for general atomic and particulate systems in three and two dimensions can be found in Refs. \cite{RN10,RN1542,RN1824}.

\subsection{Global Order: Long-range Nematic Order Parameter}

\par In the present work the degree of long-range orientation is quantified by two order parameters \cite{RN2079}: the first one is the nematic order parameter, $q_2$, being equal to unity, when all chains are perfectly aligned parallel to the director vector $\underline{n}$, corresponding to the nematic state, as in the case of three dimensions \cite{RN2055}. The second one is the tetratic order parameter, $q_4$, which is equal to unity when each pair of chains is either mutually parallel or perpendicular \cite{RN2079}. 
\par To calculate the nematic order parameter, $q_2$, we adopt the formalism of the order tensor, which in the case of oriented molecules in two dimensions, can be written in the form \cite{RN2056}: 
\begin{align}
\ojoo{ \mathbf{Q}= \left\langle \boldsymbol{u} \, \boldsymbol{u}  - \frac{1}{2}\mathbf{\delta} \right\rangle }
\label{Q_tensor}
\end{align}
where, \ojoo{$\boldsymbol{u}$} is the unit vector along the largest semiaxis of the inertia ellipsoid of a chain (considering all of its monomers as points of unit mass), \ojoo{$\mathbf{\delta}$} is the isotropic second order tensor, and $\langle\ \ \rangle$ denote average over all chains. 
For a system of molecules perfectly aligned in one direction (e.g. direction $x$), leading to a perfectly nematic mesophase we have:
\begin{align}
\llbracket \ojoo{\mathbf{Q}} \rrbracket = 
\begin{bmatrix}
\frac{1}{2} & 0 \\
0 & -\frac{1}{2}
\end{bmatrix}
\label{Q_nematic}
\end{align}
expressed in the coordinate system of its eigenvectors. 
\par To identify the direction of alignment and to quantify the level of global ordering, we compute the order tensor \ojoo{$\mathbf{Q}$} from equation (\Ref{Q_tensor}) and subsequently we diagonalize the tensor leading to its diagonal form \ojoo{$\mathbf{Q}^{\prime}$}. The resulting eigenvector corresponding to the largest eigenvalue defines the director \ojoo{$\boldsymbol{n}$ and the other one, the direction perpendicular to the director}. Moreover, by comparing the eigenvalues with the expected ones for the ideal nematic case (Eq. (\ref{Q_nematic})), we calculate the nematic order parameter $q_2$ from:
\begin{align}
\llbracket \ojoo{\mathbf{Q}^{\prime}} \rrbracket = 
\begin{bmatrix}
\lambda_1 & 0 \\
0 & -\lambda_1
\end{bmatrix} = q_2
\begin{bmatrix}
\frac{1}{2} & 0 \\
0 & -\frac{1}{2}
\end{bmatrix}
\label{q2_parameter}
\end{align}

\par In the following, the scalar parameter $q_2$ will be used to quantify long-range (global) nematic ordering, $0 \leq q_2 \leq 1$ with 0 and 1 corresponding to the limits of isotropic and perfectly aligned, nematic ordering, respectively.

\par \ojoo{Alternatively, the scalar parameter $q_2$, which is related to the nematic order parameter usually denoted as $S_2$ \cite{RN2056} and calculated by the 2\textsuperscript{nd} order Lengendre polynomial, could be derived directly from tensor $\mathbf{Q}$, without the need of diagonalizing the tensor.}
\begin{align}
    \ojoo{ q_2 = \sqrt{\frac{1}{2} \mathrm{tr}(\mathbf{Q} \cdot \mathbf{Q})} }
    \label{q2_alternative}
\end{align}

\par To calculate the tetratic order parameter, $q_4$, we adopt the formalism of the fourth order tetratic tensor \cite{RN2077}:
\begin{align}
\ojoo{
\mathbf{T} = 4 \left[ \left\langle \boldsymbol{u} \, \boldsymbol{u} \, \boldsymbol{u} \, \boldsymbol{u} \right\rangle - \mathbf{I} \right]
}
\label{T_tetratic}
\end{align}
where \ojoo{the components of the tensor $\mathbf{T}$ are given by:}
\begin{align}
\ojoo{
    \mathrm{T}_{ijkl} = 4 \left[ \left\langle u_i\,u_j\,u_k\,u_l \right\rangle - \frac{1}{8}\left( \delta_{ij}\,\delta_{kl} + \delta_{ik}\,\delta_{jl} + \delta_{il}\,\delta_{jk} \right) \right]
}
\label{I_equation}
\end{align}

\par The tetratic order tensor, given in Eq. \ref{T_tetratic} has major and minor symmetries: \ojoo{$\mathrm{T}_{ijkl}=\mathrm{T}_{jikl}=\mathrm{T}_{ijlk}=\mathrm{T}_{klij}$}. The \ojoo{4\textsuperscript{th}} order tensor components are represented in a 4x4 matrix \ojoo{$\mathbf{\undertilde{T}}$} following the steps of Ref. \cite{RN2077}, condensing both pairs of sub-indices $i$, $j$ and $k$, $l$ ($1 \leq i,j,k,l \leq 2$) of the \ojoo{$\mathbf{T}$} tensor to two sub-indices $\alpha$ and $\beta$ ($1 \leq \alpha,\beta \leq 4$) according to the scheme:
\begin{align}
\alpha, \beta = {1,2,3,4} \rightarrow ij, kl = {11, 12, 21, 22}
\label{condensation_scheme}
\end{align}

\par The \ojoo{$\mathbf{\undertilde{T}}$} matrix is subsequently diagonalized and its eigenvalues are related to the order parameters $q_2$ and $q_4$ according to \cite{RN2077}:
\begin{align} 
\begin{split}
m_1 = \frac{1}{2} \left[q_4 + \left(15 q_2^2 + q_4^2 \right)^{\frac{1}{2}} \right] \\
m_2 = 0 \\
m_3 = -q_4 \\
m_4 = \frac{1}{2} \left[q_4 - \left(15 q_2^2 + q_4^2 \right)^{\frac{1}{2}} \right]
\end{split}
\label{q4_equation}
\end{align}
\ojoo{As it is shown in Eq. \ref{q4_equation}, the sum of the eigenvalues of $\mathbf{\undertilde{T}}$ matrix is zero.}

\par The nematic order parameter, $q_2$ is derived from the eigenvalues of the \ojoo{$\mathbf{Q}$} tensor, while the eigenvalues of the \ojoo{$\mathbf{\undertilde{T}}$} matrix are used to obtain the tetratic order parameter, $q_4$ ($0 \leq q_4 \leq 1$)\ojoo{, either from the third eigenvalue or from the combination of first and fourth eigenvalues: $q_4=-m_3$ or from $q_4=m_1+m_4$.}
\ojoo{The tetratic order parameter $q_4$ should not be confused with the fourth order nematic order parameter, it is an additional metric to count for the appearance of tetratic phase.}
As discussed earlier, a nematic order parameter close to unity is strongly indicative of a nematic phase. In fact, the ideal nematic case, where all chains are oriented parallel to the director,  corresponds to both $q_2 = q_4 = 1$. The tetratic order parameter, on its own, is not sufficient to identify a tetratic phase and thus requires the nematic order parameter to be computed. The ideal tetratic phase, where chains show two perpendicular alignments, corresponds to the limit of $q_2 = 0$ and $q_4 = 1$. Finally, the isotropic state of random chain alignment corresponds to $q_2=q_4=0$.

\section{Results}

\subsection{Packing Ability}

\begin{figure*}[ht]
\centering
\includegraphics[scale=0.45]{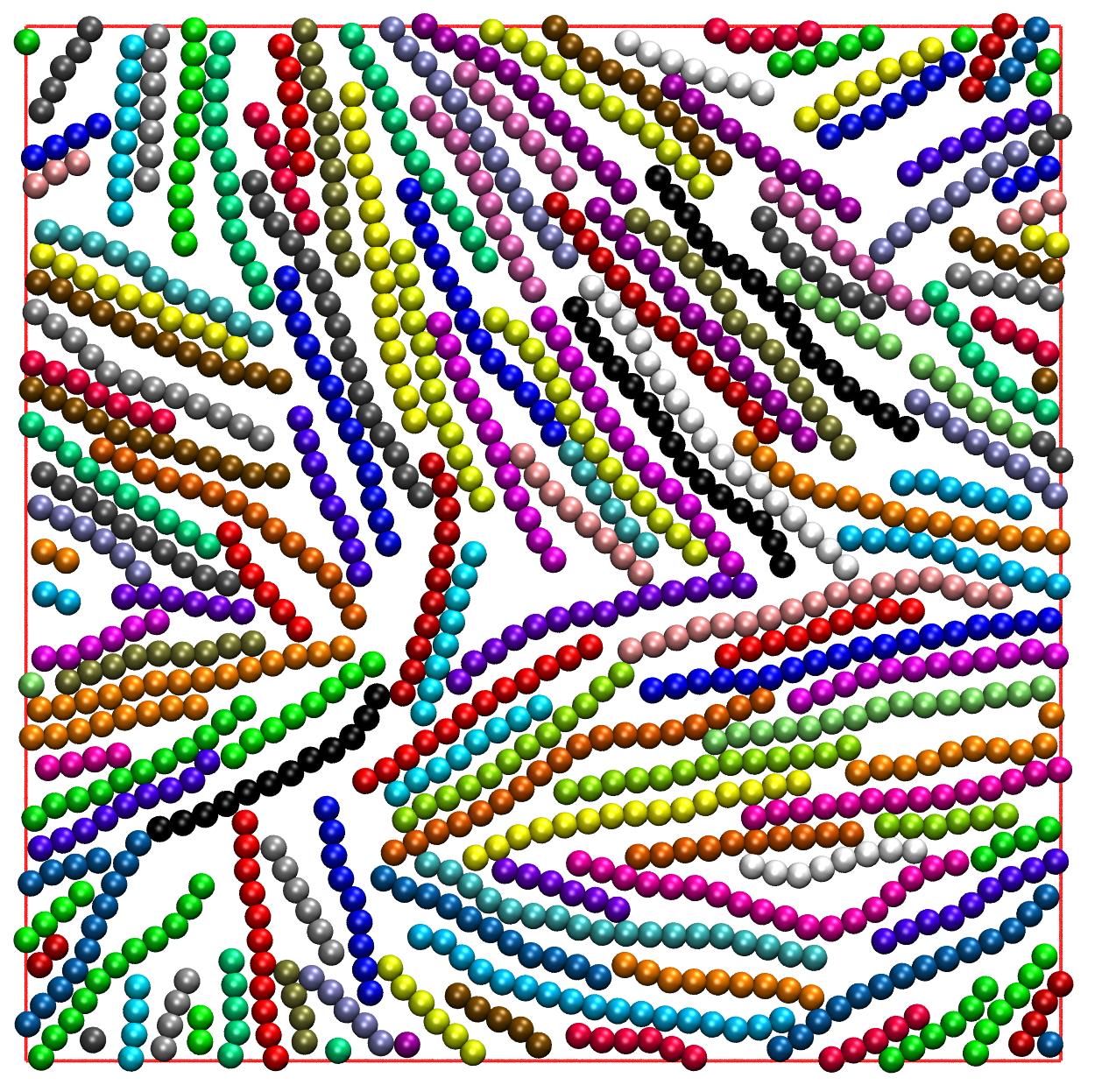}
\includegraphics[scale=0.45]{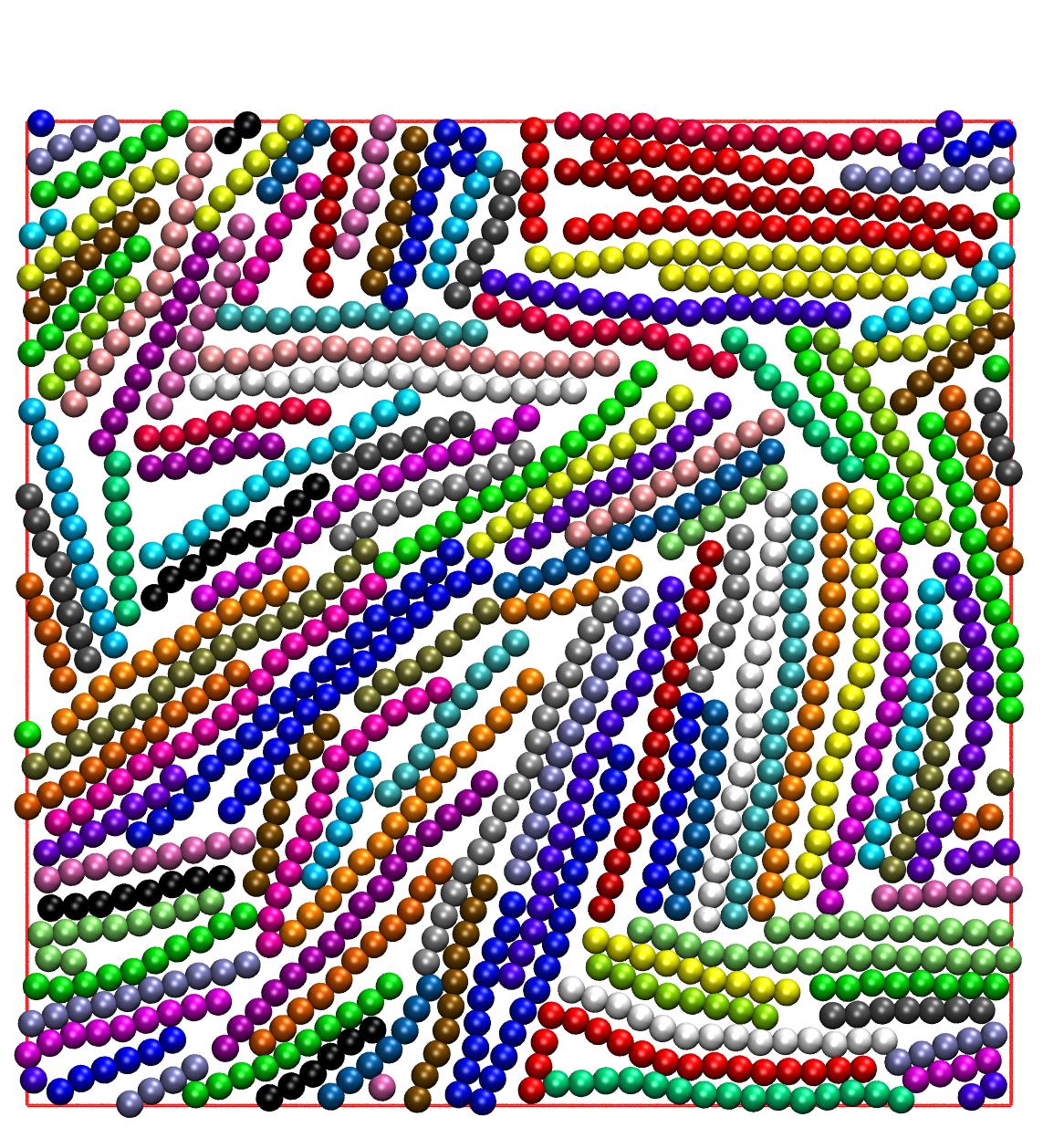}
\includegraphics[scale=0.45]{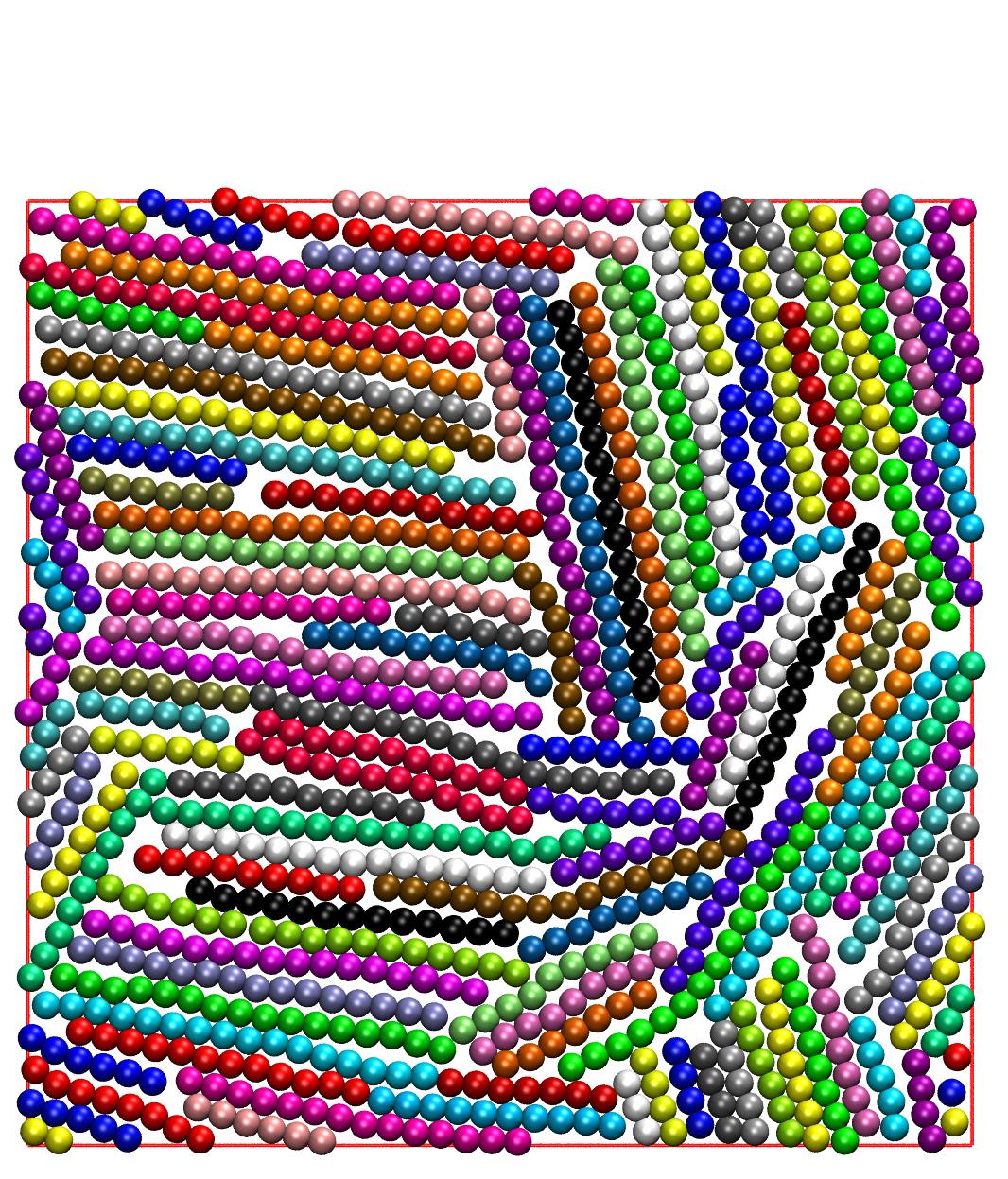}
\includegraphics[scale=0.45]{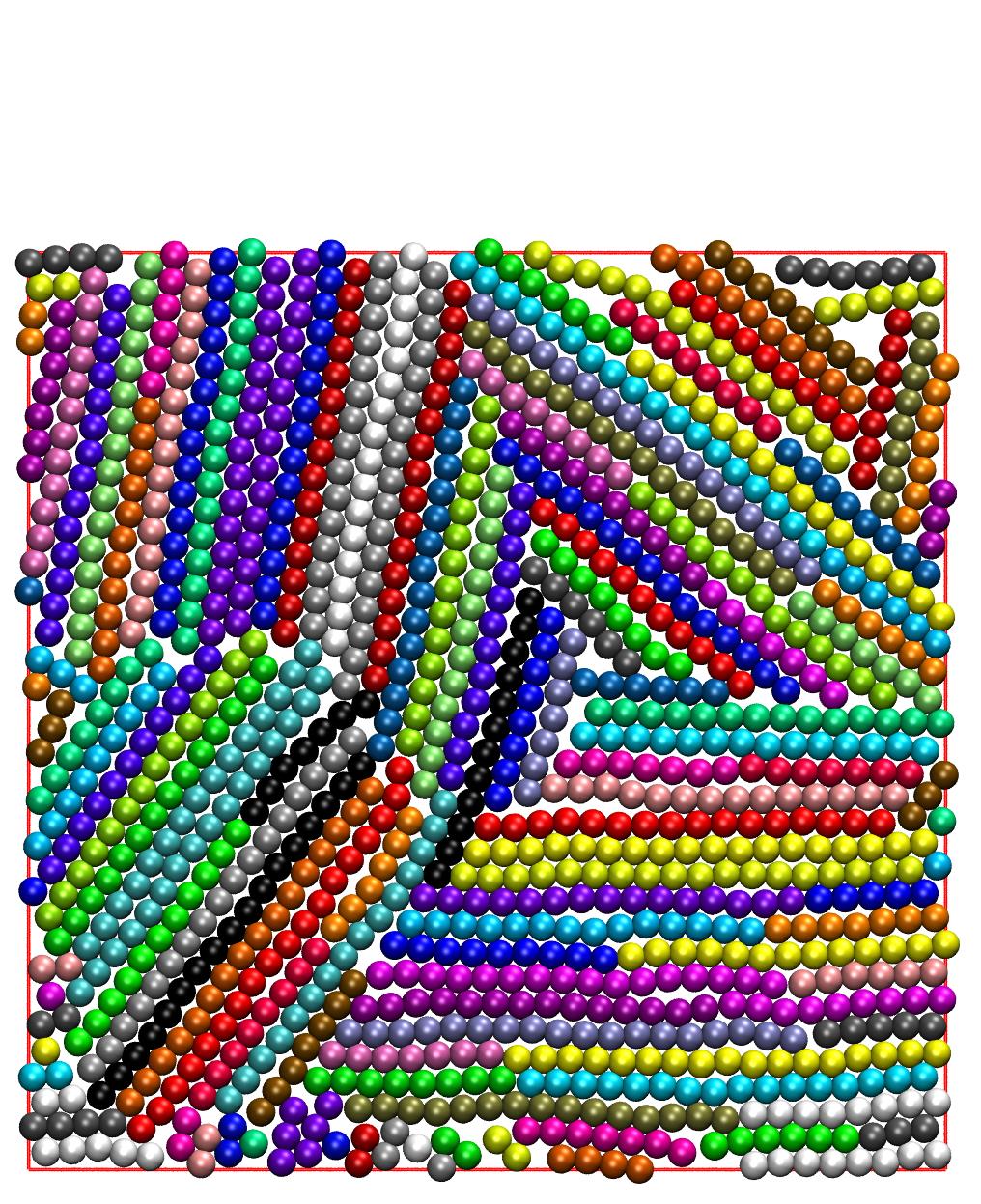}
\centering
\includegraphics[scale=0.45]{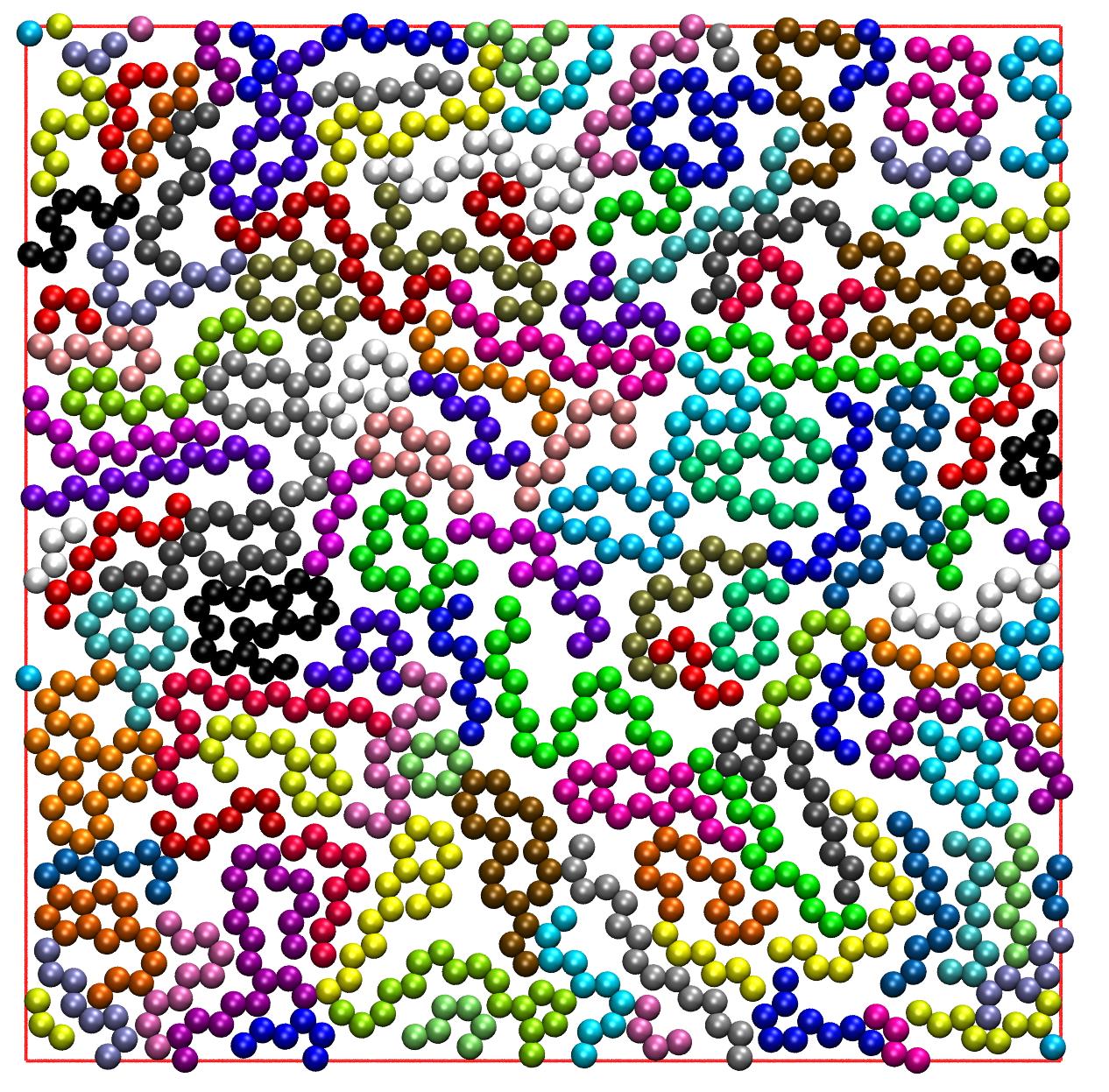}
\includegraphics[scale=0.45]{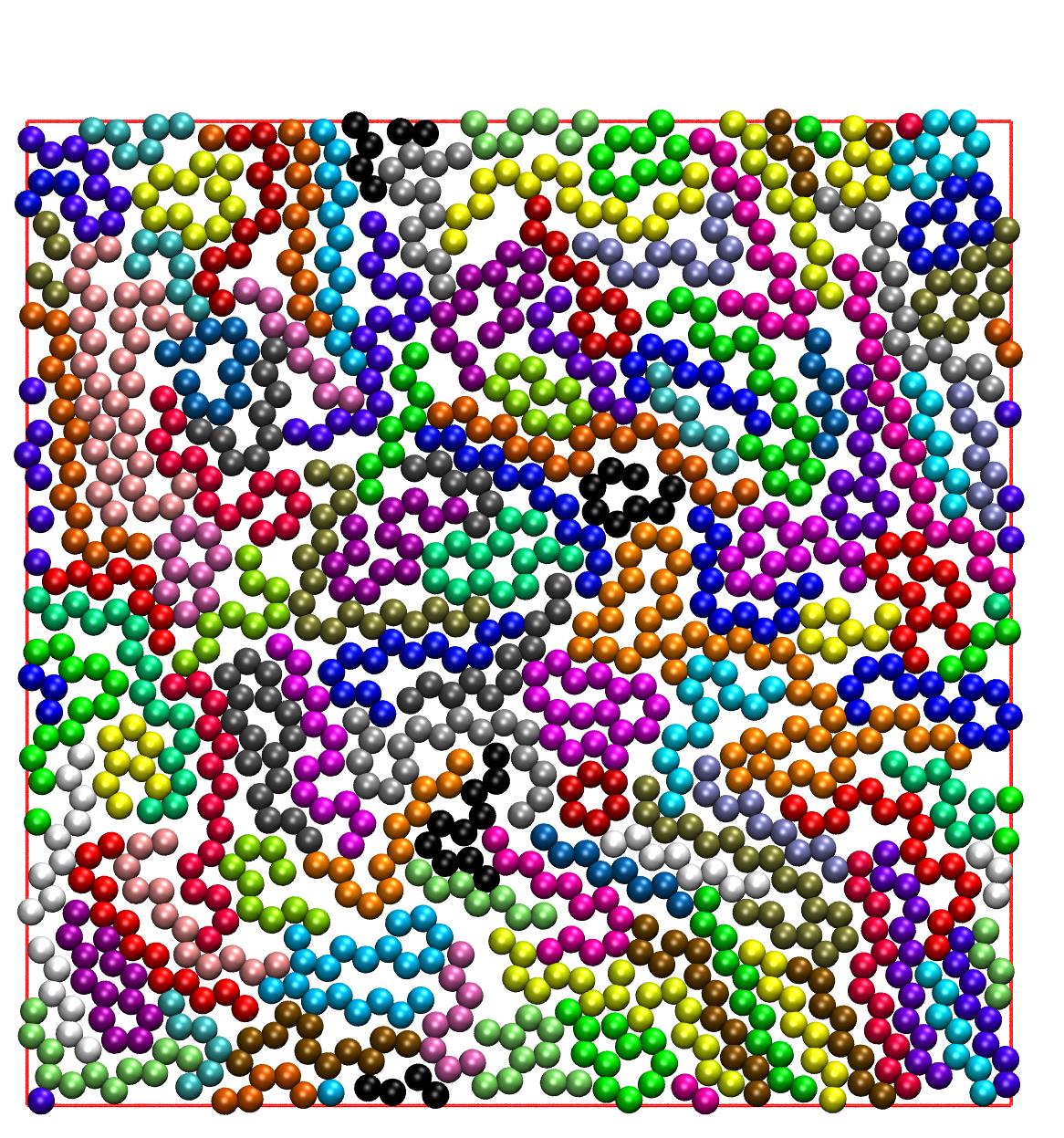}
\includegraphics[scale=0.45]{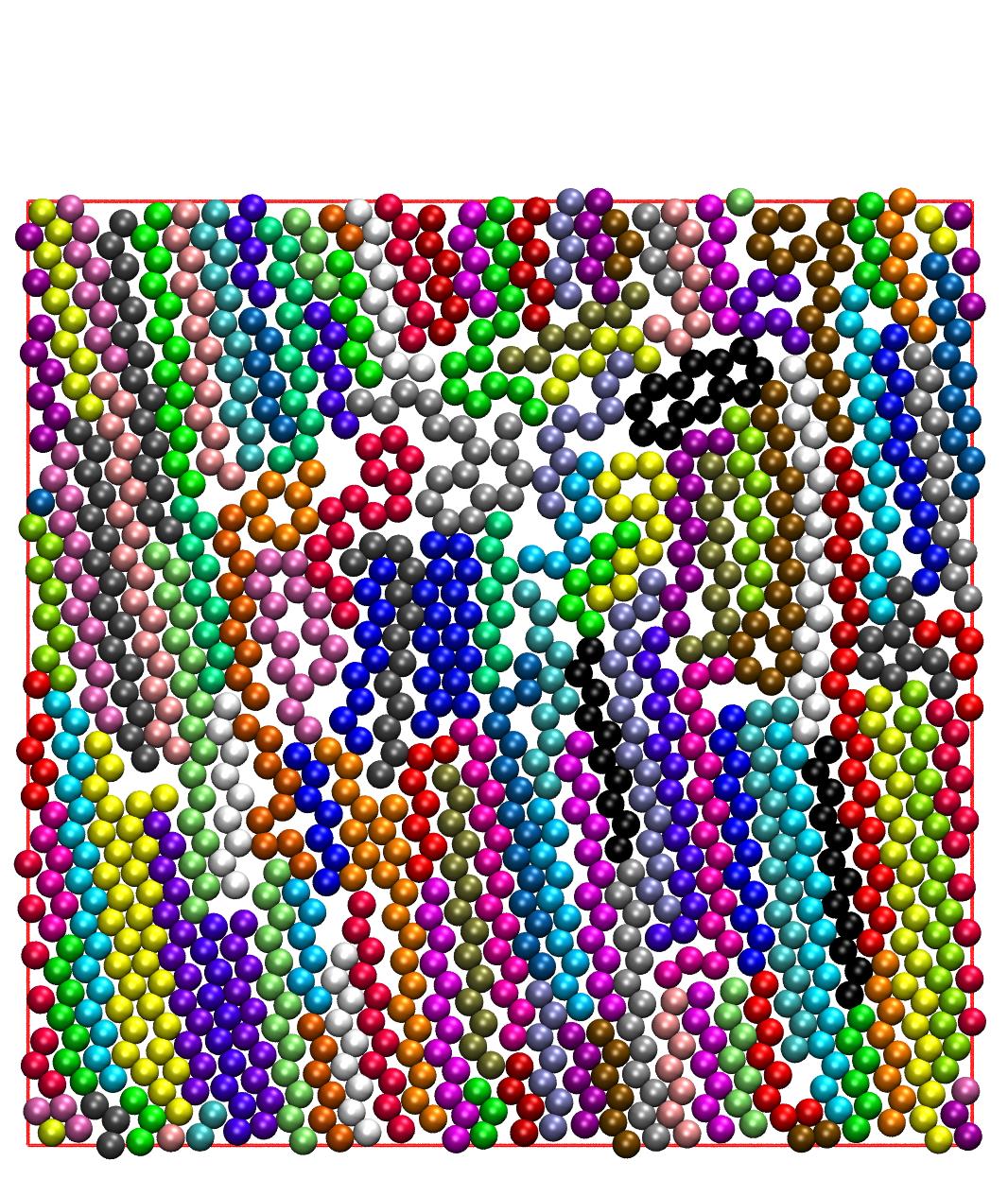}
\includegraphics[scale=0.45]{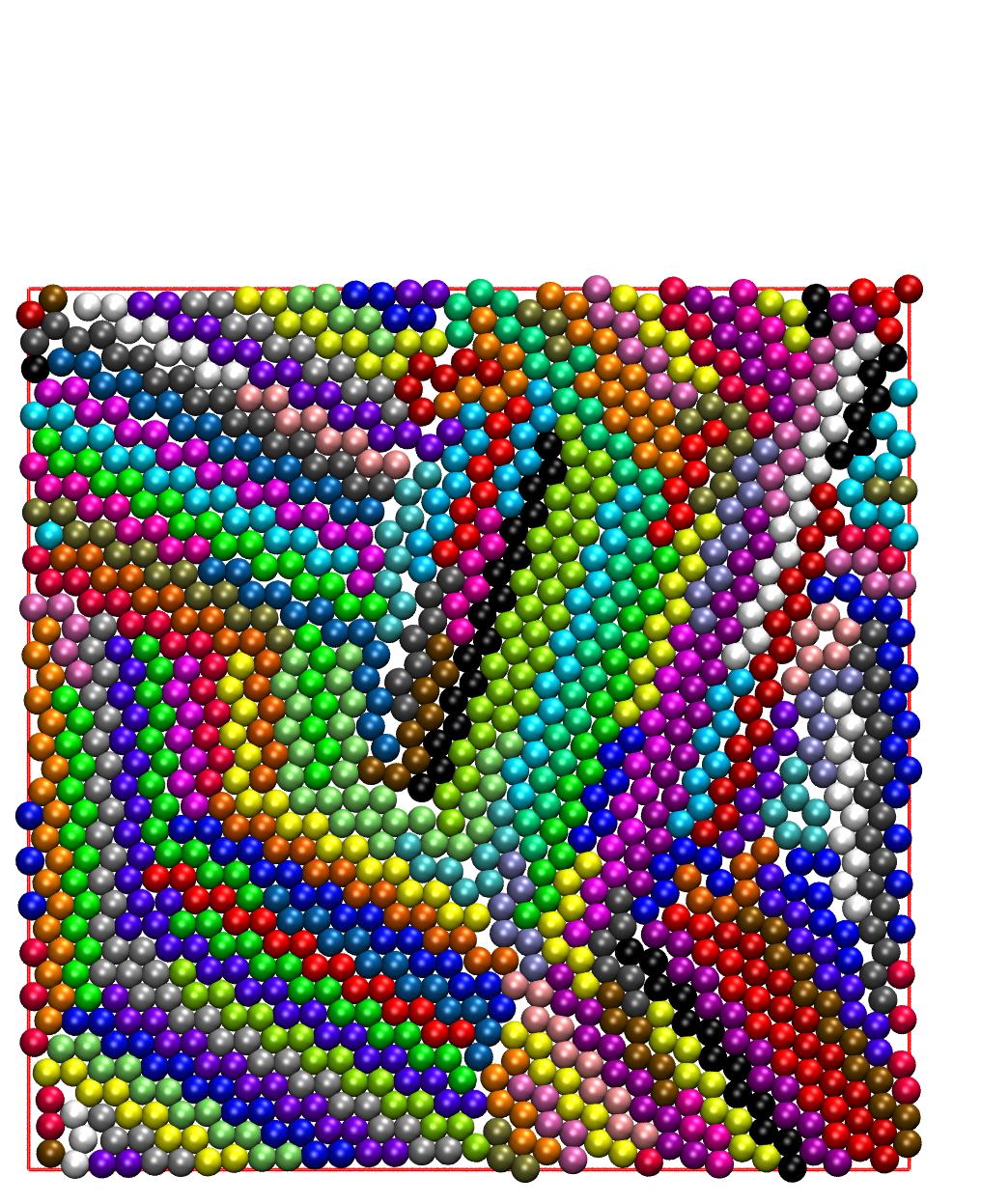}
\centering
\includegraphics[scale=0.45]{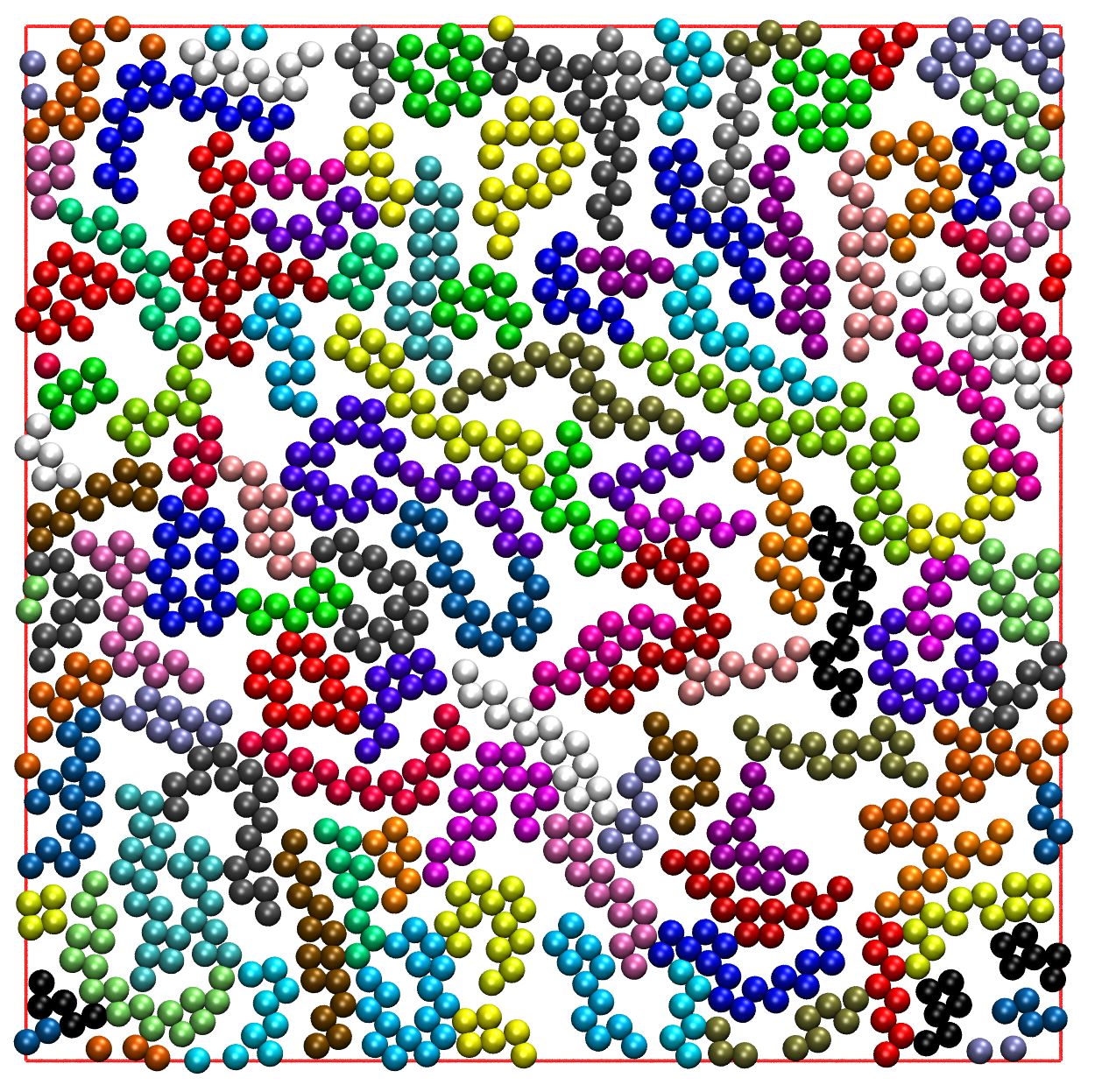}
\includegraphics[scale=0.45]{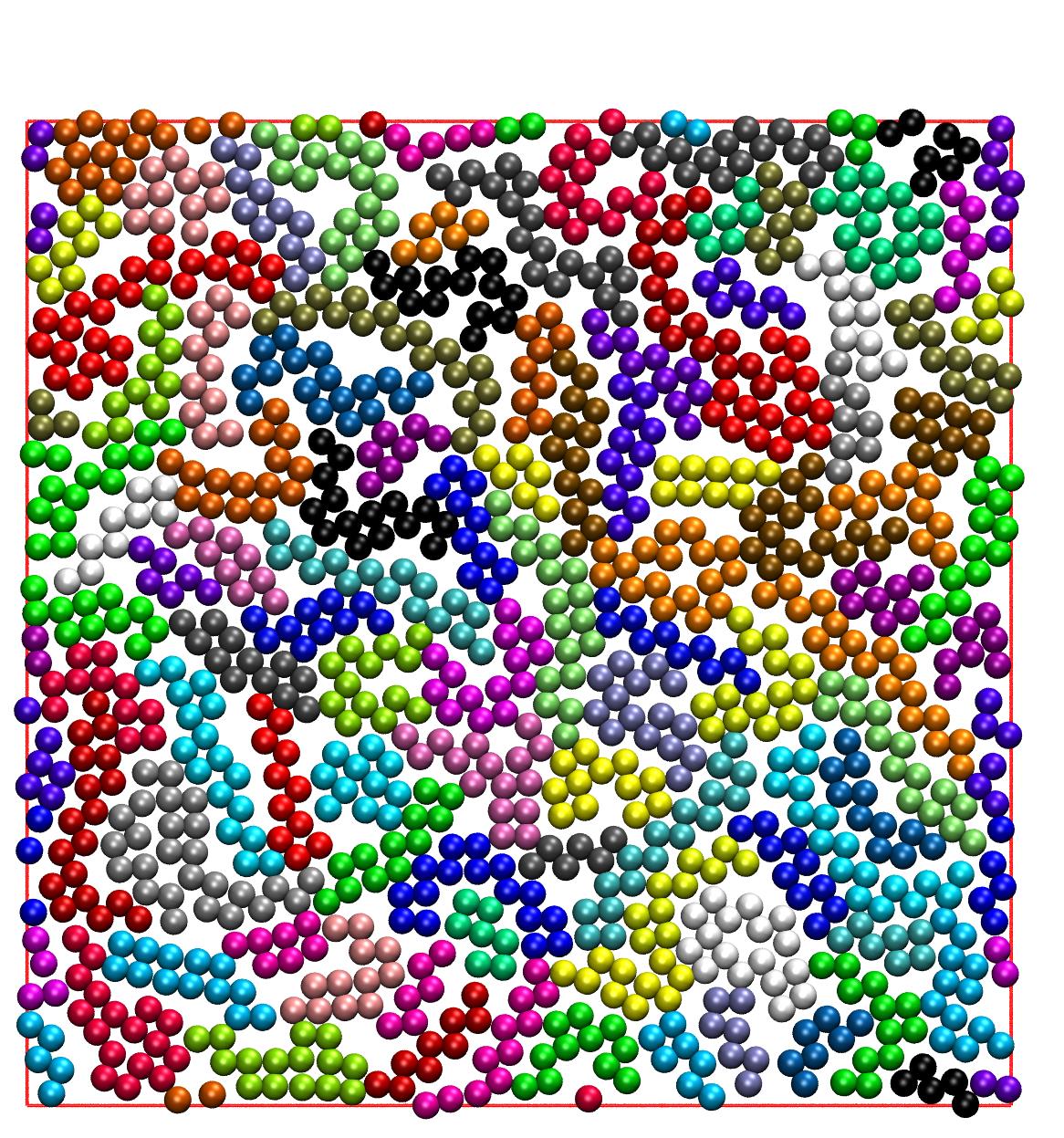}
\includegraphics[scale=0.45]{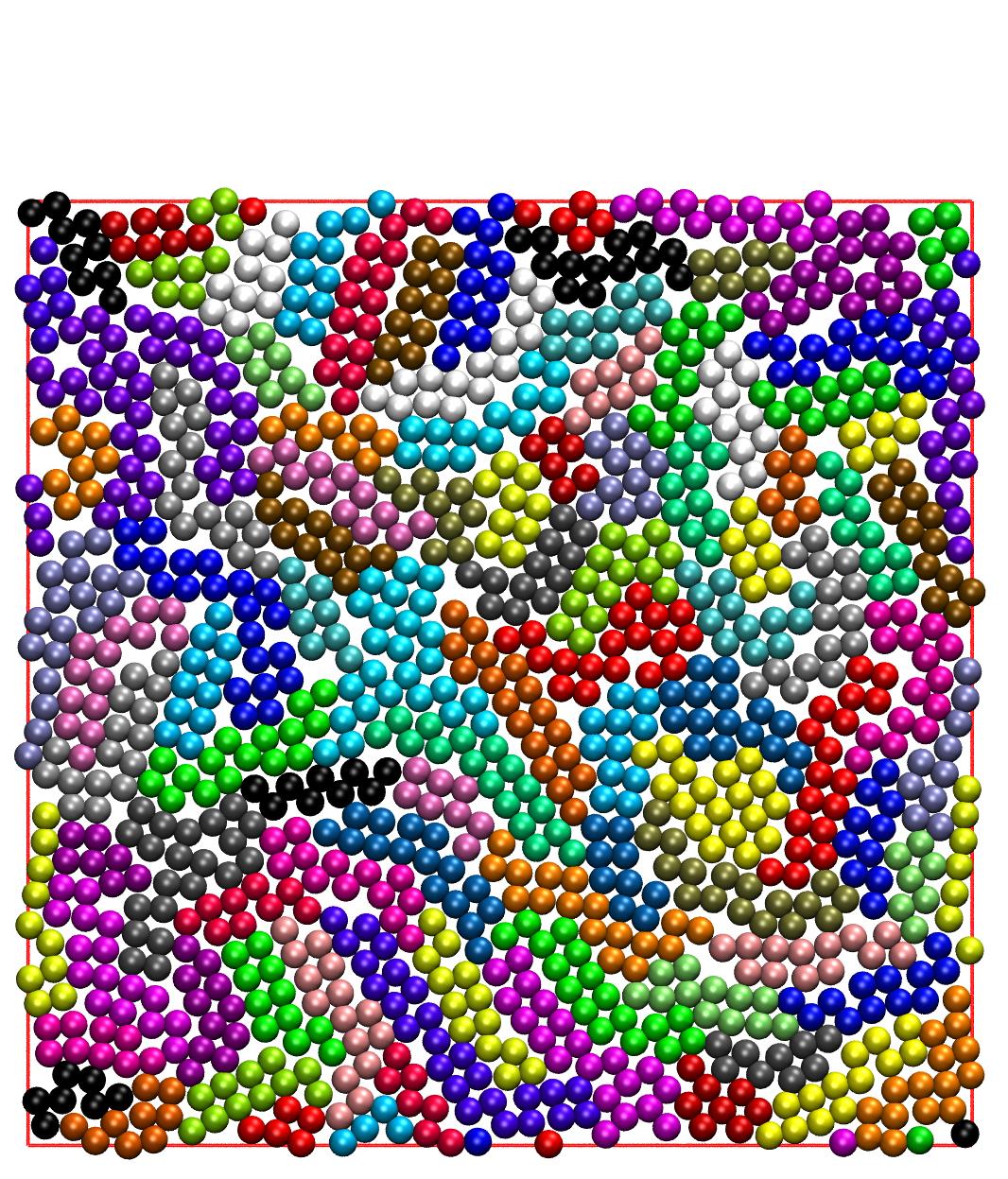}
\includegraphics[scale=0.45]{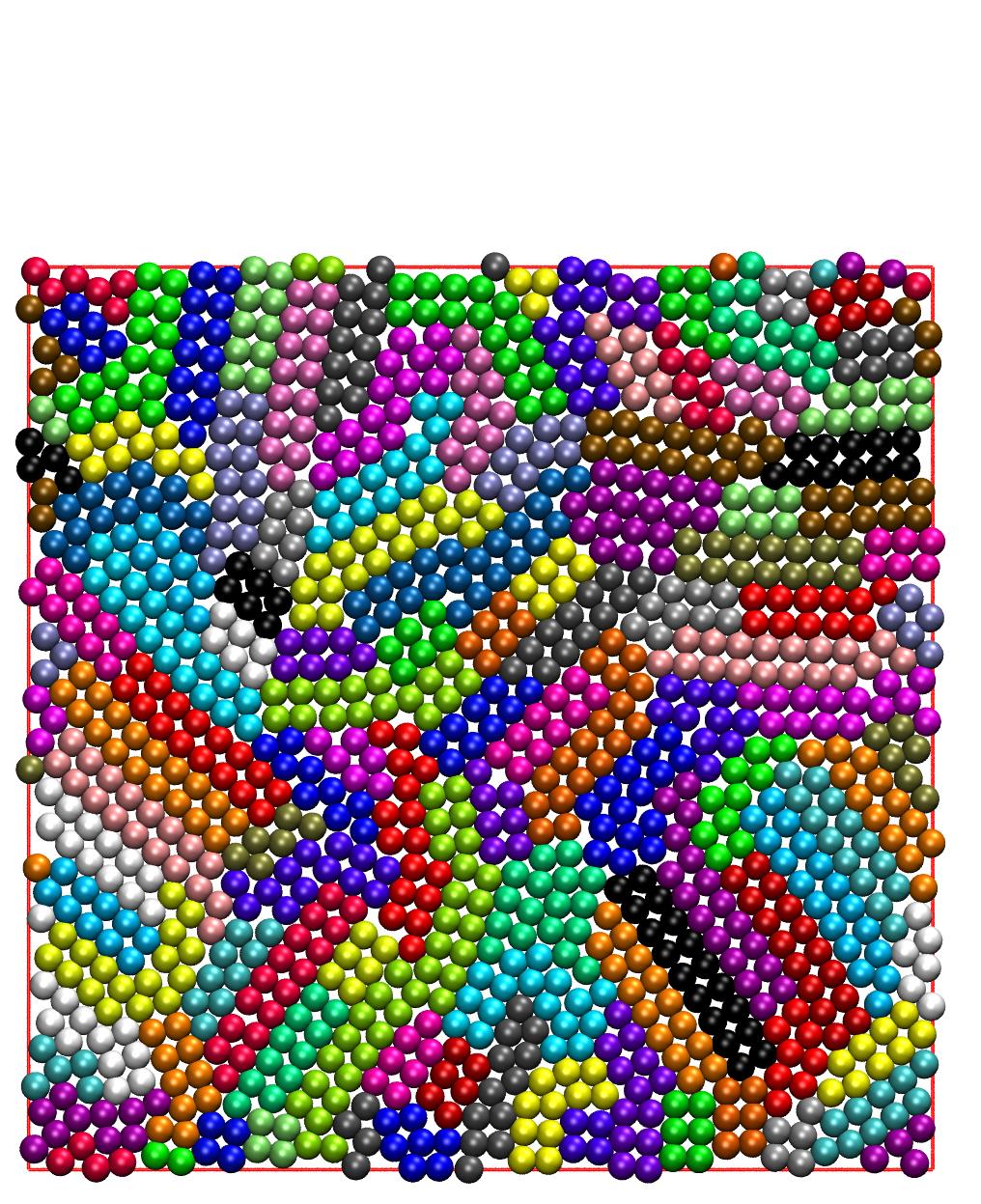}
\centering
\includegraphics[scale=0.45]{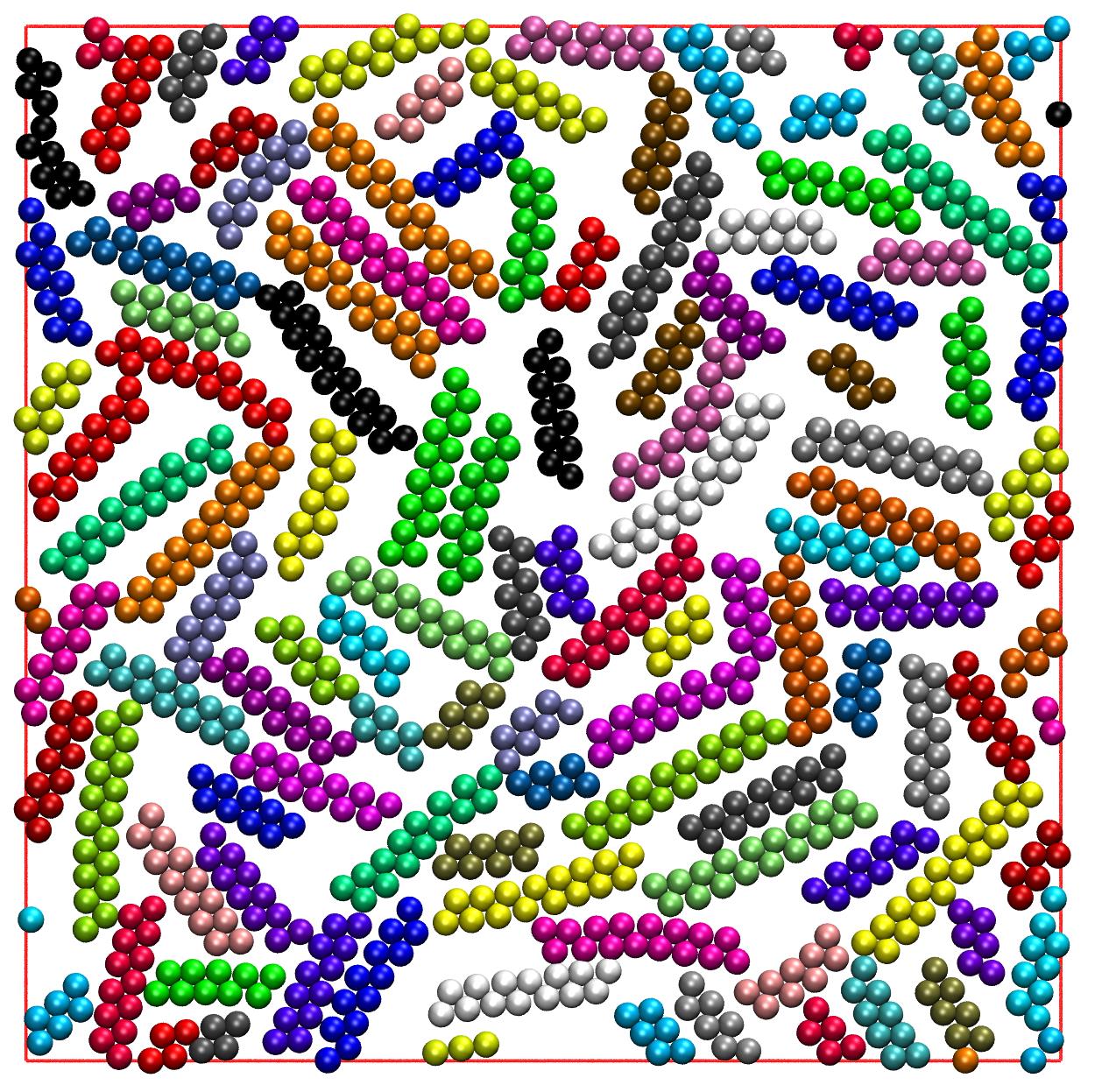}
\includegraphics[scale=0.45]{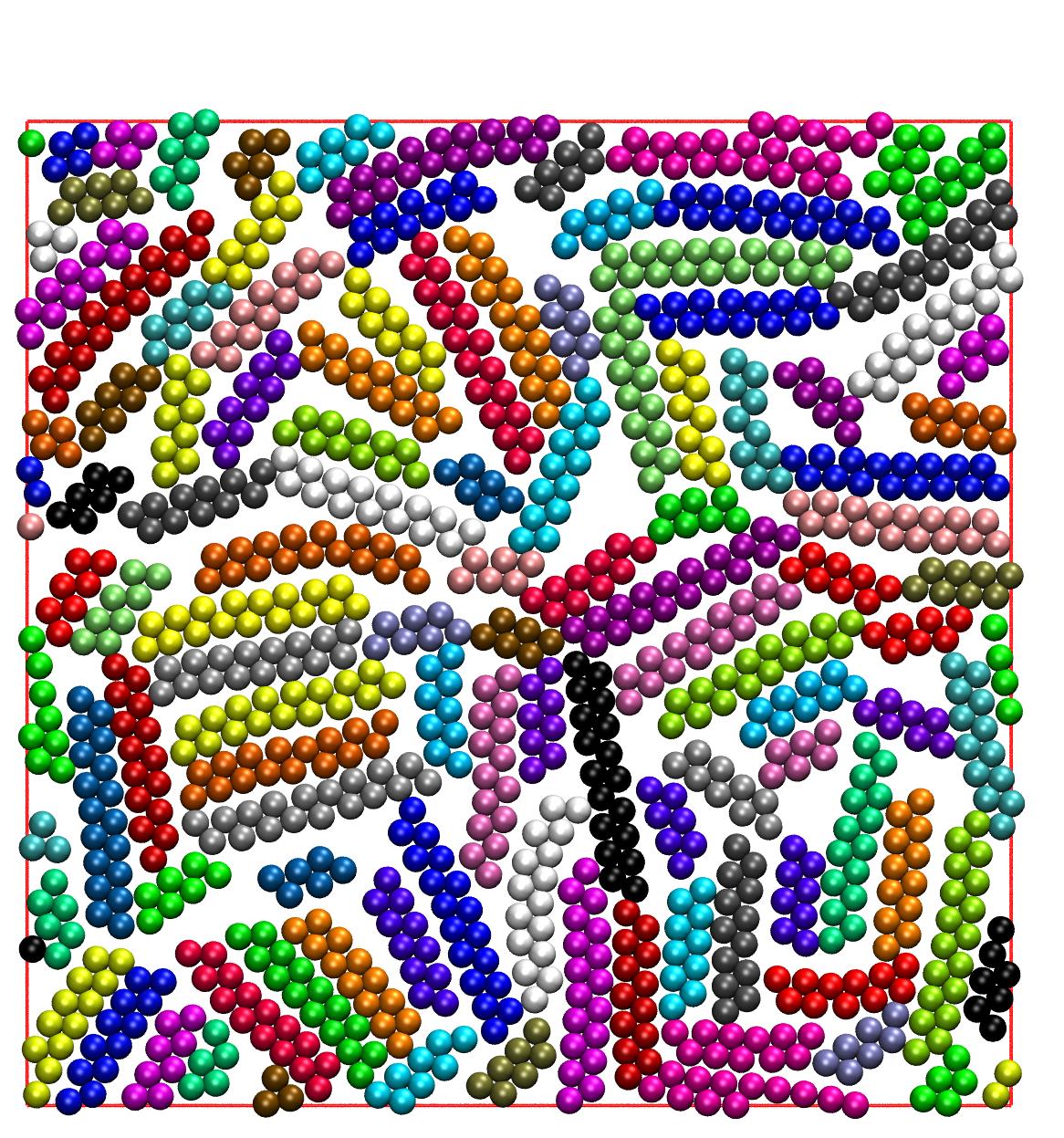}
\includegraphics[scale=0.45]{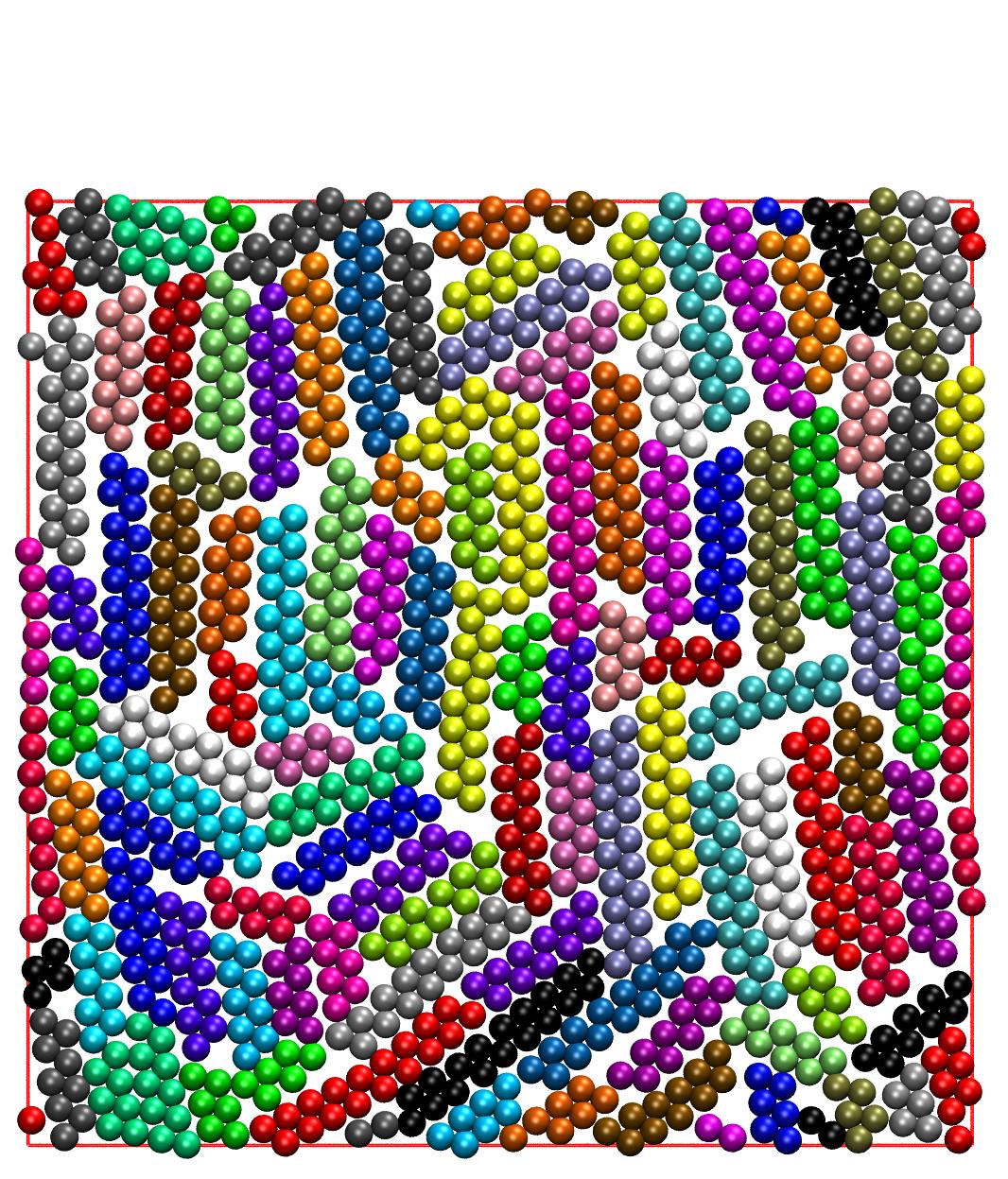}
\includegraphics[scale=0.45]{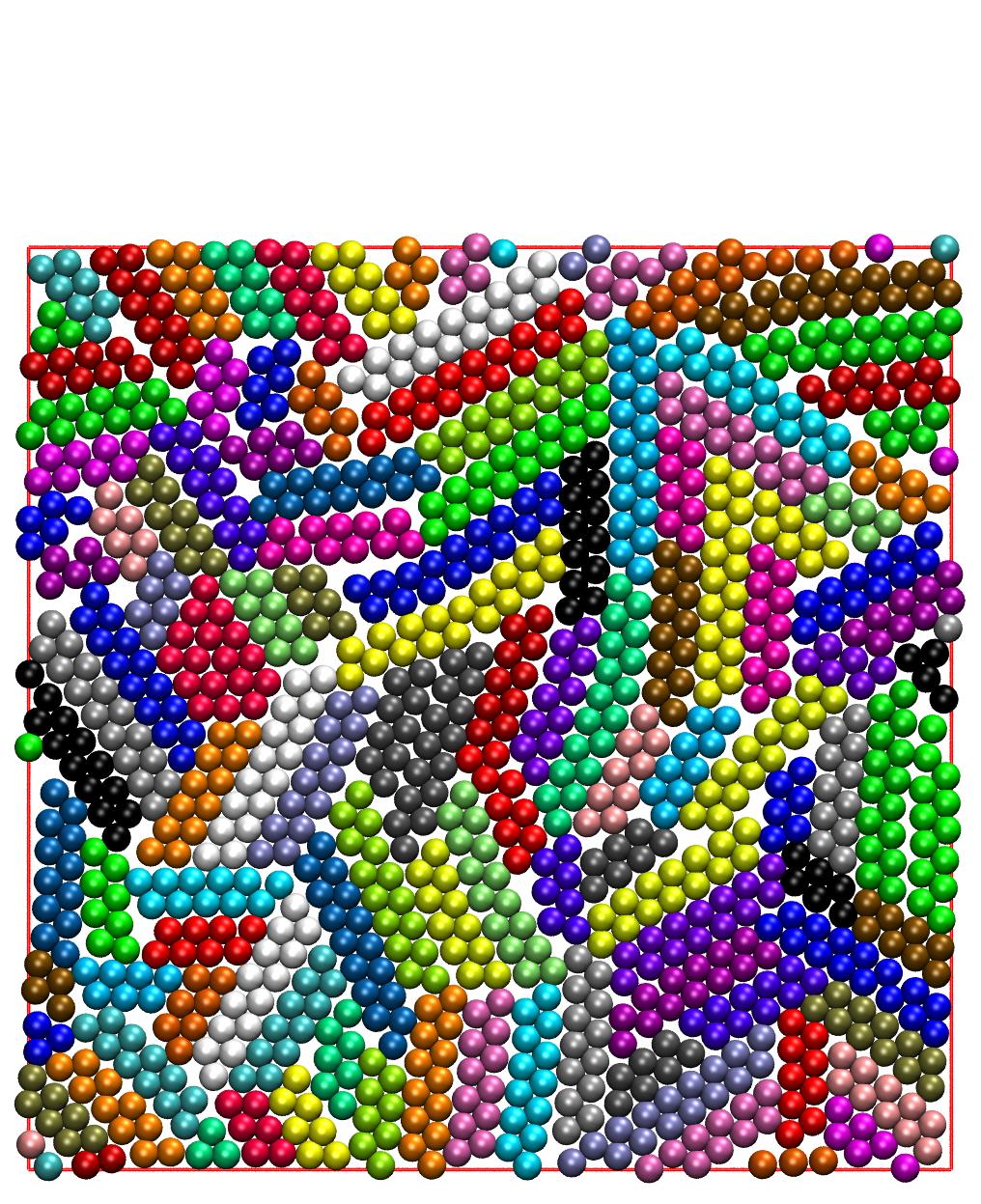}
\caption{Snapshots at the end of the constant-volume MC simulations on polymer packings in monolayers for different equilibrium bending angles, $\theta_{0}$, and at progressively higher surface coverage, $\varphi^{*}$. From top to bottom: $\theta_0 = 0$, 60, 90 and $120^{\circ}$. From left to right: $\varphi^* = 0.50$, 0.60, 0.70 and $\varphi^{*,RCP}_{2D}(\theta_0)$. The maximum achieved surface coverages can be consulted in Table \ref{table_summary}. Monomers are colored according to the parent chains and the coordinates of their centers are subjected to periodic boundary conditions in the two long dimensions of the simulation cell. Image created with the VMD software \cite{RN250}.}
\label{MC_snapshots}
\end{figure*}

Snapshots at the end of the constant-volume, production MC simulations for all different equilibrium bending angles and at progressively higher surface coverage  (packing density) are displayed in Fig. \ref{MC_snapshots}, where monomers are colored according to the parent chain and with the coordinates of their centers being subjected to periodic boundary conditions in the two long dimensions of the simulation cell. 

\par Table \ref{table_summary} summarizes the maximum achievable coverage, $\varphi^{*,RCP}_{2D}(\theta_{0})$, and packing density, $\varphi^{RCP}_{2D}(\theta_{0})$, the ratio  $n = \varphi^{*,RCP}(\theta_{0}) / \varphi^{*,max}_{2D}$, average crystallinity, $\tau^c(\theta_{0})$, and long-range nematic and tetratic order parameters, $q_{2}(\theta_{0})$ and $q_{4}(\theta_{0})$, corresponding to the RCP limit for a given equilibrium bending angle, $\theta_{0}$, in two dimensions. As a reference the surface coverage for the ideal TRI, SQU and HON crystals is approximately 0.907, 0.785 and 0.604, respectively. The latter is too dilute compared to the surface coverage achieved here over the whole range of equilibrium bending angles so that the HON crystal, even in the form of isolated ordered sites, is very rarely encountered.
\par From the data on surface coverage the following trend is established: $\varphi^{*,RCP}(FJ) > \varphi^{*,RCP}(60^{\circ}) > \varphi^{*,RCP}(90^{\circ}) > \varphi^{*,RCP}(0^{\circ}) > \varphi^{*,RCP}(120^{\circ})$. The fully flexible chains can be packed in the most efficient pattern, whose density lies within 1.4\% of the densest limit of the perfect TRI crystal. The least dense RCP limit corresponds to $\theta_{0} =120^{\circ}$. This is somewhat surprising at first sight given that the $\theta_{0} = 90^{\circ}$ is the only angle not encountered between the nearest neighbor sites in the TRI crystal. Given that i) the average chain length studied here is $N_{av} = 12$, ii) bonded spheres are tangent and thus by construction also nearest neighbors and iii) chain ends are free of bending constraints, approximately 83.3\% of the spheres of the right-angle systems have, at least, two nearest neighbors whose bending conformation is not compatible with the geometry of the TRI crystal.
\par Thus, this geometric incompatibility for $\theta_{0} = 90^{\circ}$ frustrates the formation of the TRI crystal in a similar fashion as the formation of close packed HCP and FCC crystallites in three dimensions can be hindered by bond tangency near the melting point \cite{RN319,RN497}. Hence, it is expected that the $\theta_{0} = 90^{\circ}$ chains cannot achieve surface coverages as high as polymers whose bending angles are geometrically compatible with TRI. In parallel, one should consider that inter-molecular alignment of chains (see discussion on long-range order) further affect the packing ability of semi-flexible polymers. This statement is also relevant in the case of the right-angle chains. The RCP limit of $\theta_{0} = 90^{\circ}$, as calculated from the present MC simulations ($\varphi^{*}(90^{\circ}) \approx 0.797$), is higher than the surface coverage of the ideal SQU crystal ($\varphi^{*}(SQU) \approx 0.785$). Accordingly, the expected sphere arrangement in the monolayer should correspond to a blend of square conformations built through intra-chain arrangements, because of the imposed bending constraints, and more compact local packings achieved through inter-chain conformations. As a consequence in the densest possible state, $\theta_{0} = 90^{\circ}$ chains should maximize their external contour (their 2D non-convex hull) available for inter-chain arrangements. In doing so, chains should adopt extended conformations and the higher the concentration the more elongated the corresponding configuration. This is verified by comparing for example the snapshots of the left- and right-most panels in Fig. \ref{MC_snapshots} for $\theta_{0} = 90^{\circ}$. The interplay between inter- and intra-chain arrangements has a major impact in the local and global structure of the chains, as will be demonstrated in the continuation. 

\begin{table*}
\begin{ruledtabular}
\centering
\caption{Maximum achieved surface coverage, $\varphi^*$, packing density, $\varphi$, and reduced surface coverage, $\varphi^* / \varphi^{*,max}_{2D}$, along with the corresponding average crystallinity, $\tau^c$, and the nematic and tetratic long-range order parameter, $q_2$ and $q_4$, as a function of equilibrium bending angle, $\theta_0$, at the RCP limit. Also reported are corresponding data for the freely-jointed (FJ) chains after Ref. \cite{RN2034}.}
\begin{tabular}{c c c c c c c} 
 $\theta_0$ & $\varphi^*$ & $\varphi$ & $\varphi^* / \varphi^{*,max}_{2D}$ & $\tau^{c}$ & $q_2$ & $q_4$ \\ [0.5ex] 
 \hline
 $0^{\circ}$   & 0.774 & 0.516 & 0.853 & 0.723 & 0.113  & 0.517 \rule{0pt}{2.5ex}\\ 
 $60^{\circ}$  & 0.839 & 0.559 & 0.925 & 0.867 & 0.358  & 0.890 \\
 $90^{\circ}$  & 0.797 & 0.532 & 0.879 & 0.130 & 0.0517 & 0.113 \\
 $120^{\circ}$ & 0.764 & 0.510 & 0.842 & 0.562 & 0.144  & 0.343 \\
 FJ            & 0.895 & 0.597 & 0.987 & 0.945 & 0.396  & 0.187
 
\end{tabular}
\label{table_summary}
\end{ruledtabular}
\end{table*}

\subsection{Polymer Size}

\begin{figure}
\centering
\includegraphics[scale=0.40]{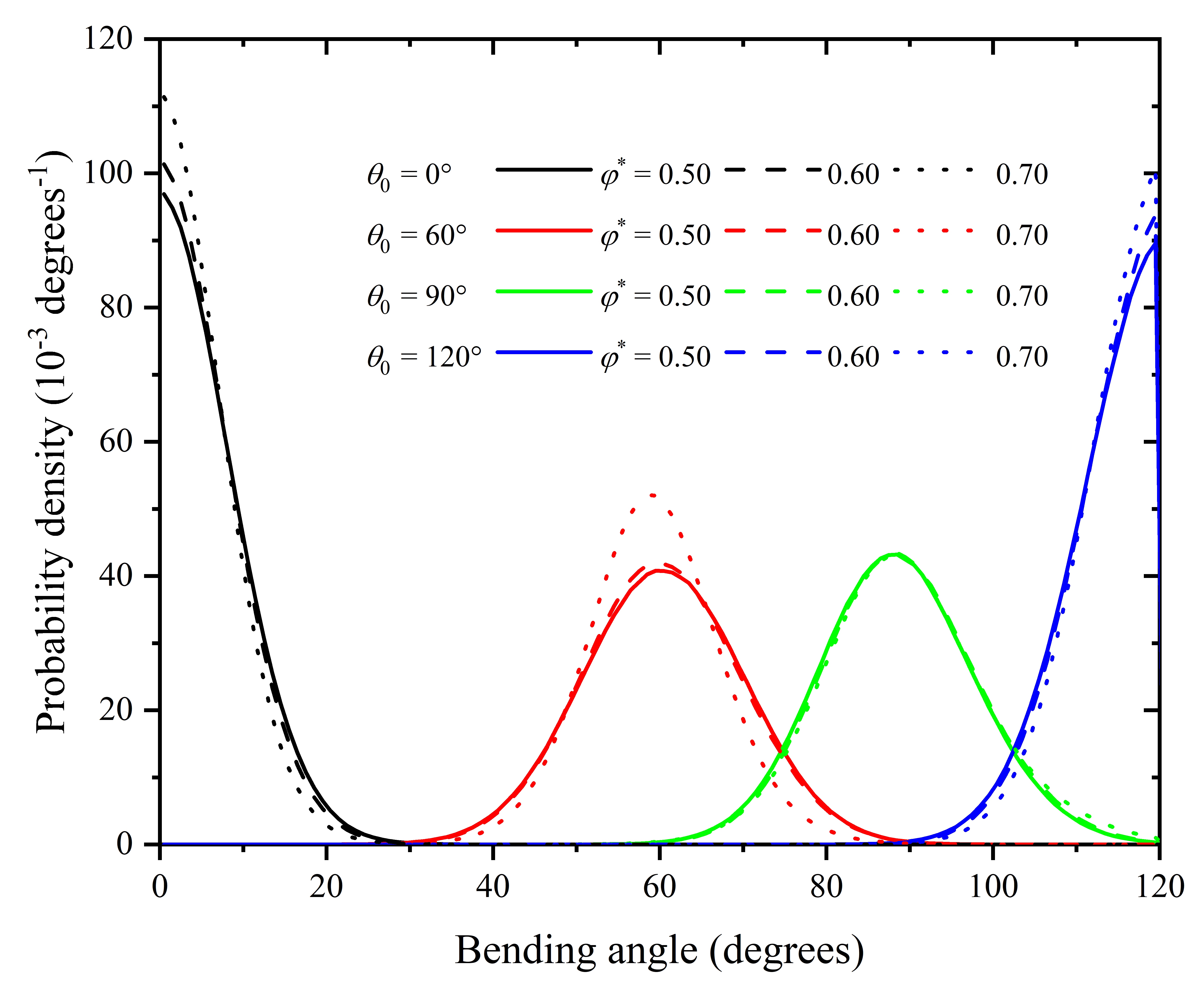}
\caption{Bending angle distribution at surface coverage of $\phi^* = 0.50$, 0.60 and 0.70 for different equilibrium bending angles, $\theta_0$. In all cases $k_{\theta} / k_{B}T= 9$ rad$^{-2}$.}
\label{bending_distr}
\end{figure}

Fig. \ref{bending_distr} shows the distribution of bending angles at various surface coverages. The harmonic spring constant of $k_{\theta} / k_{B}T= 9$ rad$^{-2}$ allows for the bending angles to thermally fluctuate in an interval of approximately $\pm 20^{\circ}$ around the equilibrium angle, as in \cite{RN695,RN1339} and in our \cite{RN2010} past work on the crystallization of athermal, semi-flexible polymer chains in three dimensions. As shown in Fig. \ref{bending_distr}, the distribution becomes progressively narrower from $\varphi^* = 0.50$ to 0.70 for all equilibrium angles except for $90^{\circ}$ where the distribution remains practically unaffected by the change in the concentration. 

\par For semi-flexible polymers their size and shape are both dominated by the constraints imposed by chain connectivity and to a lesser degree by packing density. Such constraints are absent in freely-jointed chains and thus their size only depends on volume fraction. Accordingly, for fully flexible athermal polymers four distinct scaling regimes can be identified on the dependence of chain size on volume fraction from dilute to nearly jammed packings \cite{RN13,RN12}. Bond lengths are  kept at $l = \sigma$ within a very small numerical tolerance, while the formation of a single layer, whose thickness is equal to the monomer diameter, forces all groups of four consecutive bonds to adopt planar configurations. The distribution of bending angles, which are dominated by the harmonic potential of Eq. \ref{Bending_Eq} has been presented in Fig. \ref{bending_distr}. This combination of chain geometry, along with surface coverage, leads to a non-trivial effect on polymer size, considering also the different equilibrium bending angles adopted here. 
\par
\begin{figure}
\centering
\includegraphics[scale=0.35]{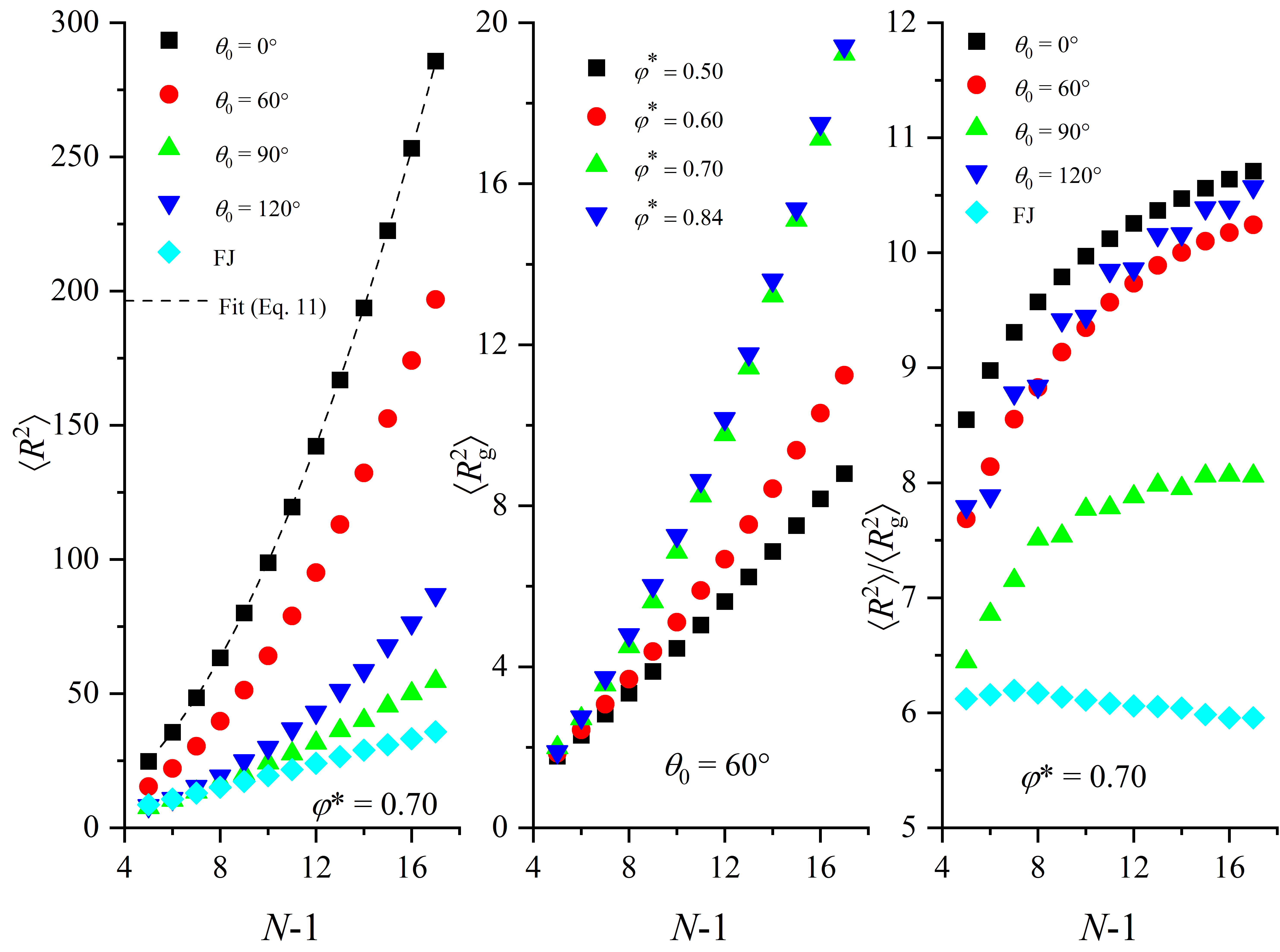}
\caption{Left panel: Mean square end-to-end distance, $\langle R^2 \rangle$, as function of number of bonds, $N-1$, for different equilibrium bending angles, $\theta_{0}$, at a surface coverage of $\varphi^* = 0.70$. Also shown with the dashed black line is a best fit using the analytic expression of the worm-like chain model (Eq. \ref{R2_wormlikechains}) on simulation data for $\theta_{0} = 0$. Middle panel: mean square radius of gyration, $\langle R^2_{g} \rangle$ as function of $N-1$ at different surface coverages for $\theta_{0} = 60^{\circ}$. Right panel: ratio $\langle R^2 \rangle / \langle R^2_{g} \rangle$ as a function of $N-1$ for different equilibrium bending angles at $\varphi^* = 0.70$.}
\label{chain_size}
\end{figure}

 Chain size is quantified through the mean square end-to-end distance, $\langle R^2 \rangle$, and the mean square radius of gyration, $\langle R^2_{g} \rangle$, where brackets denote averaging over all chains and system configurations. Furthermore, as simulations are conducted on systems with chain length dispersity and because chain-connectivity-altering moves guarantee the robust sampling of the long-range polymer conformations, the effect of chain size on chain length in the range $N \in [6, 18]$ (Fig. \ref{chain_size}) can be obtained straight from the MC frames.  Left and right panels present the dependence of $\langle R^2 \rangle$ and of the ratio $\langle R^2 \rangle / \langle R^2_g \rangle$, respectively, on the number of bonds per chain, $N-1$, at a surface coverage of $\varphi^* = 0.70$ for all equilibrium bending angles studied here, including a comparison with the freely-jointed (FJ) chains. 
\ojoo{Given that the rod-like chains simulated here correspond to very small value of $\theta_{0} \rightarrow 0$ we can fit the simulation data with the analytical predictions of the worm-like chain model \cite{RN1921} according to which \cite{RN1477,RN2118}:}

\begin{align}
\ojoo{
\langle R^2 \rangle = 2 l_p R_{max}-2l_p^2 \left[1-exp\left(-\frac{R_{max}}{l_p}\right)\right]
}
\label{R2_wormlikechains}
\end{align}

\ojoo{where $R_{max}$ is the maximum chain length ($R_{max}=(N-1)$) and $l_p$ the persistence length. In the rod-like limit ($R_{max}<< l_p$), as in here, we have $\langle R^2 \rangle \approx R_{max}^2$. Such a fitting is also shown along with the corresponding simulation data ($\theta_{0} = 0$) on the left panel of Fig. \ref{chain_size}.}

\par The middle panel shows the dependence of $\langle R^2_g \rangle$ on $N-1$ for $\theta_{0} = 60^{\circ}$ at various surface coverages. Acute bending angles of $\theta_{0} = 0$ and $60^{\circ}$ lead to more elongated chains while freely-jointed systems are the most compact. As packing density increases,  chain size increases, but there is no appreciable difference in the radius of gyration between $\varphi^* = 0.70$ and $\varphi^{*,RCP}_{2D}(60^{\circ}) \approx 0.839$. With respect to the ratio $\langle R^2 \rangle / \langle R^2_g \rangle$ the freely-jointed chain system reaches a plateau value very close to 6, followed by the right-angle system for which $\langle R^2 \rangle / \langle R^2_g \rangle \approx 8$. The remaining systems show a systematic increase of the ratio with chain length, suggesting that the simulation of longer chains is required to observe ideal polymeric behavior. For comparison, the limits of fully flexible and very stiff chains in the worm-like (Porod-Kratky) model correspond to $\langle R^2 \rangle / \langle R^2_g \rangle = 6$ and 12, respectively \cite{RN1477}.  

\par
\begin{figure}
\centering
\includegraphics[scale=0.40]{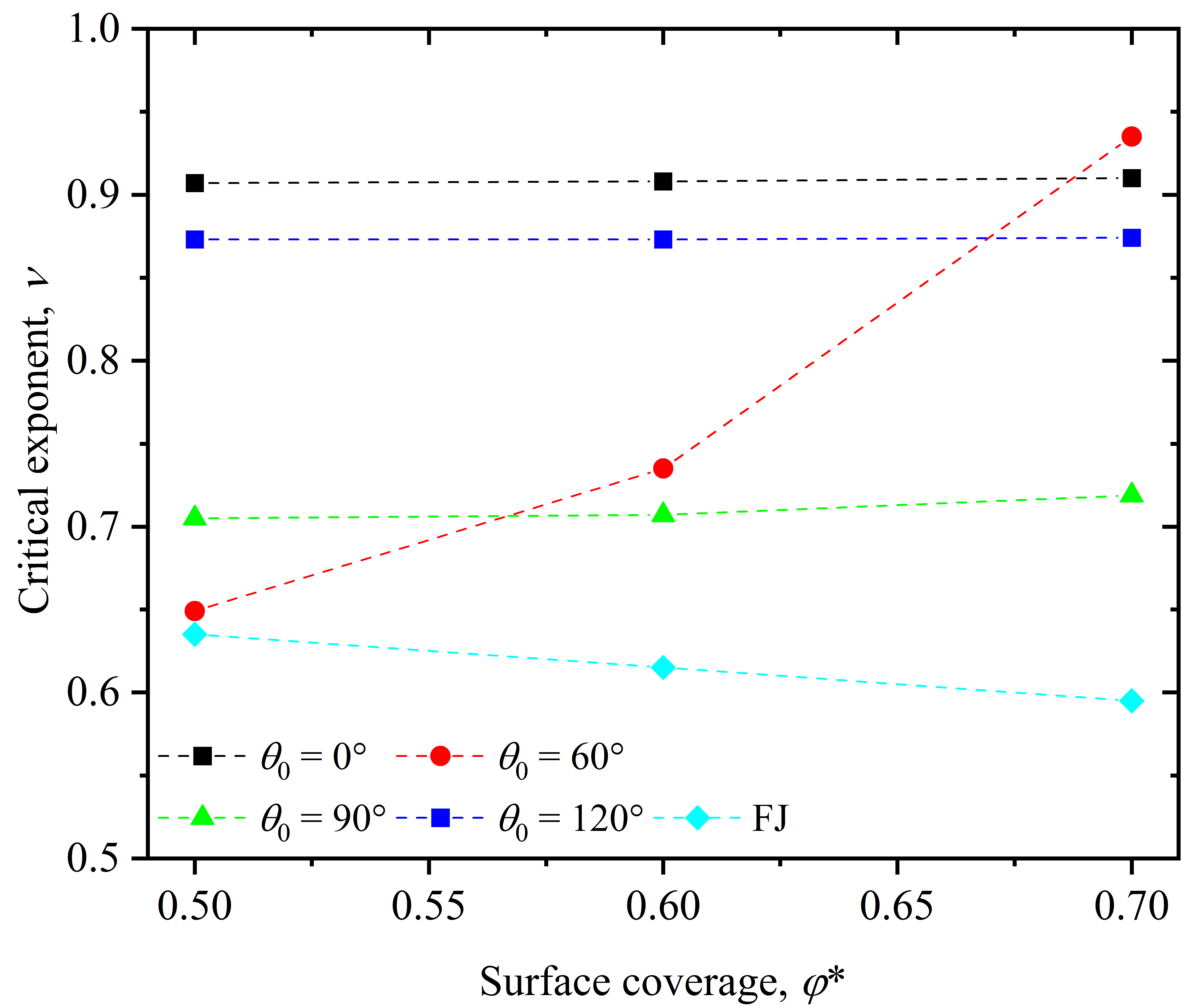}
\caption{Flory critical exponent, $v$, as a function of surface coverage, $\varphi^*$, for the different equilibrium bending angles, $\theta_0$, studied here including a comparison with the freely-jointed (FJ) polymers. Dashed lines, connecting the scattered simulation data, serve only as guide for the eye.}
\label{Flory_exponent}
\end{figure}

 One can also extract the critical Flory exponent, $v$, from the dependence of chain size on chain length, through $\langle R^2_g \rangle \propto (N-1)^{2v}$, considering the number of bonds instead of the number of monomers, given the short chains studied here. In two dimensions and considering excluded volume interactions and chain coiling $v = 0.75$ \cite{RN297}. The maximum chain extensibility corresponds to an all-trans, rod-like or zig-zag configuration according to which $R_{max} = (N-1) \cos(\theta_0/2)$, i.e.  $v = 1$. On the other limit the densely packed and fully collapsed polymers correspond to $v = 0.5$ \cite{RN1889}. Fig. \ref{Flory_exponent} shows the dependence of $v$ on surface coverage for semi-flexible polymers of different equilibrium bending angles, together with a comparison with fully flexible chains under the same conditions. The exponent $v$ is calculated from the simulation data of $\langle R^2_g \rangle$ versus $N-1$, as for example the one presented in the middle panel of Fig. \ref{chain_size}. Rod-like systems corresponding to $\theta_0 = 0^{\circ}$ show a Flory exponent which is independent of packing density and approximately equal to 0.91. This value is slightly lower than the expected one for rods ($v = 1$). This small deviation is due to bending angles following a distribution around the equilibrium value, as seen in Fig. \ref{bending_distr}, rather than all being exactly equal to $0^{\circ}$. The locally compact system of $\theta_0 = 120^{\circ}$ shows also high value of the critical exponent, which is again unaffected by packing density.  Chains with $\theta_0 = 90^{\circ}$ are characterized by significantly lower values of $v \approx 0.7$, which do not depend on surface coverage. In contrast, the scaling exponent for the fully flexible chains shows a decrease as packing density increases as a result of the progressive chain collapse. In parallel, the most profound effect of surface coverage is encountered for  $\theta_0 = 120^{\circ}$ as $v$ increases markedly until it reaches a value of $v \approx 0.935$ at $\varphi^* = 0.70$. This trend, as will be demonstrated in the continuation, is strongly related to the formation of a nematic phase where chains adopt maximum length, zig-zag conformations.

\par Focusing on rod-like systems Ref. \cite{RN2061} has demonstrated that spherocylinders confined to a plane show a nematic phase when their aspect ratio is higher than a threshold value, $L/D > 7$, where $L$ is the length and $D$ the diameter (the limit of $D \rightarrow 0$ corresponding to needles \cite{RN2056}). Shorter rods show an isotropic phase where chains are aligned either side-by-side or perpendicular to each other \cite{RN2061}. In our rod-like case ($\theta_0 = 0^{\circ}$) we have dispersity in chain lengths combined with a distribution of bending angles around the equilibrium value resulting in a shape which is not fixed but rather changes continually over the simulation. Thus, our computer-generated samples deviate from the ideal, fixed-shape hard rods of the aforementioned studies. Still, we can attempt a comparison of the aspect ratio as a function of surface coverage and $N$ by considering that $D = \sigma = 1$ and $R(N) = L$.  For $\theta_0 = 0^{\circ}$, and at all surface coverages studied here, including the one of the RCP limit, approximately 23\% of the chain population, corresponding to lengths $N$ = 6, 7 and 8, have aspect ratio $L/D = R(N)/\sigma \leq 7$. This could potentially make the observed nematic phase, if such phase exists in our simulations, unstable compared to systems of significantly longer chains. The analysis in Ref. \cite{RN2066} for infinitely thin rods (needles) demonstrates that rod length dispersity affects significantly the phase behavior, with long and short rods stabilizing and weakening, respectively, the nematic phase.  

\subsection{Local Structure: Crystallinity}

One essential element in the identification of the crystal phase, MRJ state or RCP limit is the quantification of local order. Here we employ the CCE norm descriptor \cite{RN1542} to quantify the structural similarity of the local environment around each monomer against the triangular (TRI), square (SQU) and honeycomb (HON) crystals and the pentagonal (PEN) local symmetry in two dimensions. In all calculations to be presented below a threshold of $\epsilon^{thres} = 0.245$ is adopted. Spheres with TRI, SQU, HON and PEN similarity are colored in blue, red, purple and green, respectively. Monomers whose CCE norm is higher than the threshold value for all reference crystals and local symmetry are labeled as amorphous (AMO), or more precisely as "unidentified", and are shown in yellow.    
\par
\begin{figure*}[ht]
\centering
\includegraphics[scale=0.52]{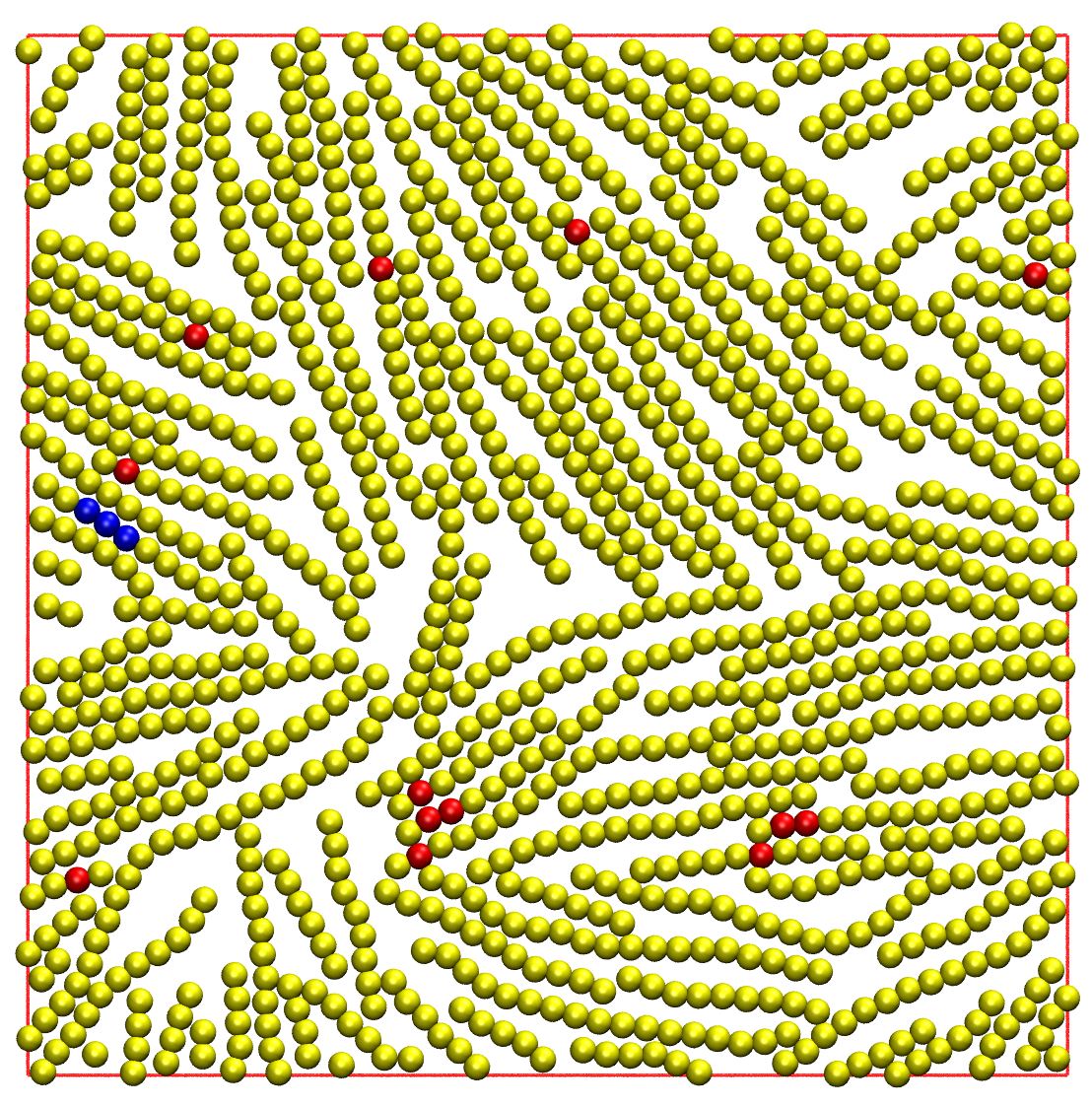}
\includegraphics[scale=0.52]{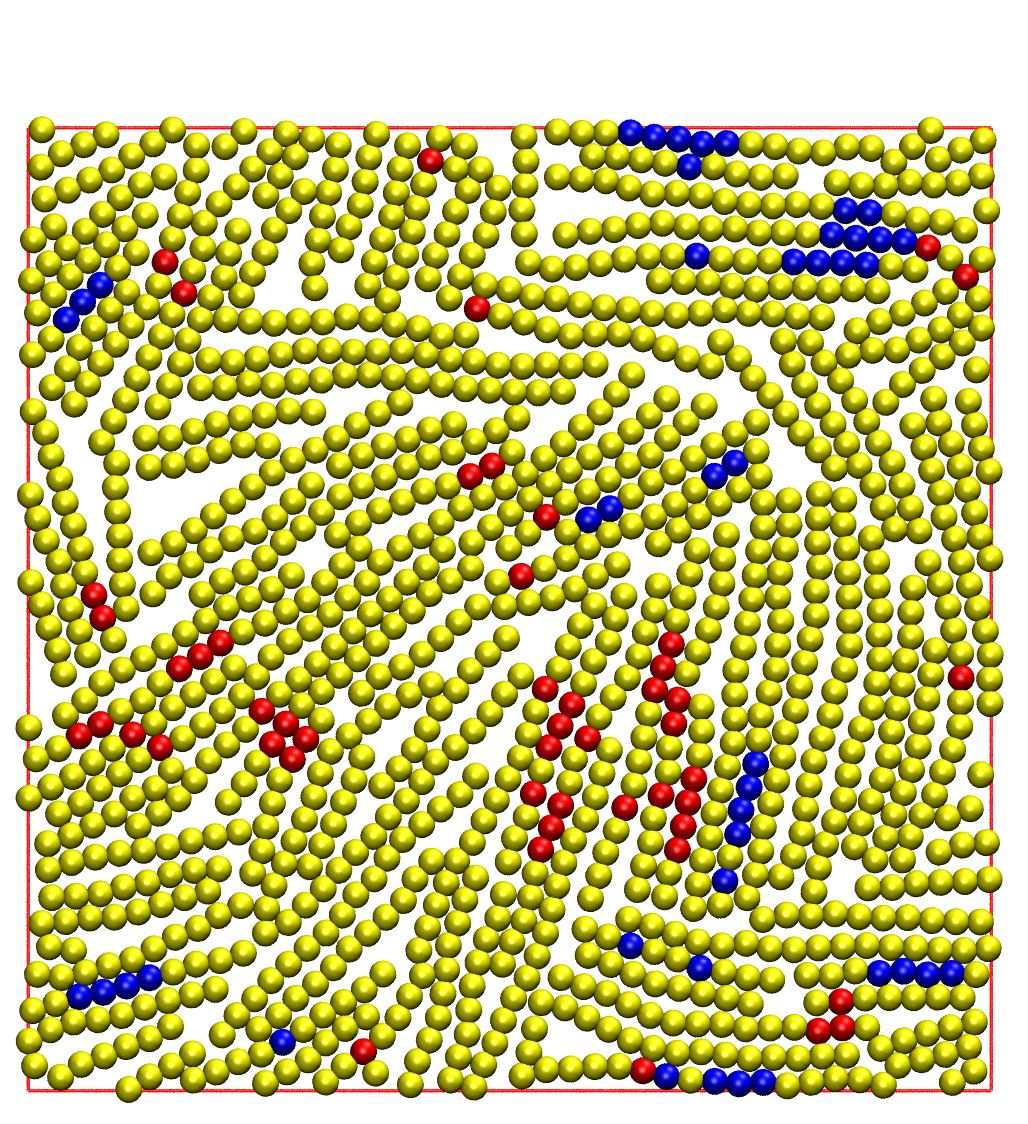}
\includegraphics[scale=0.52]{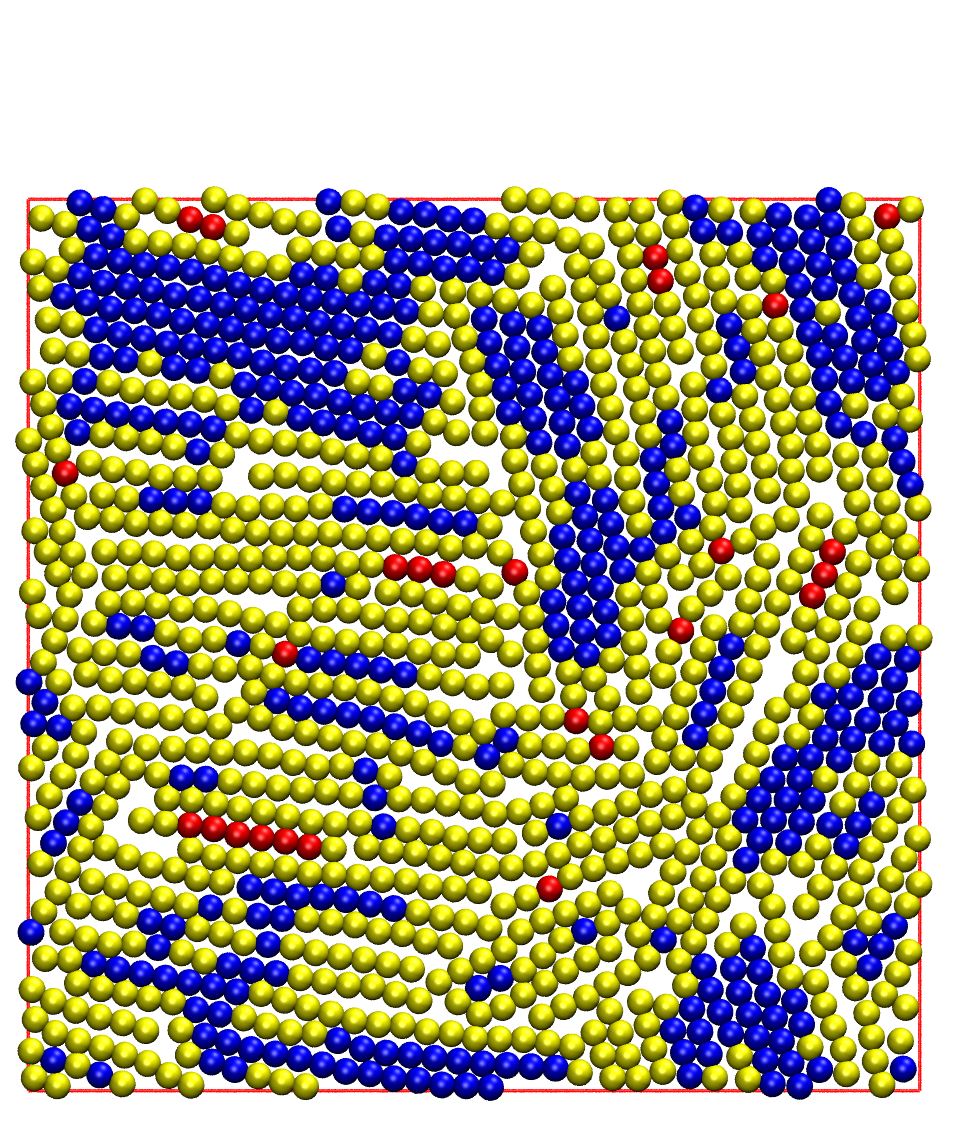}
\includegraphics[scale=0.52]{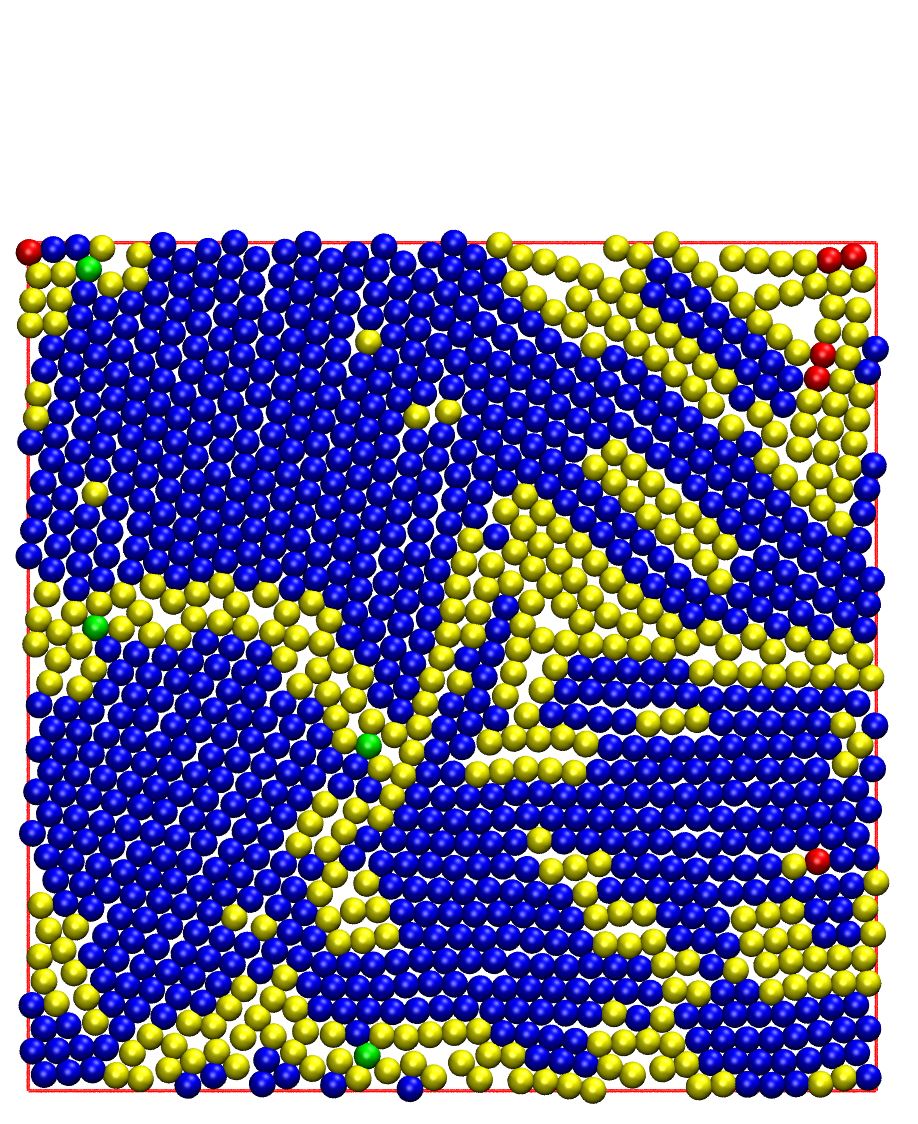}
\centering
\includegraphics[scale=0.52]{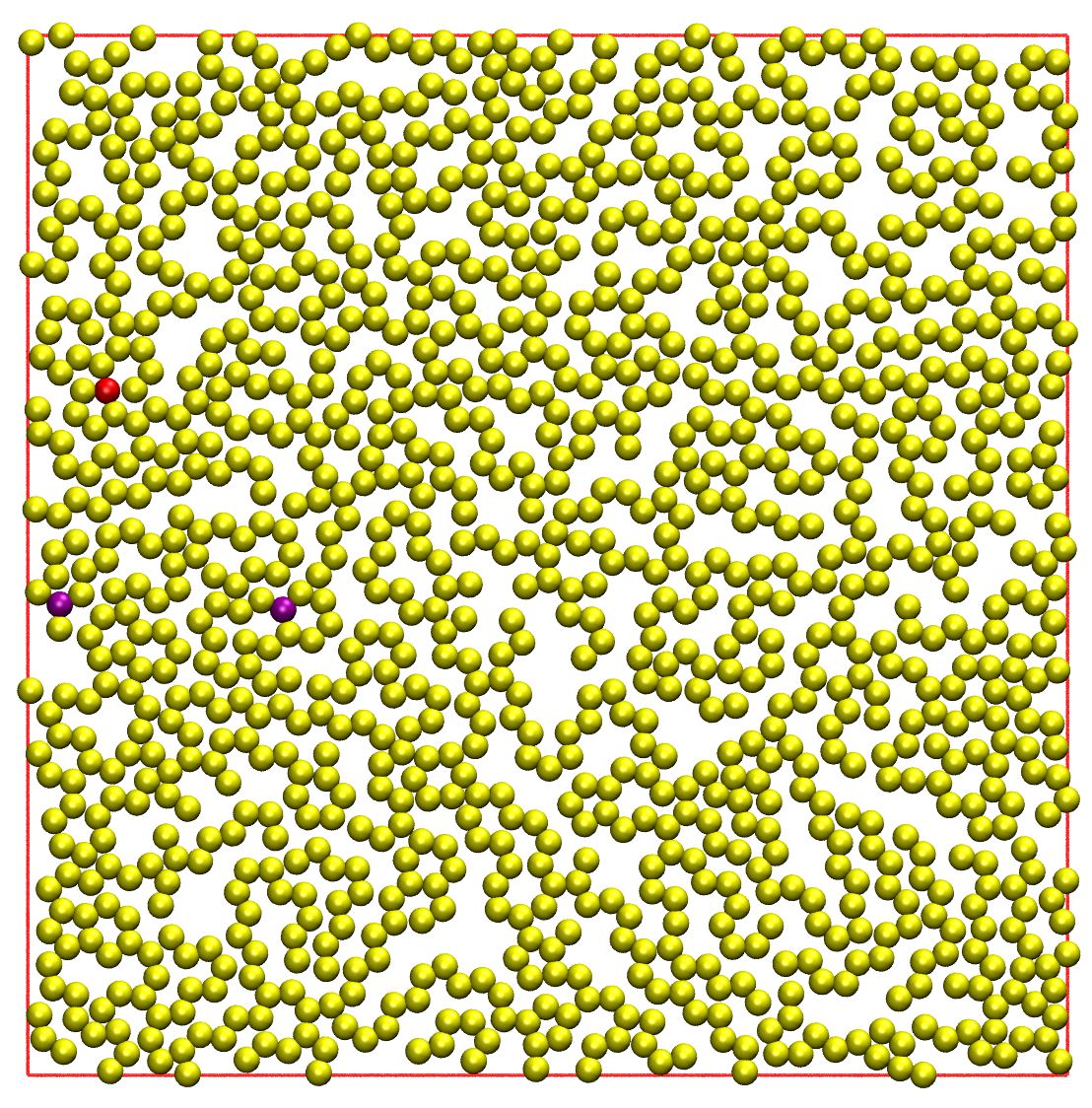}
\includegraphics[scale=0.52]{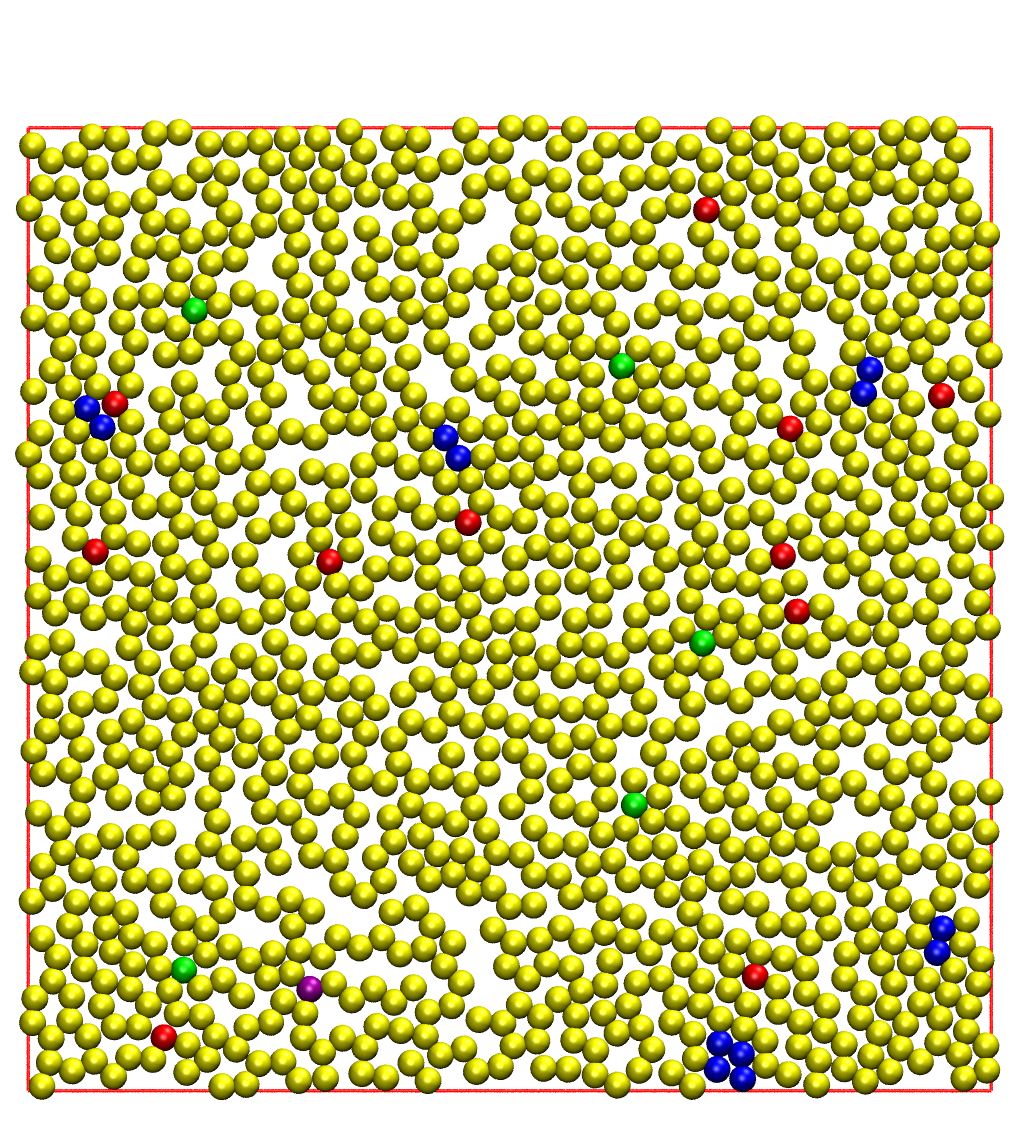}
\includegraphics[scale=0.52]{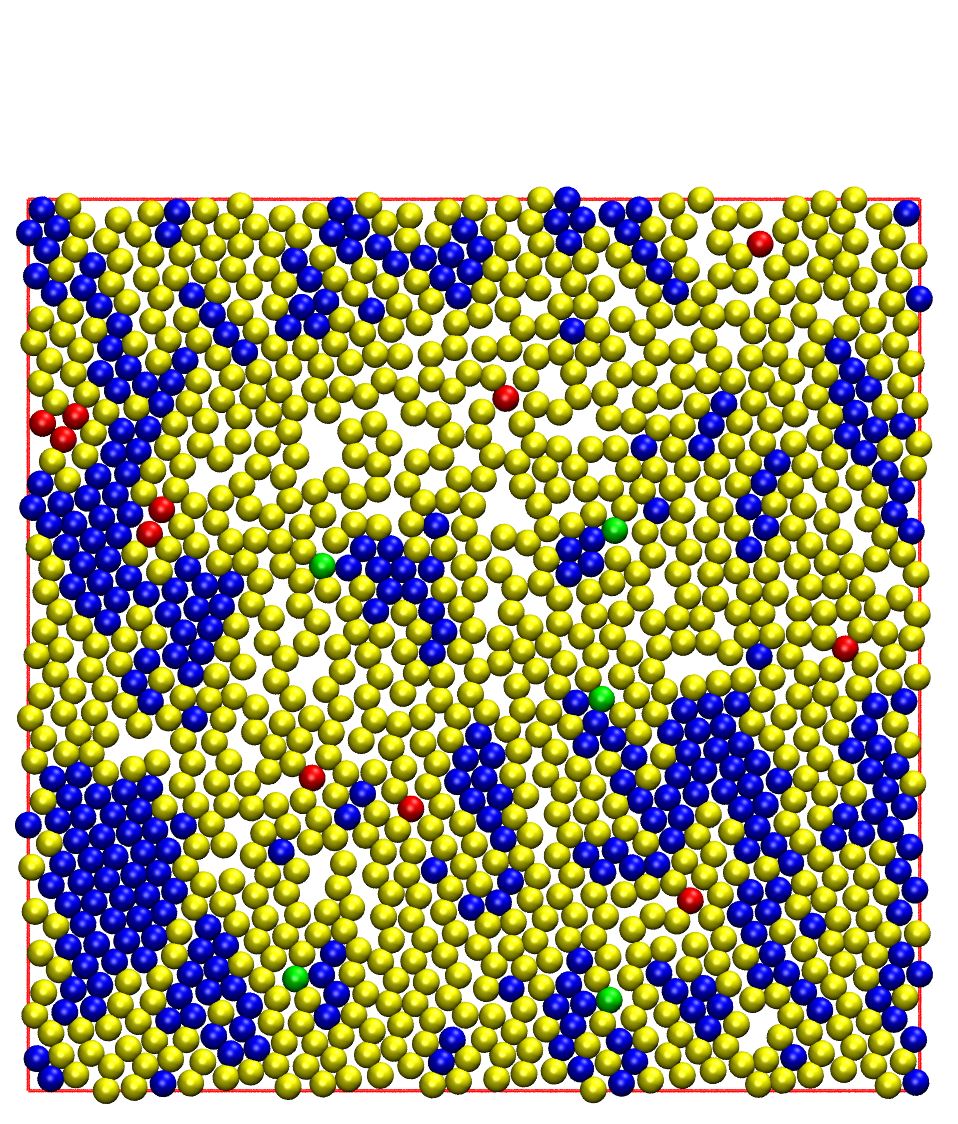}
\includegraphics[scale=0.52]{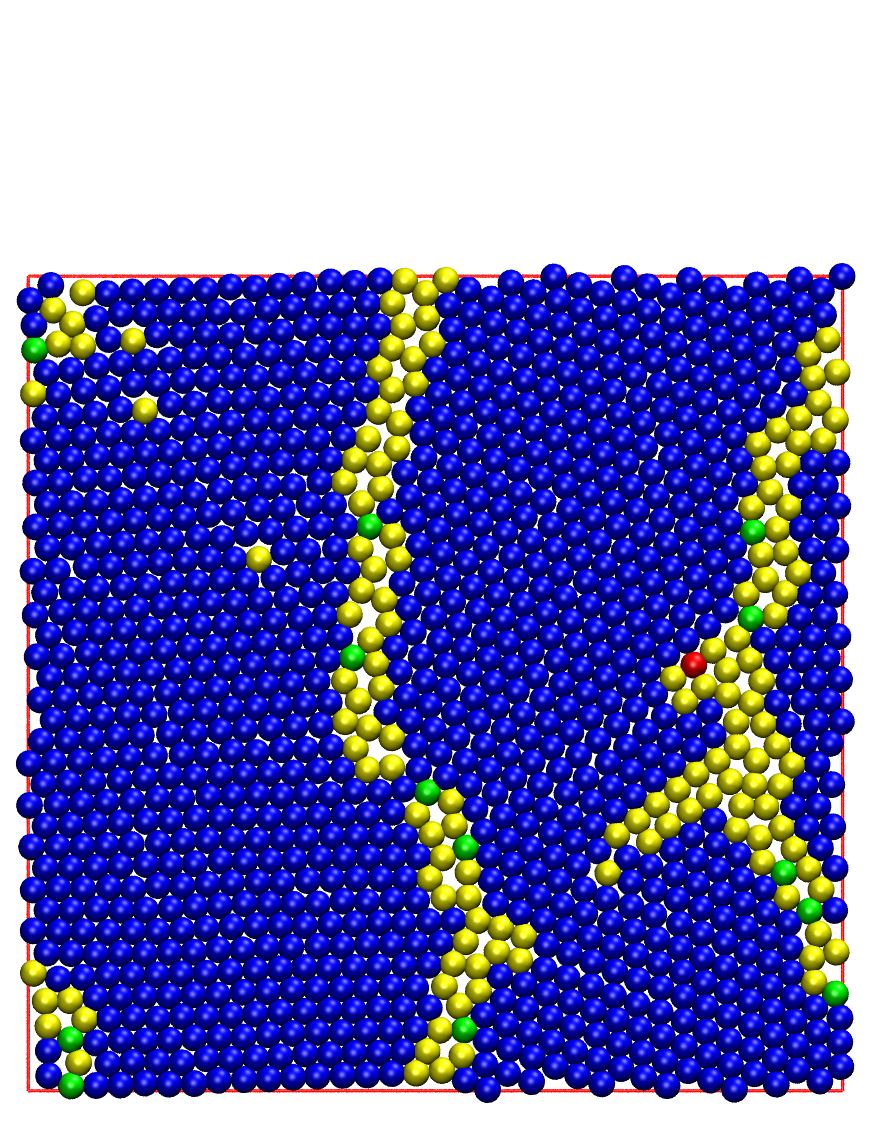}
\centering
\includegraphics[scale=0.52]{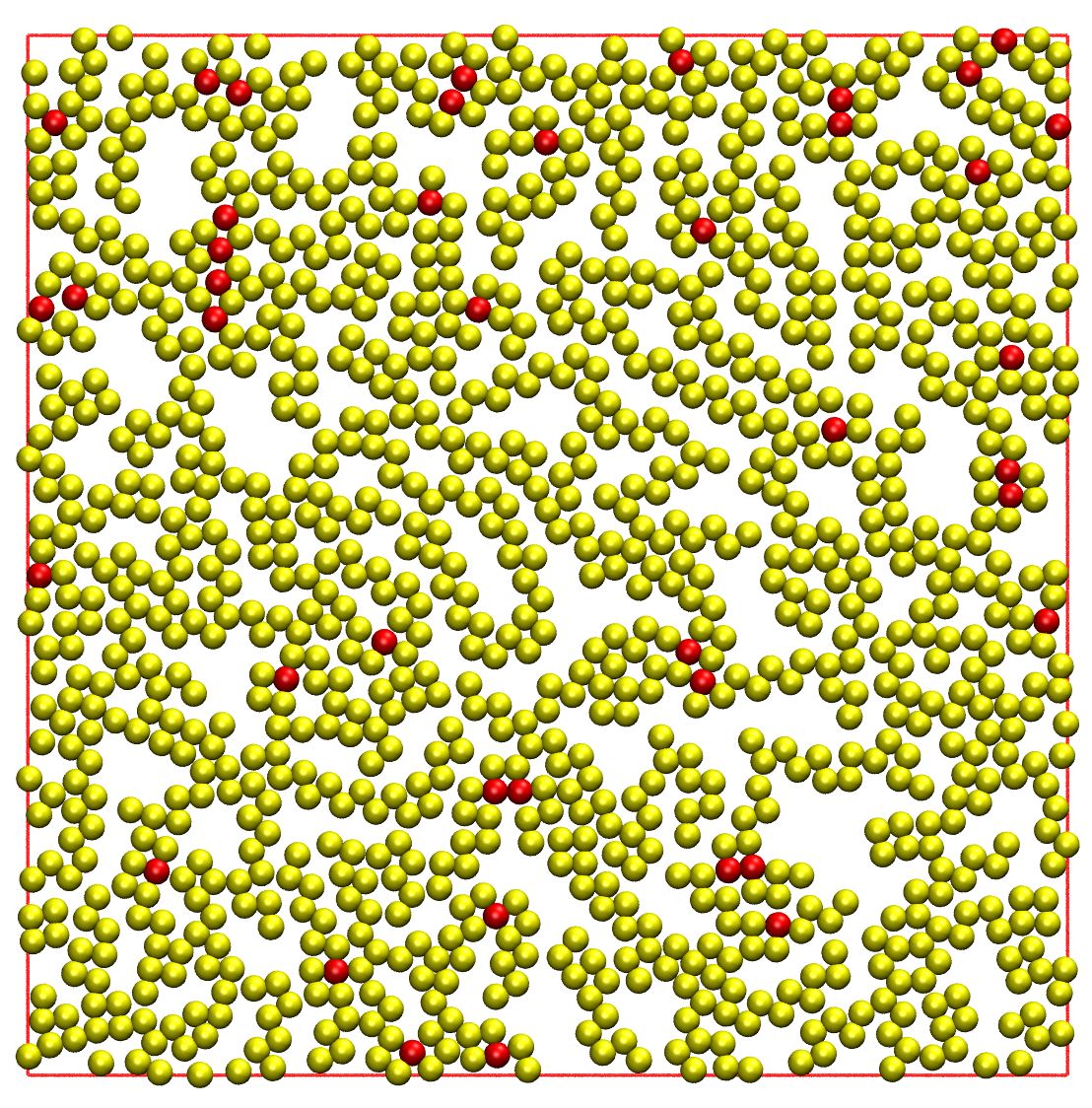}
\includegraphics[scale=0.52]{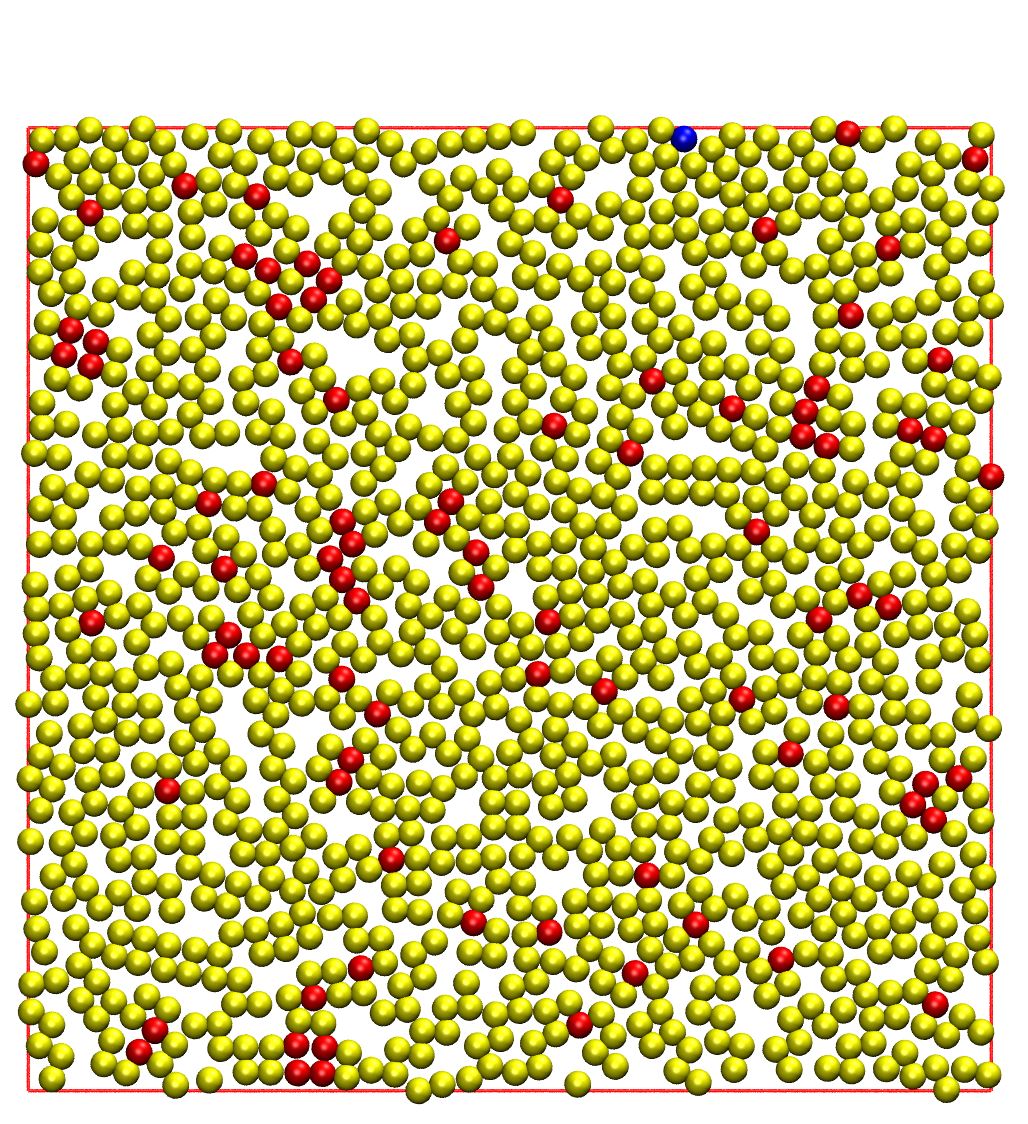}
\includegraphics[scale=0.52]{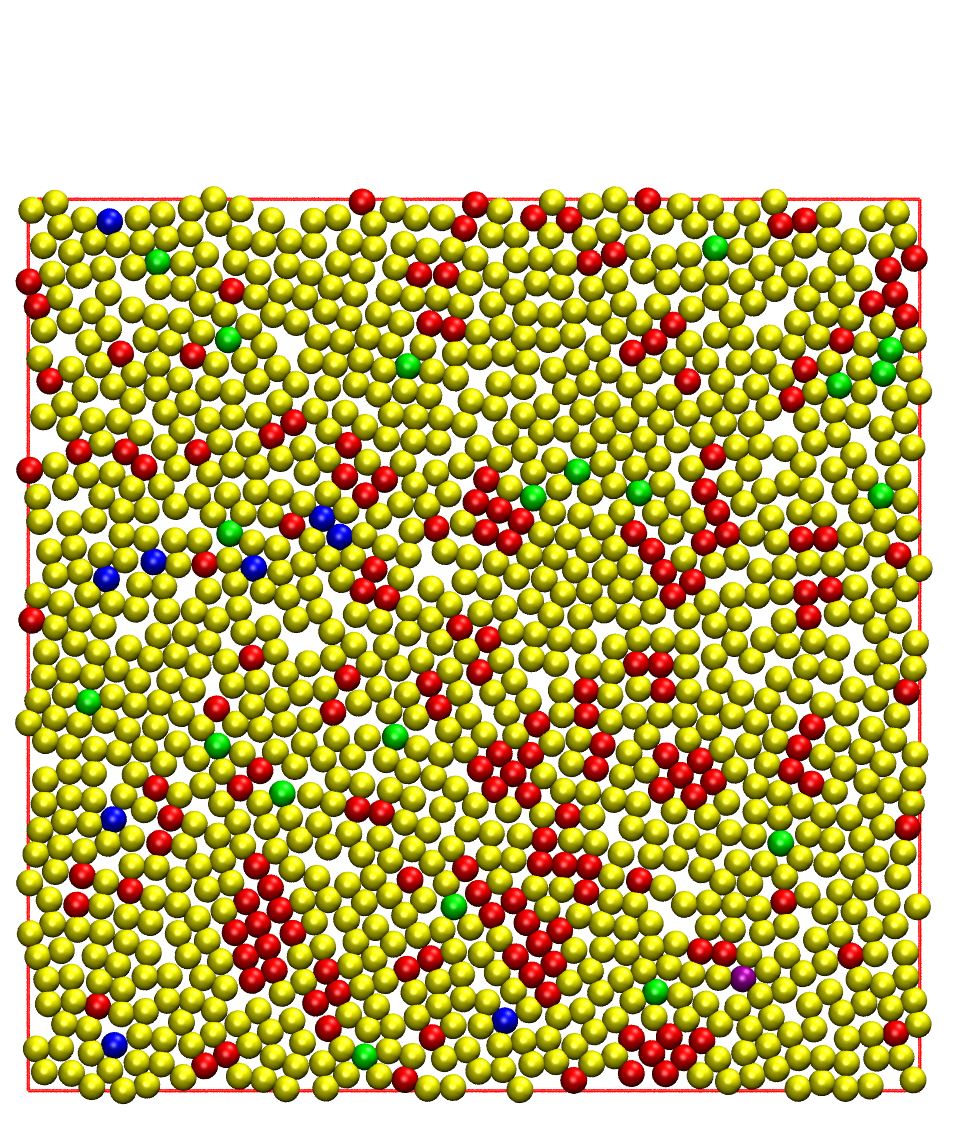}
\includegraphics[scale=0.52]{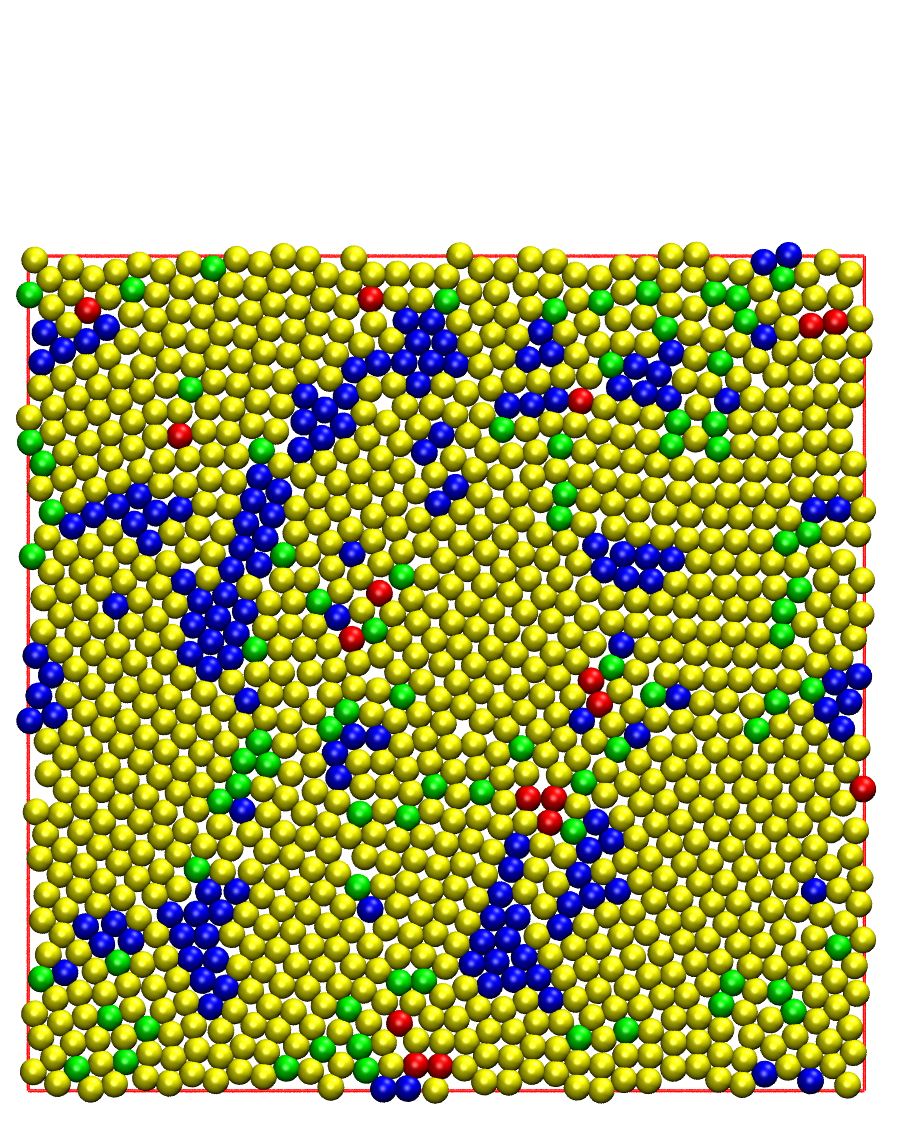}
\centering
\includegraphics[scale=0.52]{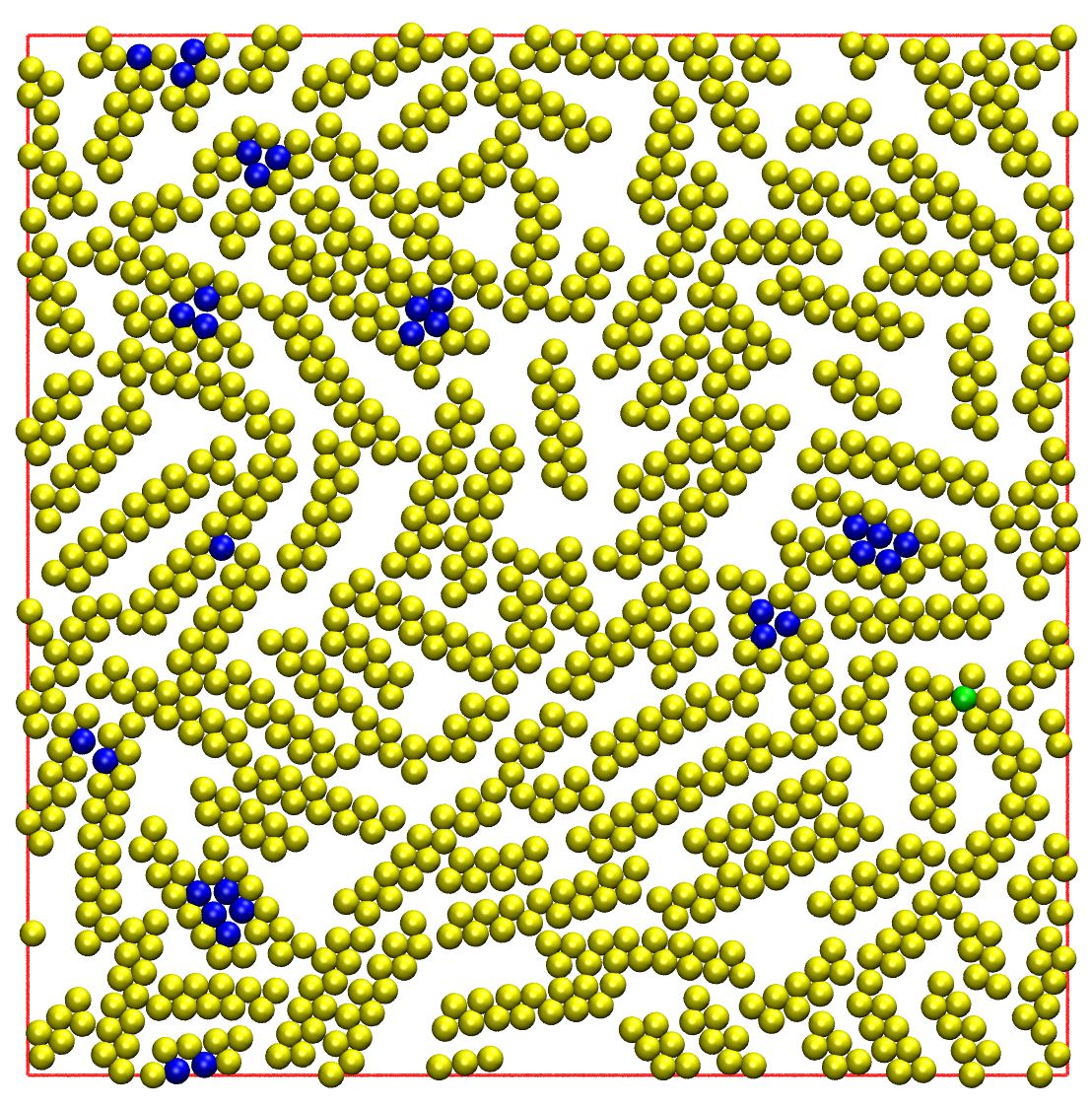}
\includegraphics[scale=0.52]{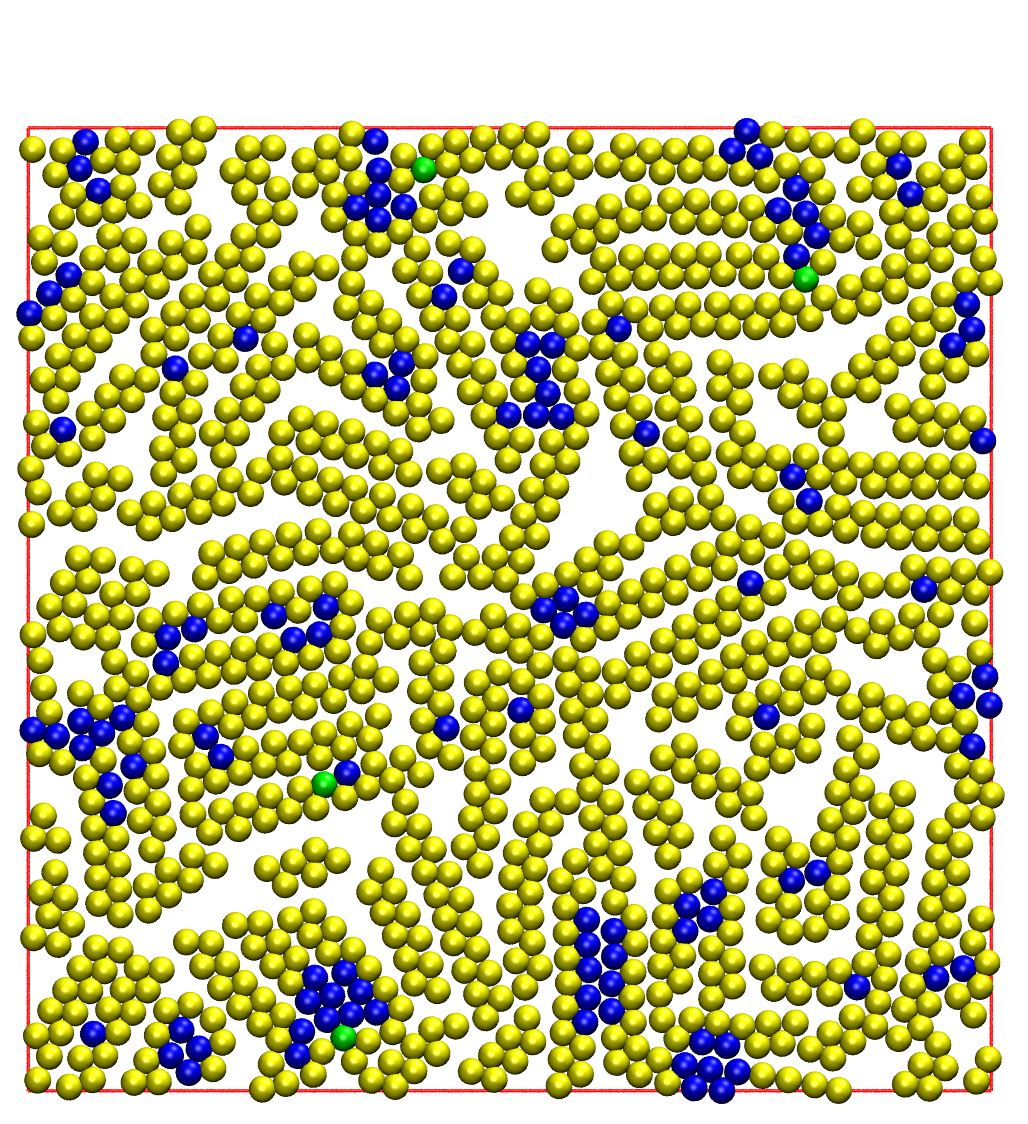}
\includegraphics[scale=0.52]{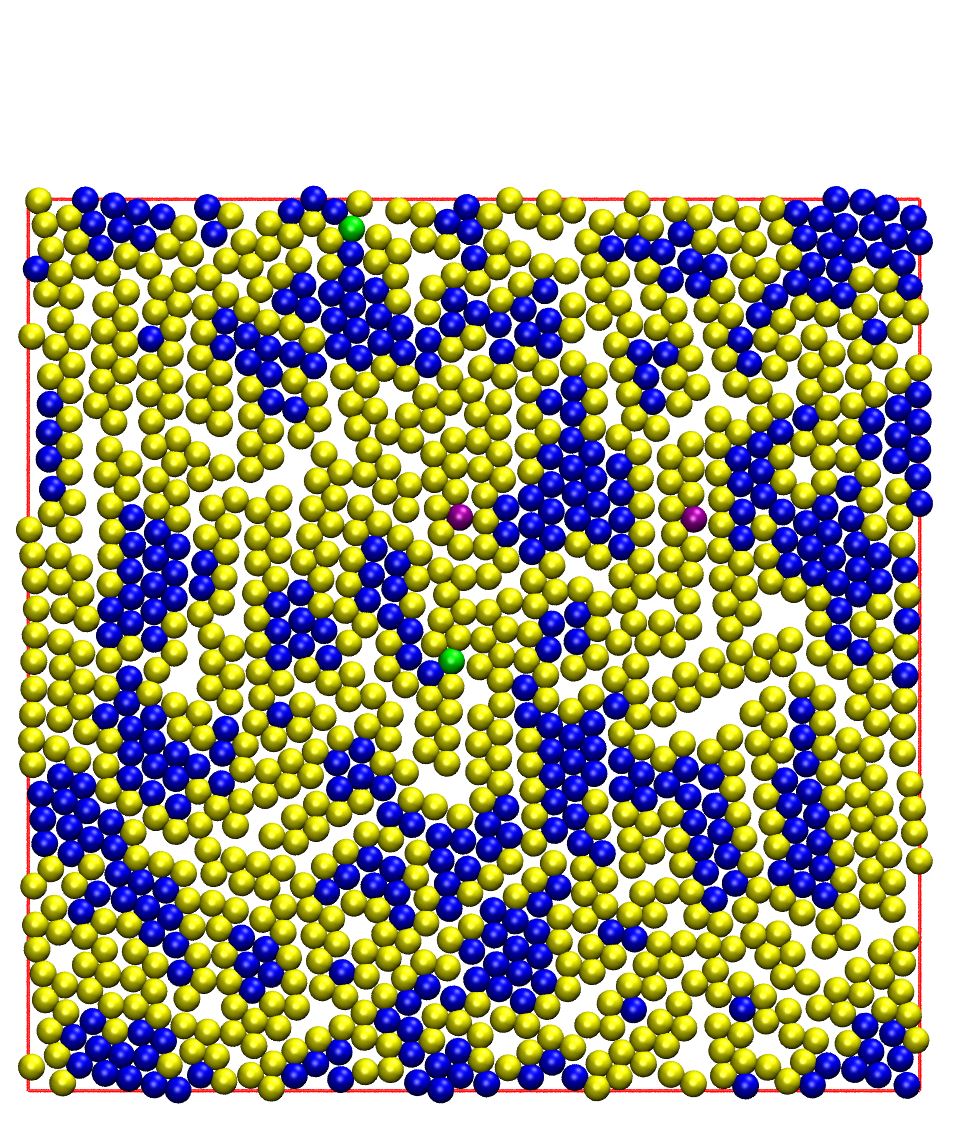}
\includegraphics[scale=0.52]{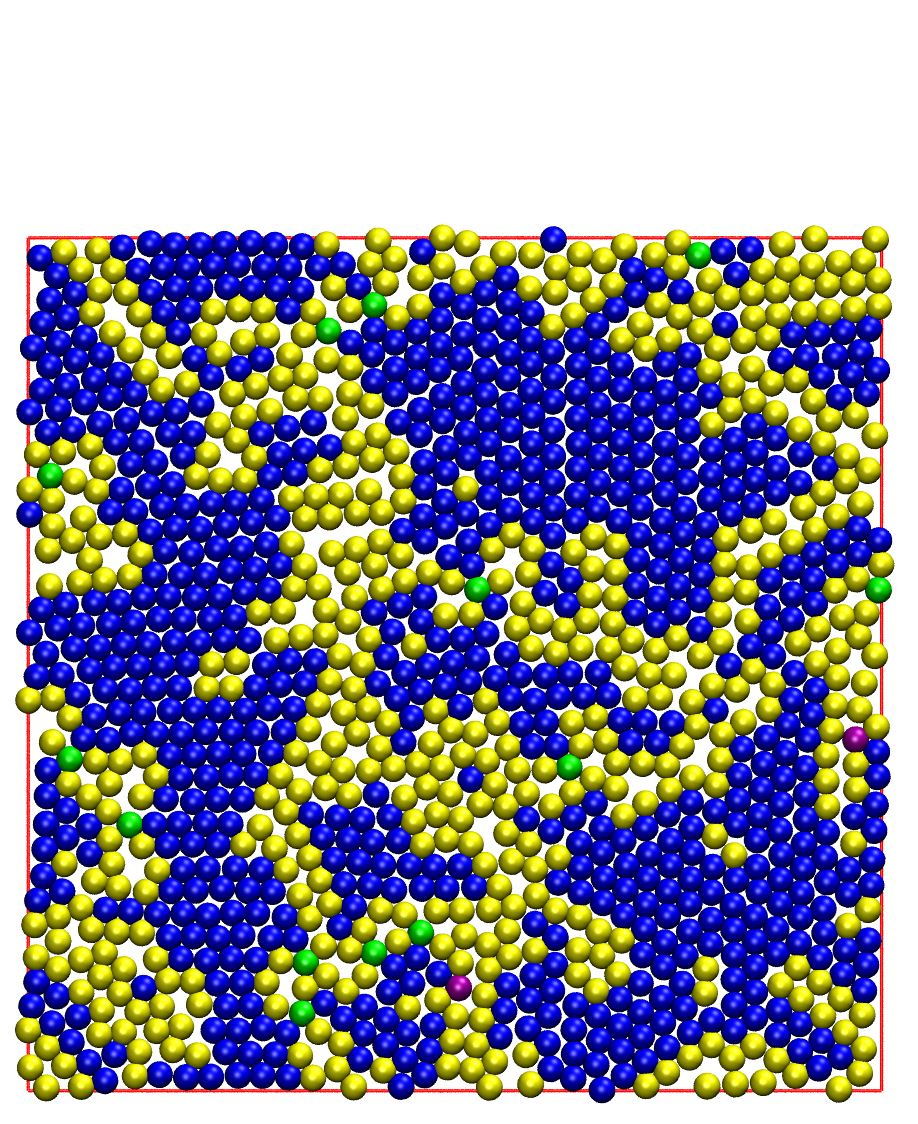}
\caption{System snapshots at the end of the MC simulation where monomers are colored according to their crystal similarity as quantified by the CCE norm descriptor: blue, red, purple and green colors correspond to TRI, SQU, HON crystals and PEN local symmetry, respectively. Amorphous or unidentified (AMO) sites are shown in yellow color. From top to bottom: $\theta_0 = 0$, 60, 90 and $120^{\circ}$. From left to right: $\varphi^* = 0.50$, 0.60, 0.70 and $\varphi^{*,RCP}_{2D}(\theta_0)$. Image created with the VMD software \cite{RN250}.}
\label{snapshots_CCE}
\end{figure*}

 Fig. \ref{snapshots_CCE} shows system snapshots at the end of the MC simulations at different packing densities, including the densest ones for each equilibrium bending angle. Monomers are colored according to the CCE norm description as explained in the previous paragraph. At the lowest surface coverage ($\varphi^* = 0.50$) all systems are predominantly amorphous and the number of sites with crystal structure is very low. As density increases crystallinity increases and the ordered sites form small, isolated clusters. These clusters grow in size as concentration increases, as can be observed at $\varphi^* = 0.70$. Finally, at RCP all systems exhibit the highest local order.  Given the very high values of surface coverage the RCP limit for  $\theta_0 = 0$, 60, 90 and $120^{\circ}$ corresponds to a TRI crystal, ridden with defects in the form of amorphous sites and free of isolated clusters of other crystal morphologies.   

\par
\begin{figure}
\centering
\includegraphics[scale=0.40]{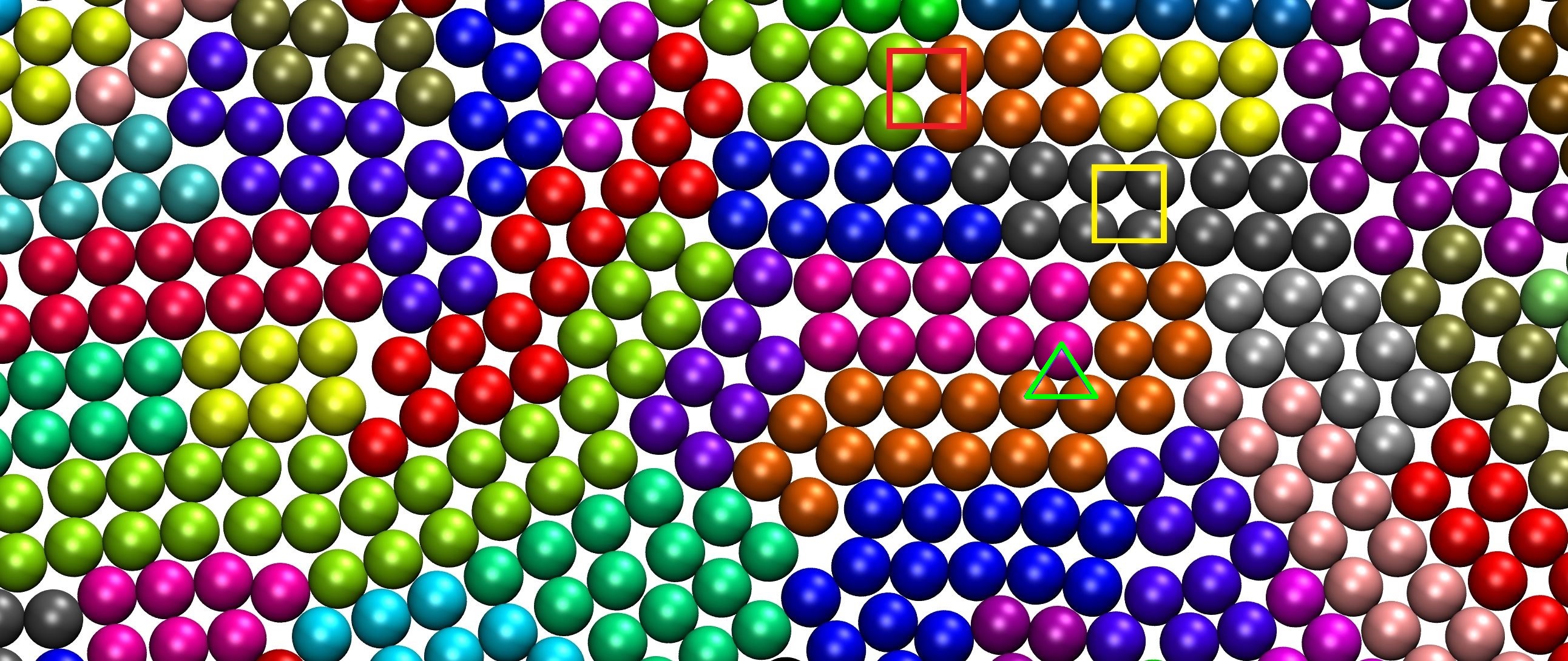}
\caption{Detail of the snapshot at the end of the MC simulation of the $\theta_0 = 90^{\circ}$ at $\varphi^{*,RCP}_{2D}(90^{\circ}) \approx 0.797$. Spheres are colored according to the parent chain. Yellow and red squares indicate intra- and inter-molecular local arrangements, respectively, leading to the formation of a square, while green triangle indicates interchain packing. Image created with the VMD software \cite{RN250}.}
\label{Snapshot_rightangle}
\end{figure}

 The only exception to the general trend dictating that RCP shows the highest local order is the system of right-angle chains. This is expected given that $\theta_0 = 90^{\circ}$ is incompatible with the geometry of the adjacent sites in the TRI crystal. Furthermore, the bending constant ($k_{\theta} / k_{B}T= 9$ rad$^{-2}$) is strong enough so that it does not allow large angle distortions which could lead to bending conformations of 60 or $120^{\circ}$, which are elements of the TRI crystal connectivity. It is interesting to notice that visual inspection of the CCE-based snapshots for the right-angle chains in Fig. \ref{snapshots_CCE} reveals that at $\varphi^* = 0.50$, 0.60 and 0.70, while the system is predominantly amorphous, the isolated ordered sites form SQU crystallites and the higher the concentration the higher the SQU order. However, as the system transits to the RCP state the pattern changes drastically, the SQU sites disappear almost completely and get replaced by small clusters of TRI character. 
As the RCP density is higher than that of the pure SQU crystal the resulting packing is necessarily a tiling pattern which is effectively a blend of intra-molecular square packing, imposed by the right-angle constraint, and inter-molecular chain alignments which fill the gaps in the square spacing to maximize local density. The net result is a configuration which is on one hand denser than the SQU crystal but on the other hand significantly less ordered, at least with respect to the tested SQU and TRI reference crystals. This can be clearly seen in the structural detail of Fig. \ref{Snapshot_rightangle} corresponding to a segment of the polymer packing at $\varphi^{*,RCP}_{2D}(90^{\circ})$. Sphere monomers are colored according to the parent polymer so as to identify the intra- and inter-chain local arrangements. Square and triangle shapes are drawn to indicate the formation of intra- and inter-molecular squares and of intermolecular triangles filling the gaps between the vertices of the squares. Also, it can be seen that a significant fraction of the squares is distorted because bending angles follow a distribution around the equilibrium right angle, as shown in Fig. \ref{bending_distr}.

\subsection{Global Structure: Nematic and Tetratic Order Parameter}

\begin{figure*}[ht]
\centering
\includegraphics[scale=0.40]{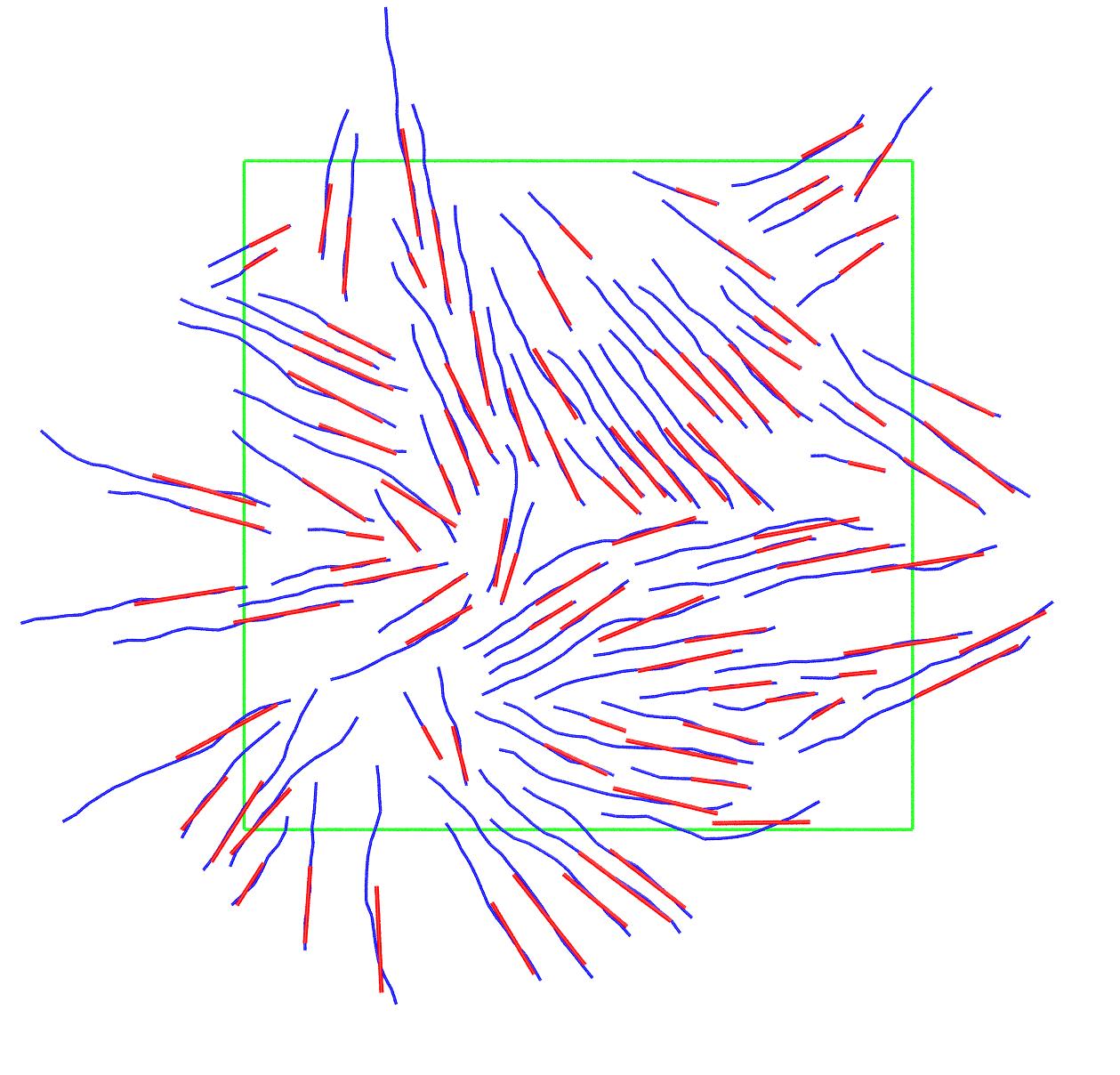}
\includegraphics[scale=0.40]{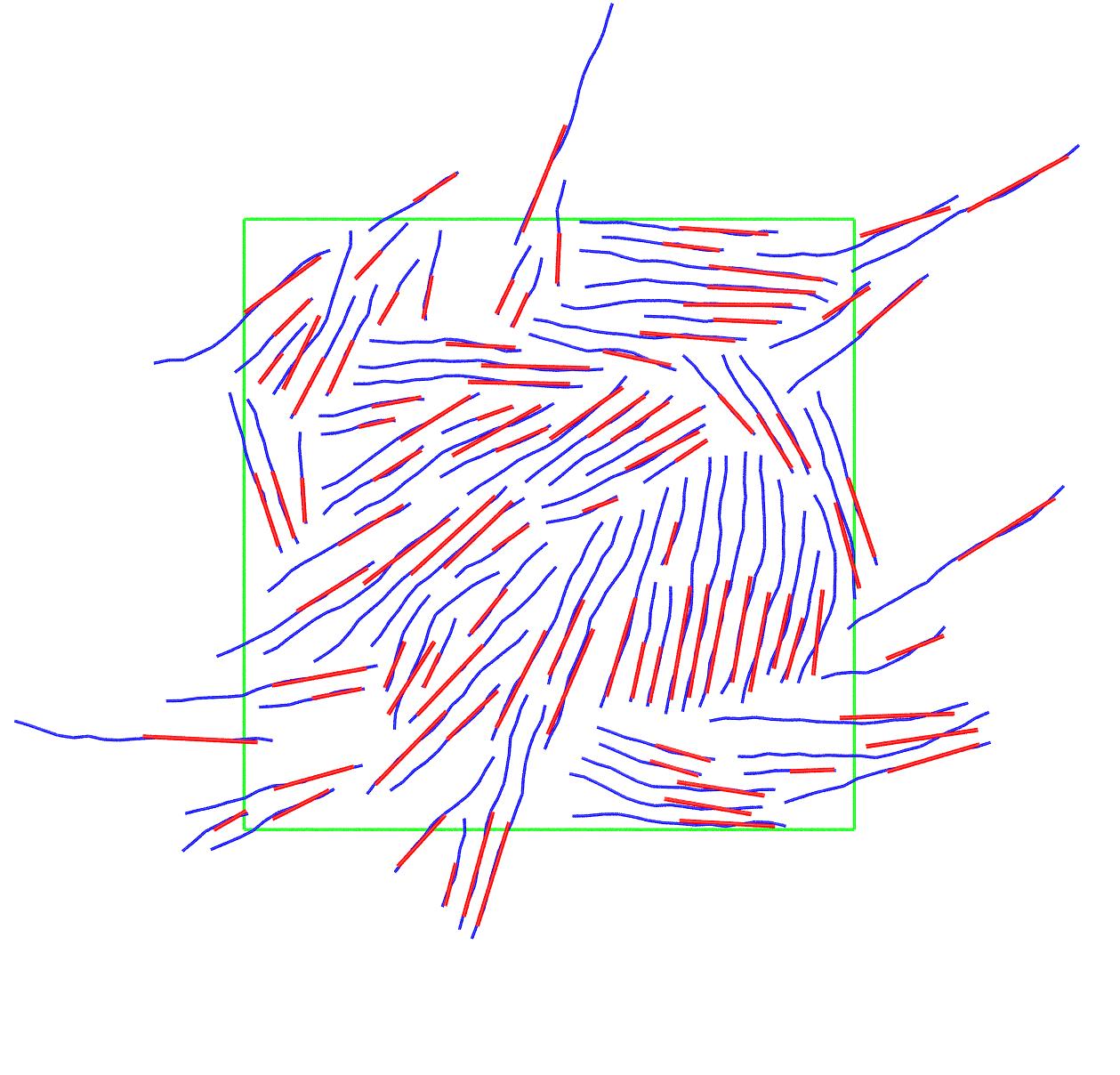}
\includegraphics[scale=0.40]{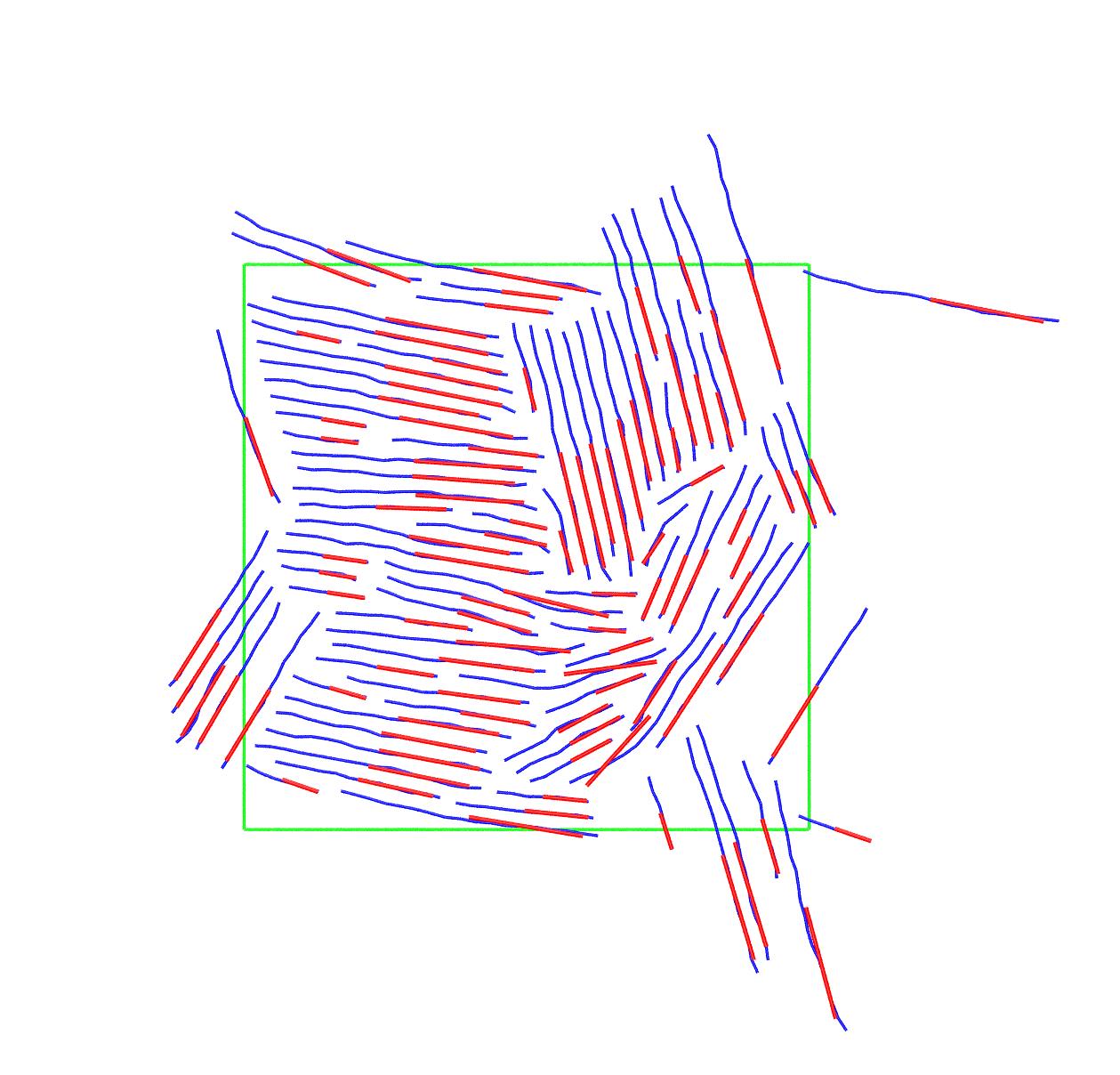}
\includegraphics[scale=0.40]{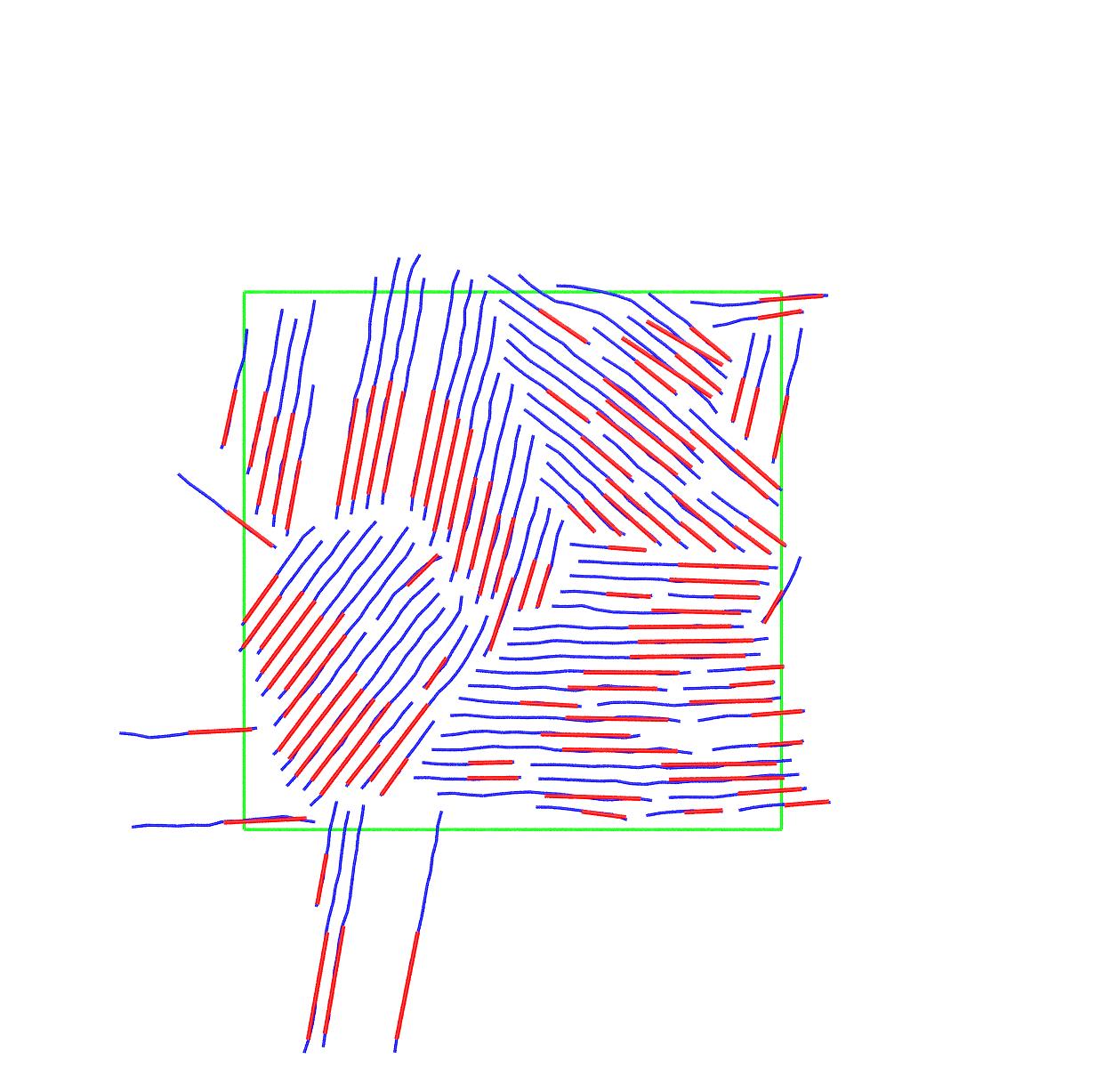}
\centering
\includegraphics[scale=0.40]{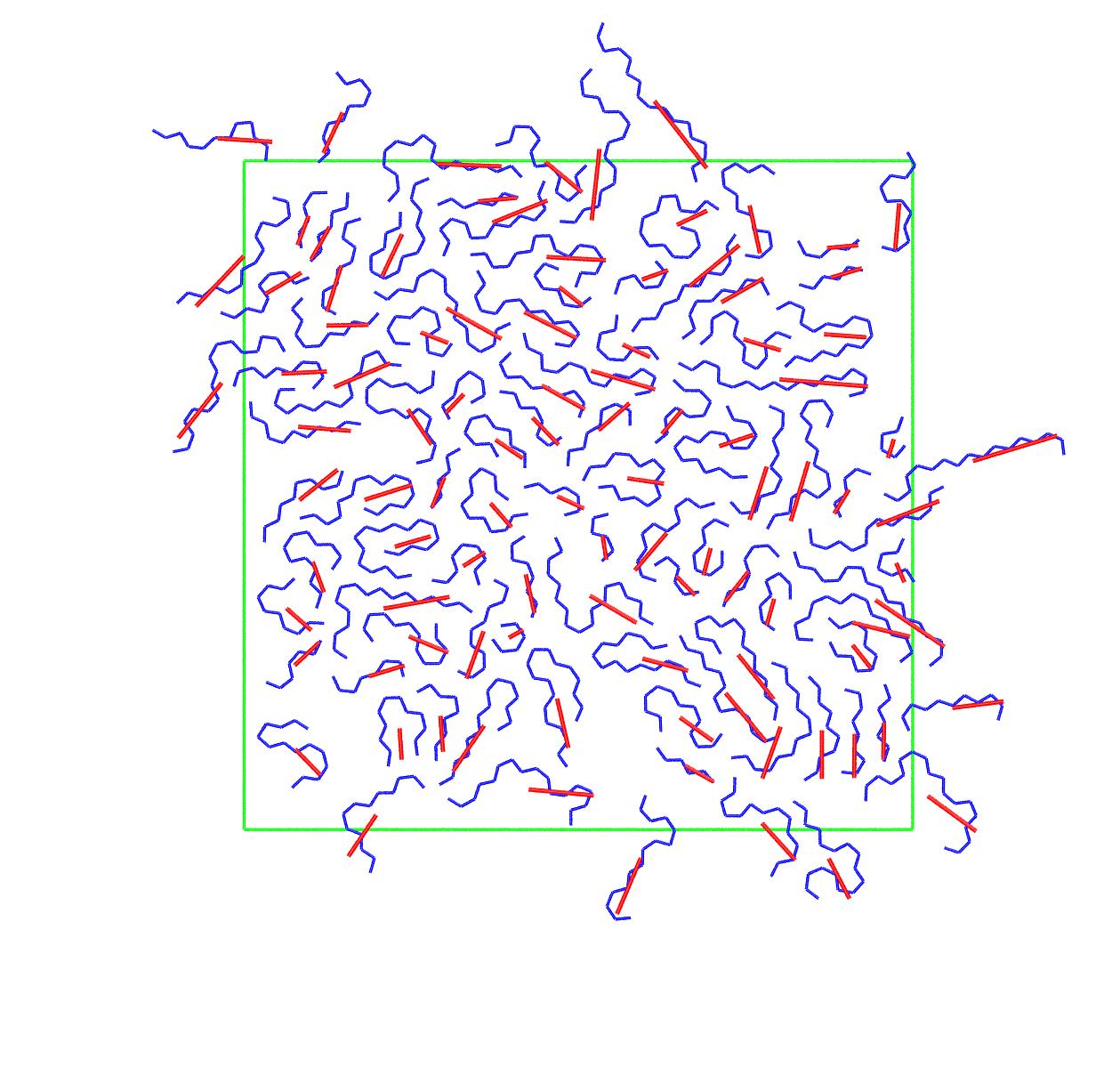}
\includegraphics[scale=0.40]{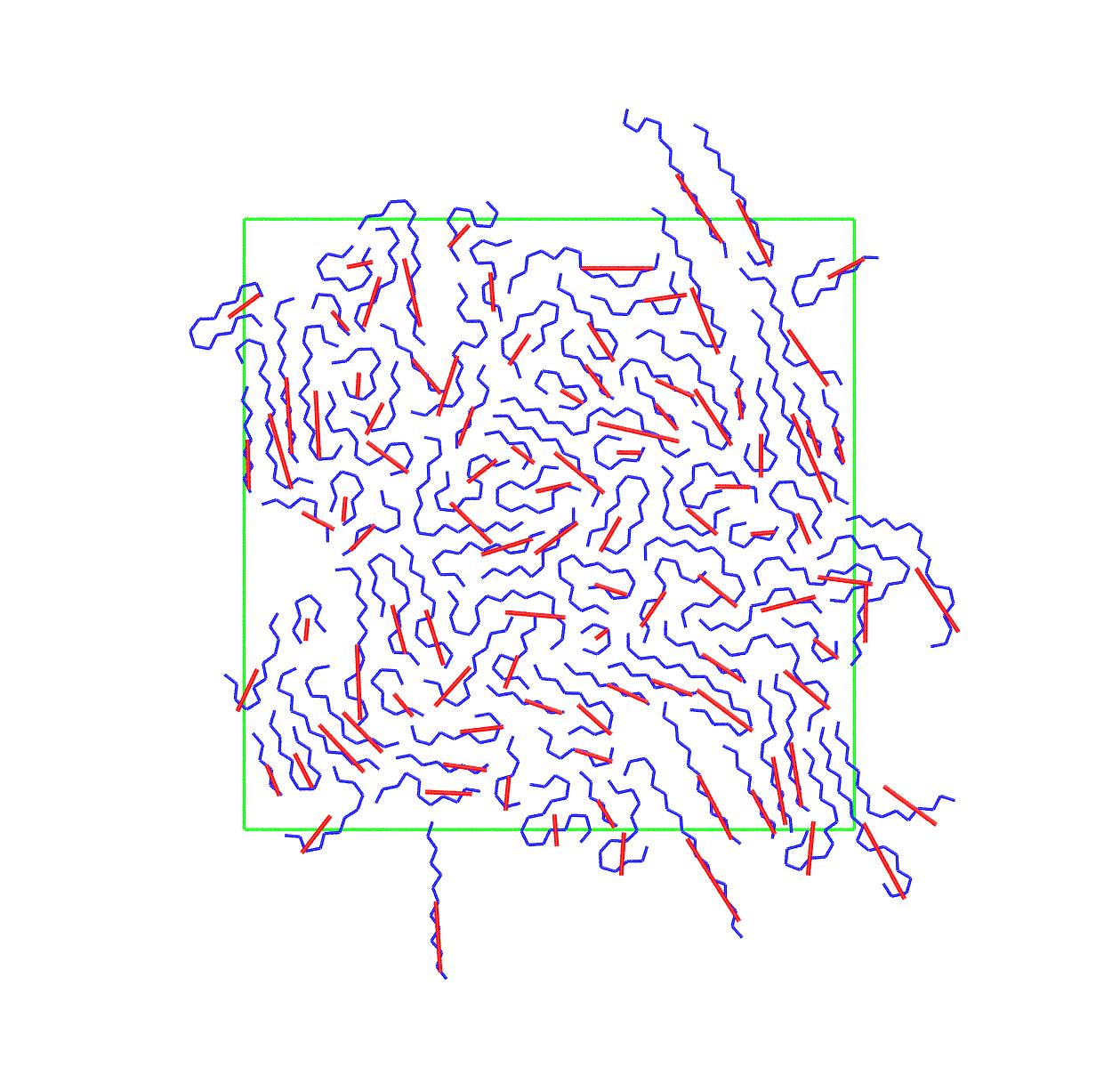}
\includegraphics[scale=0.40]{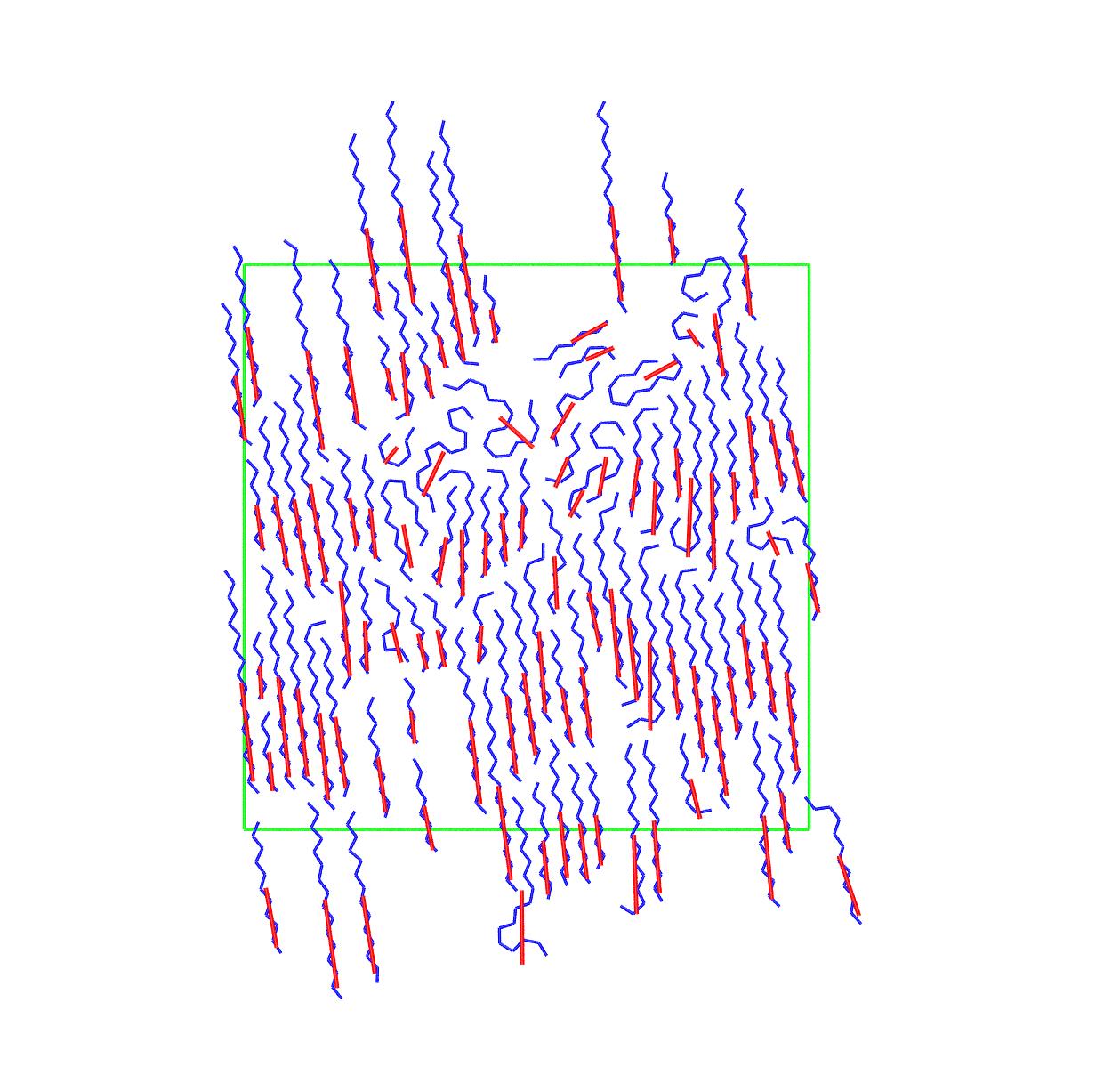}
\includegraphics[scale=0.40]{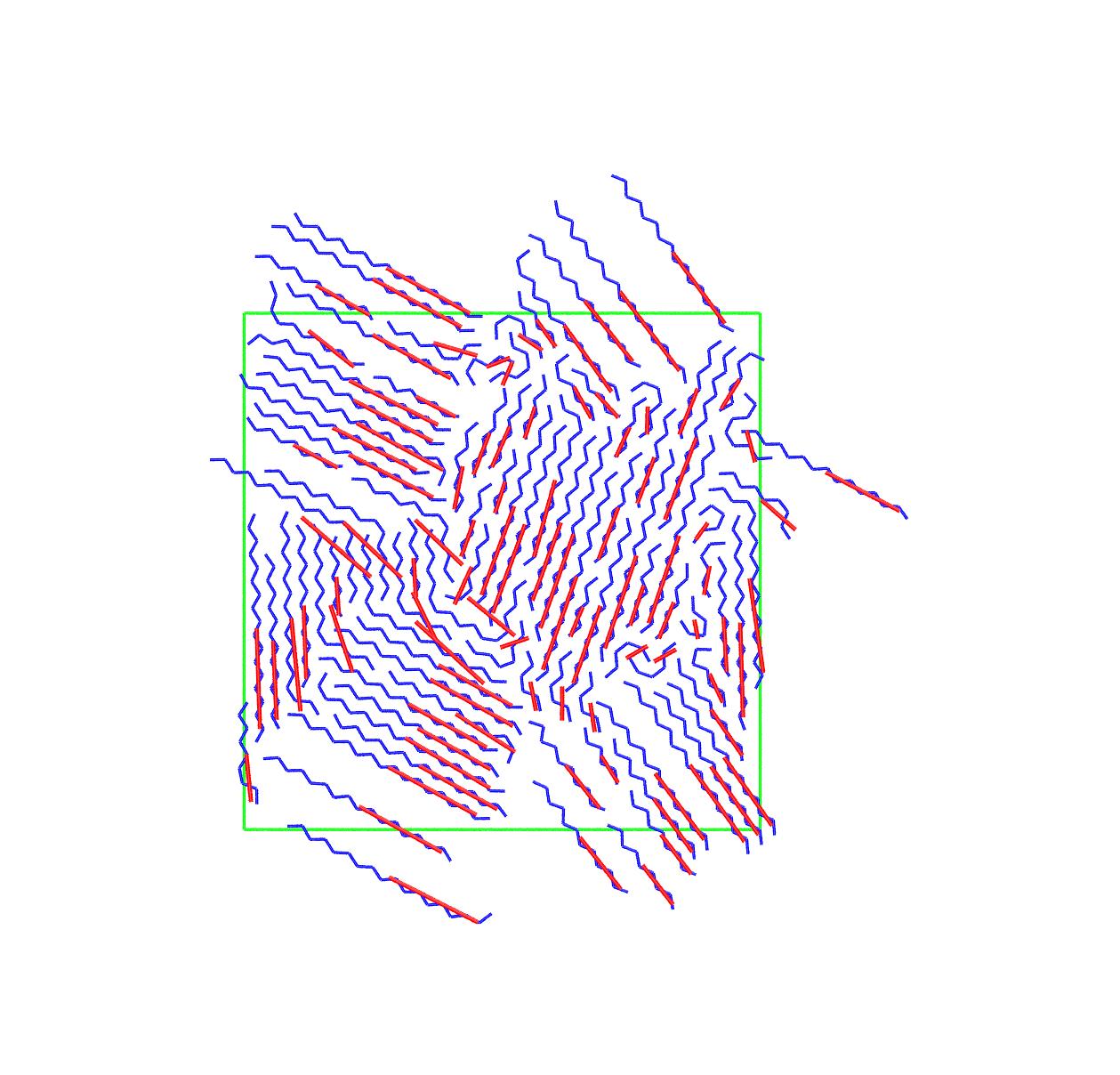}
\centering
\includegraphics[scale=0.40]{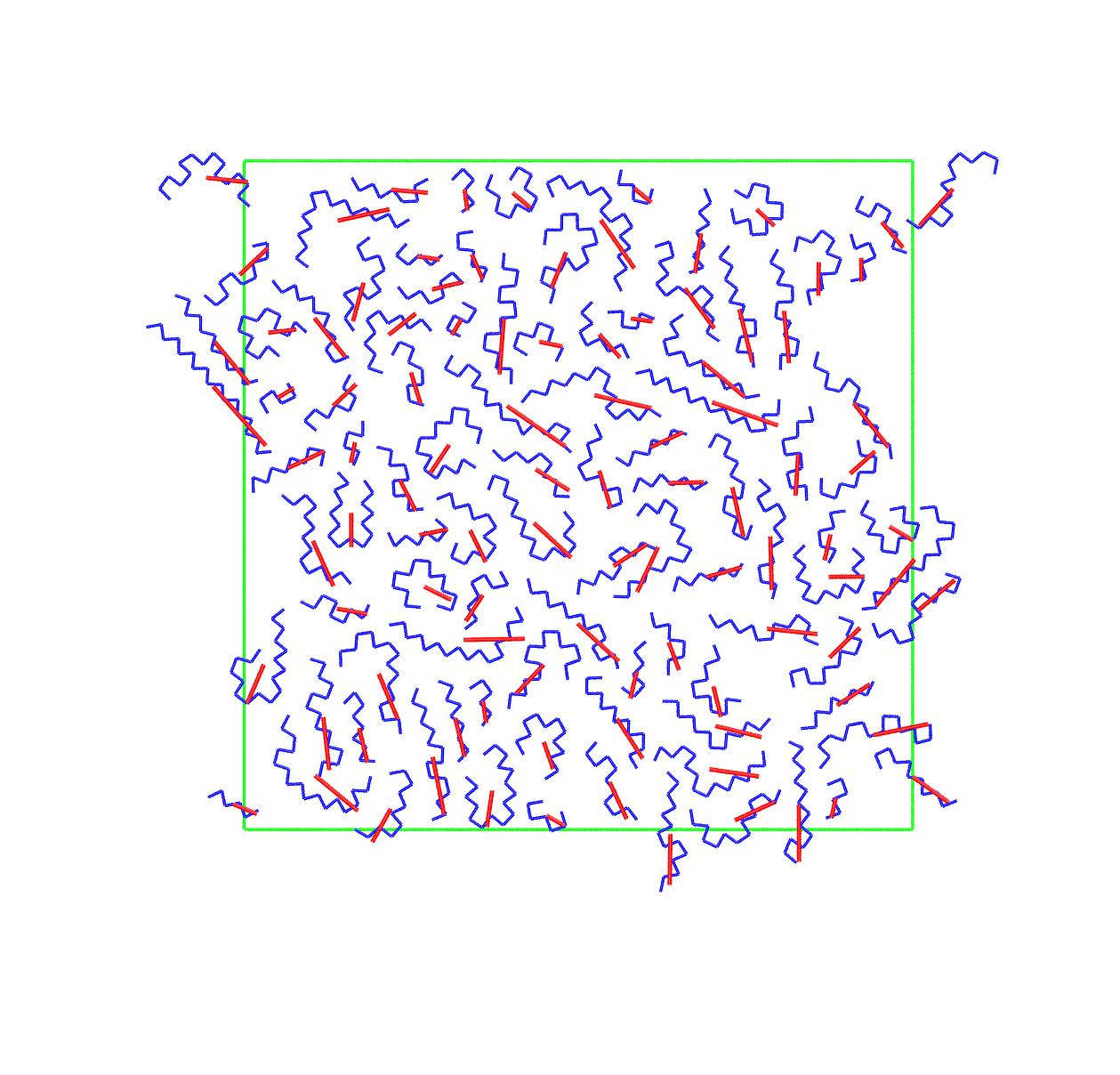}
\includegraphics[scale=0.40]{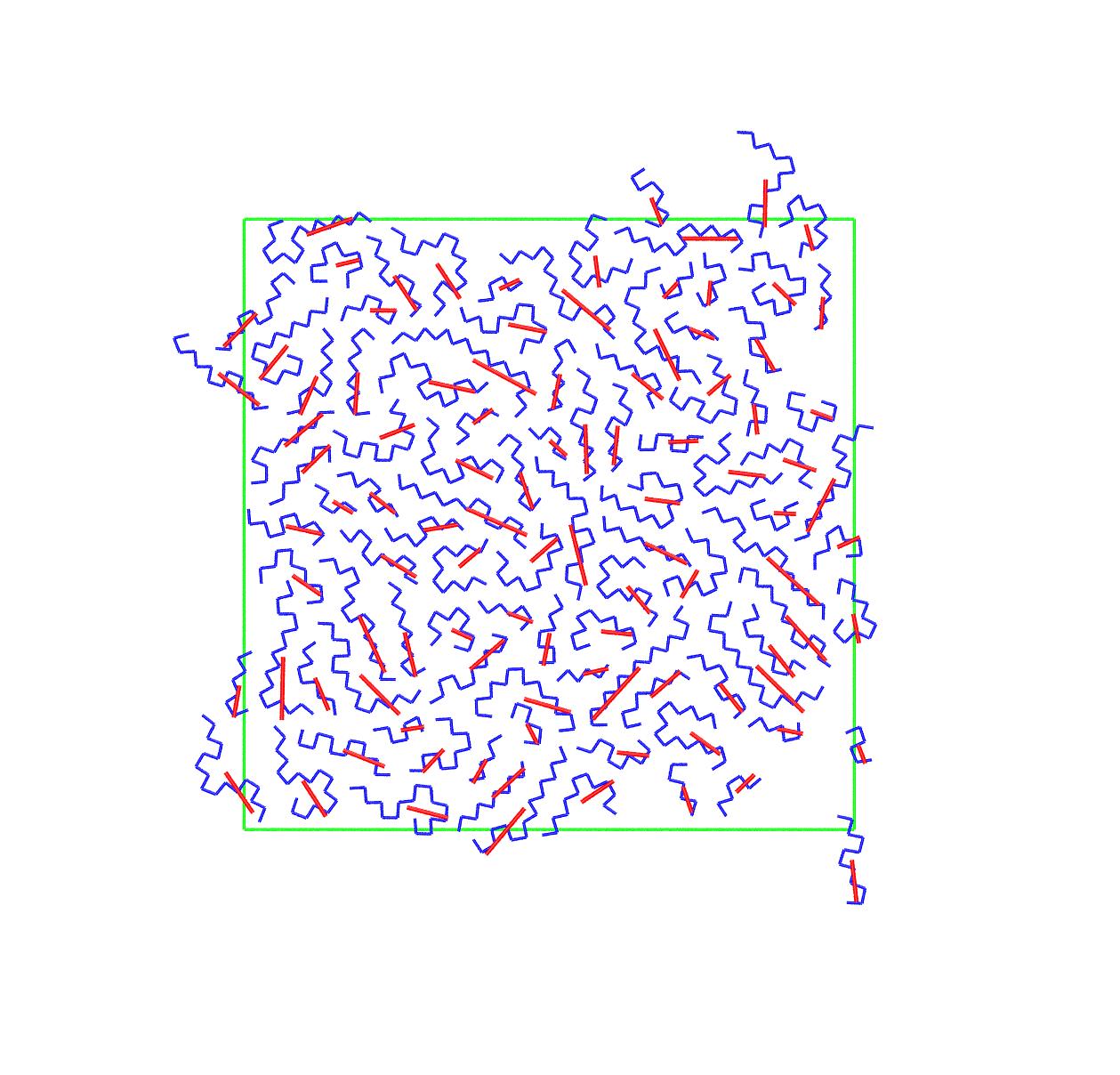}
\includegraphics[scale=0.40]{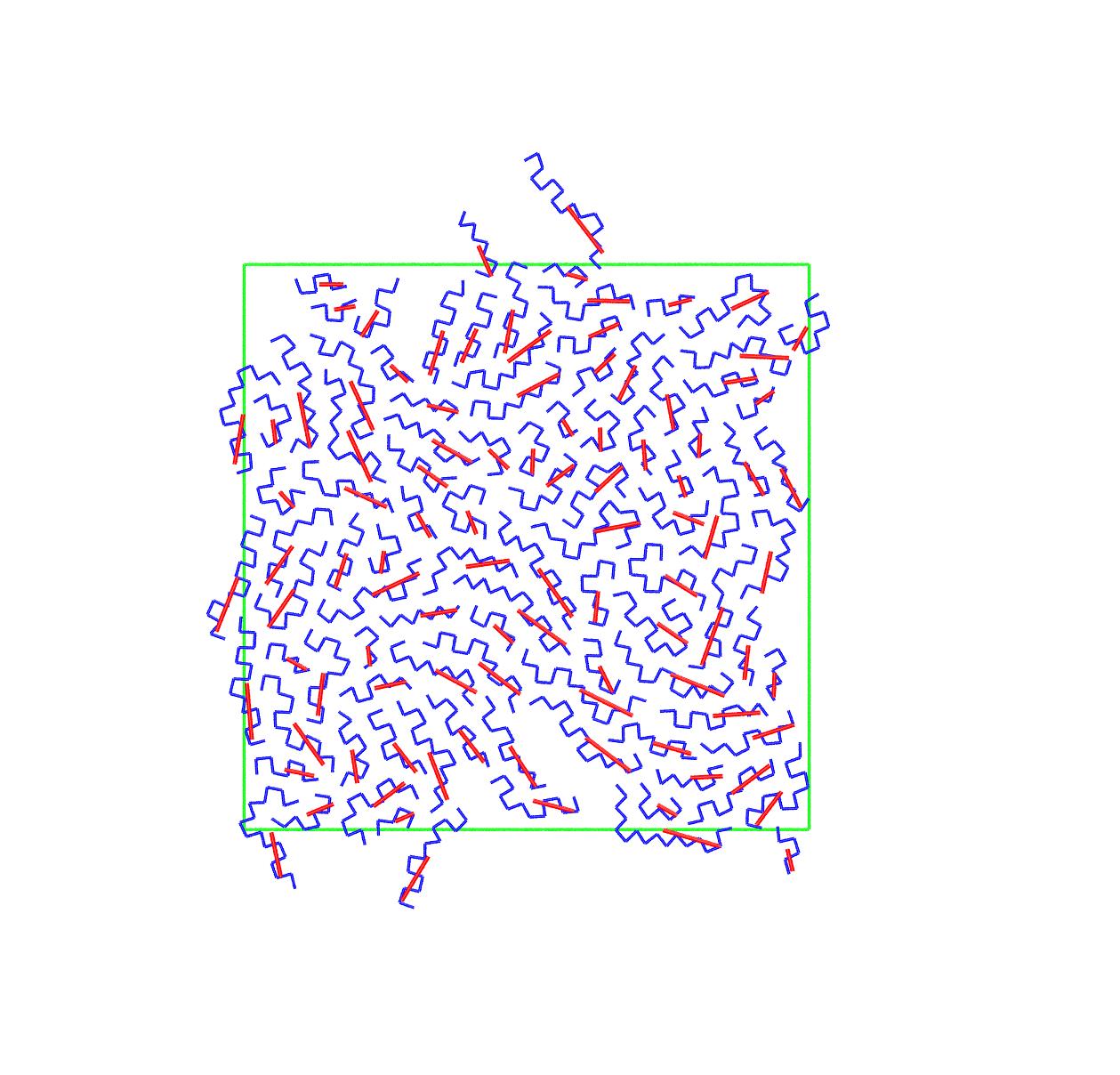}
\includegraphics[scale=0.40]{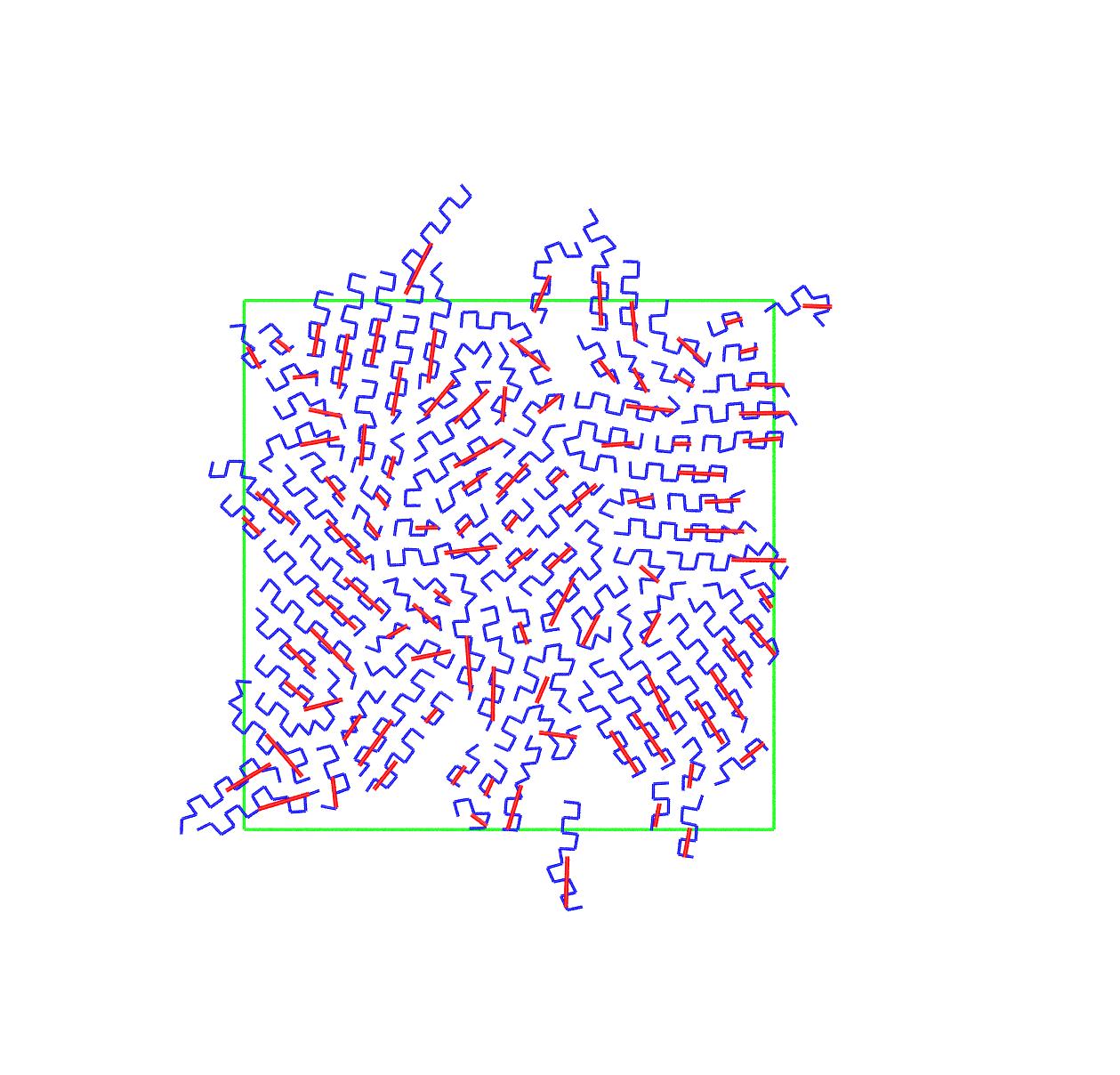}
\centering
\includegraphics[scale=0.40]{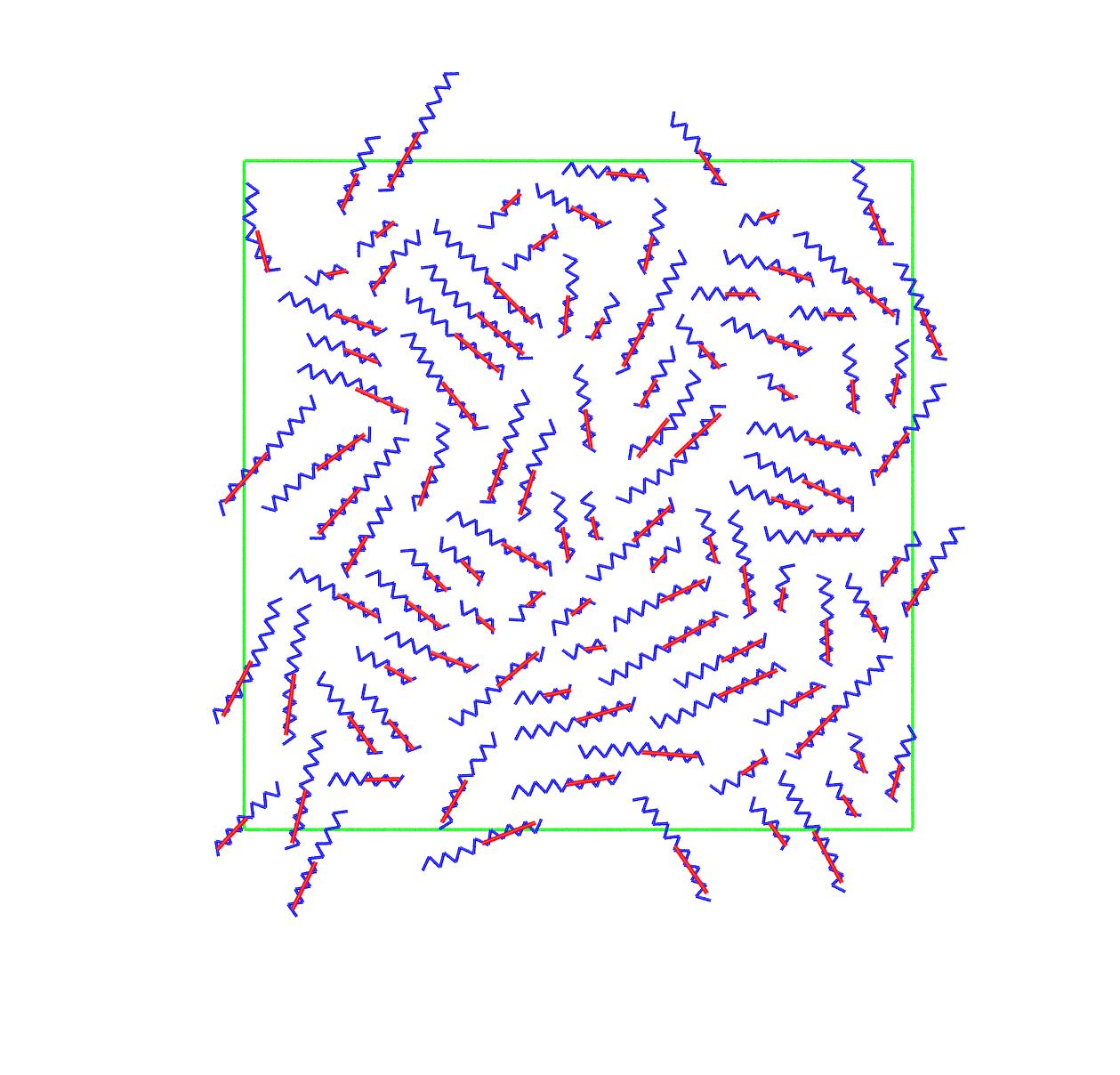}
\includegraphics[scale=0.40]{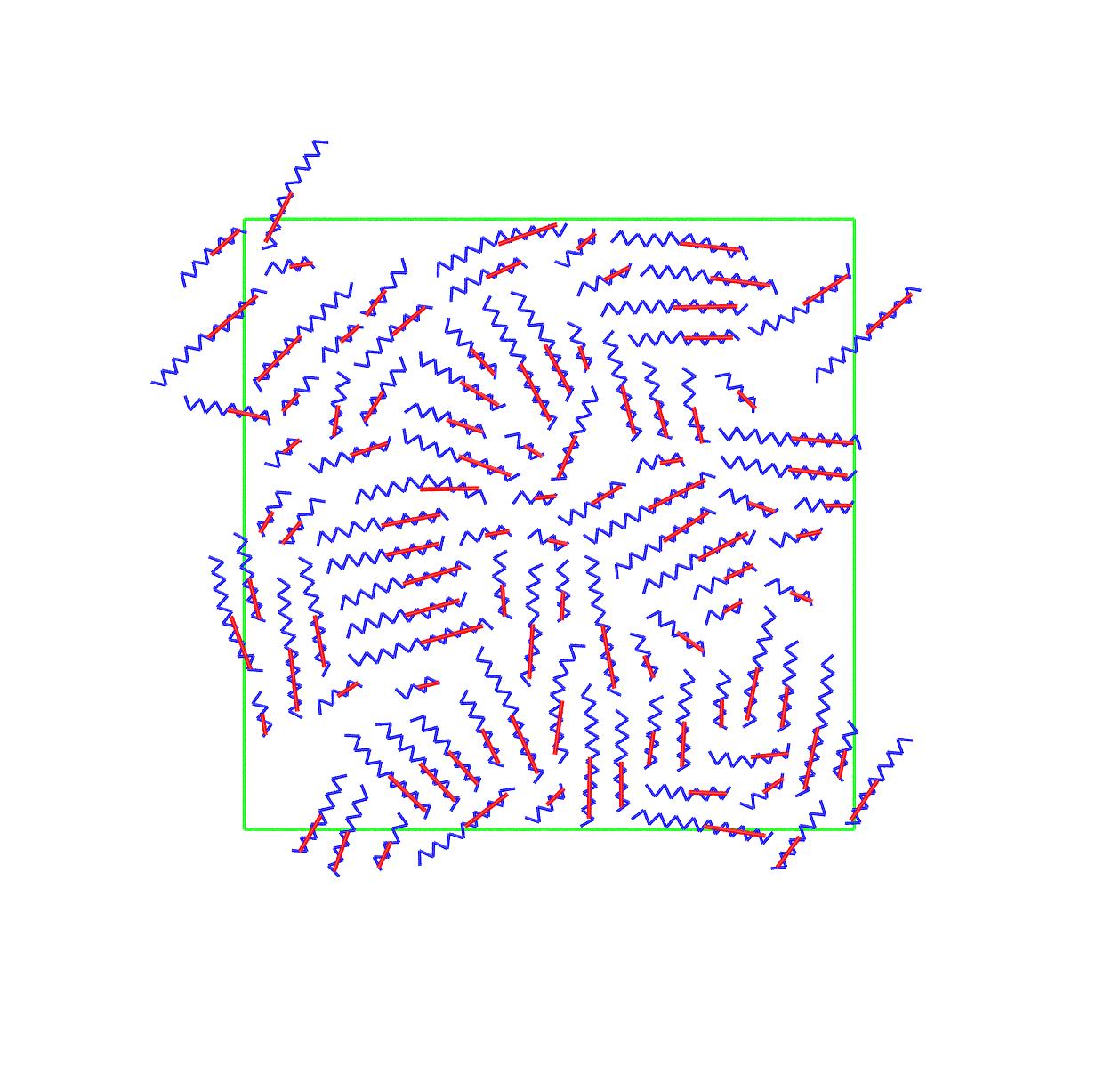}
\includegraphics[scale=0.40]{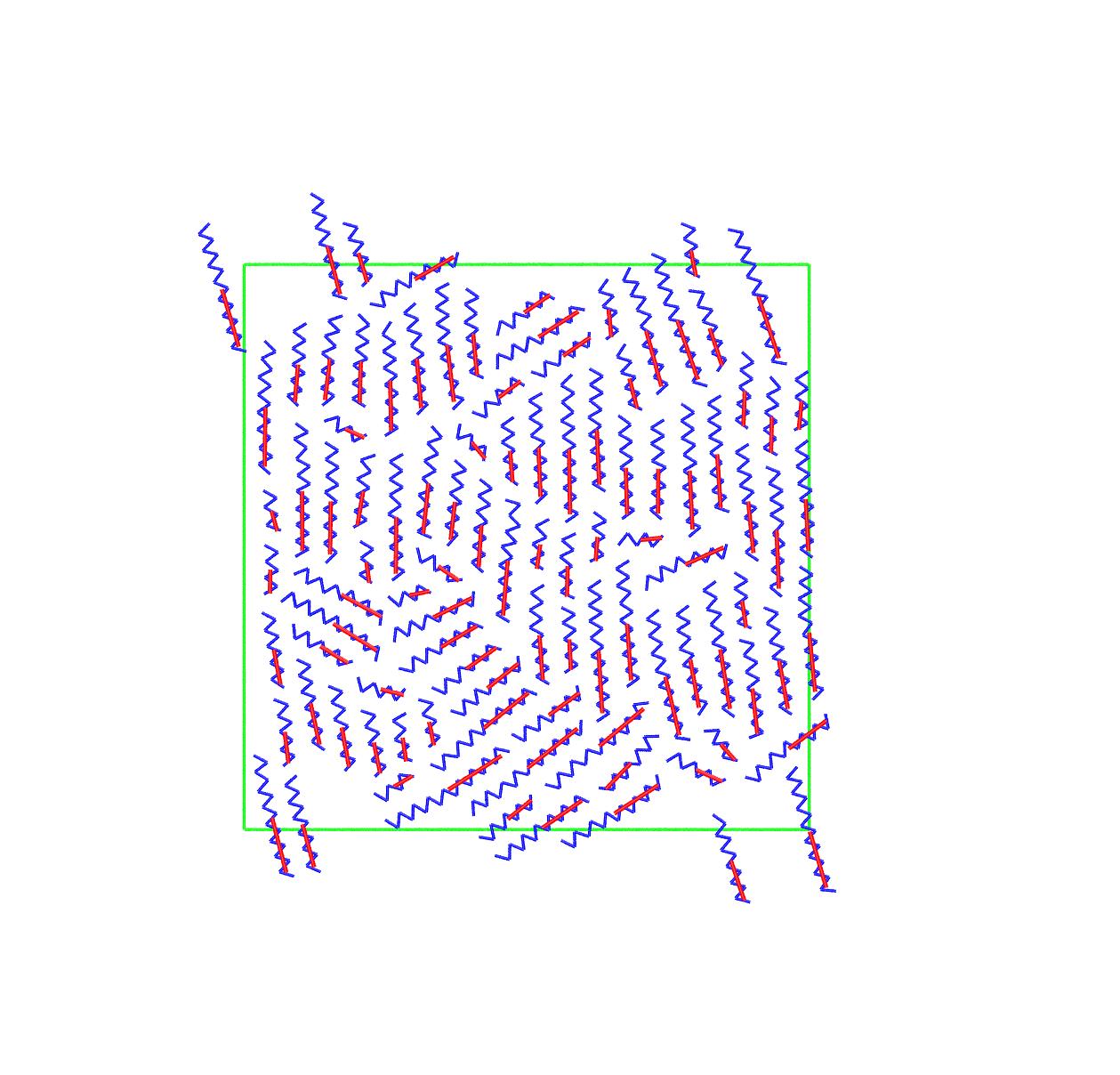}
\includegraphics[scale=0.40]{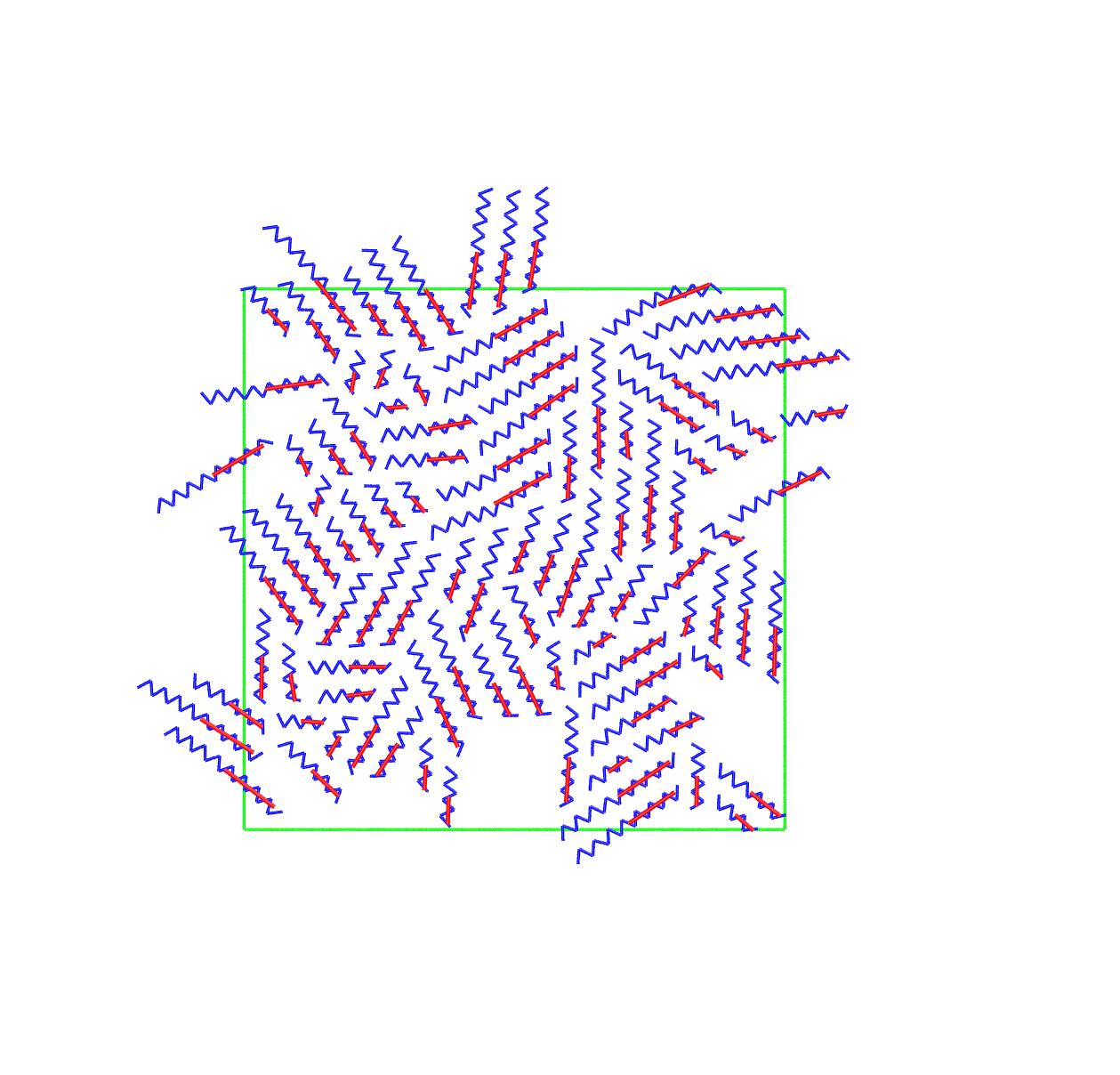}
\caption{System snapshots at the end of the MC simulation for semi-flexible polymers where chains are represented by lines and are colored in blue. Also shown in red is the vector of the largest (semi)axis of the inertia ellipsoid, with a length proportional to the corresponding length of the axis. From top to bottom: $\theta_0 = 0$, 60, 90 and $120^{\circ}$. From left to right: $\varphi^* = 0.50$, 0.60, 0.70 and $\varphi^{*,RCP}_{2D}(\theta_0)$. Image created with the VMD software \cite{RN250}.}
\label{LRO_finalsnapshots}
\end{figure*}

\begin{figure*}[ht]
\centering
\includegraphics[scale=0.40]{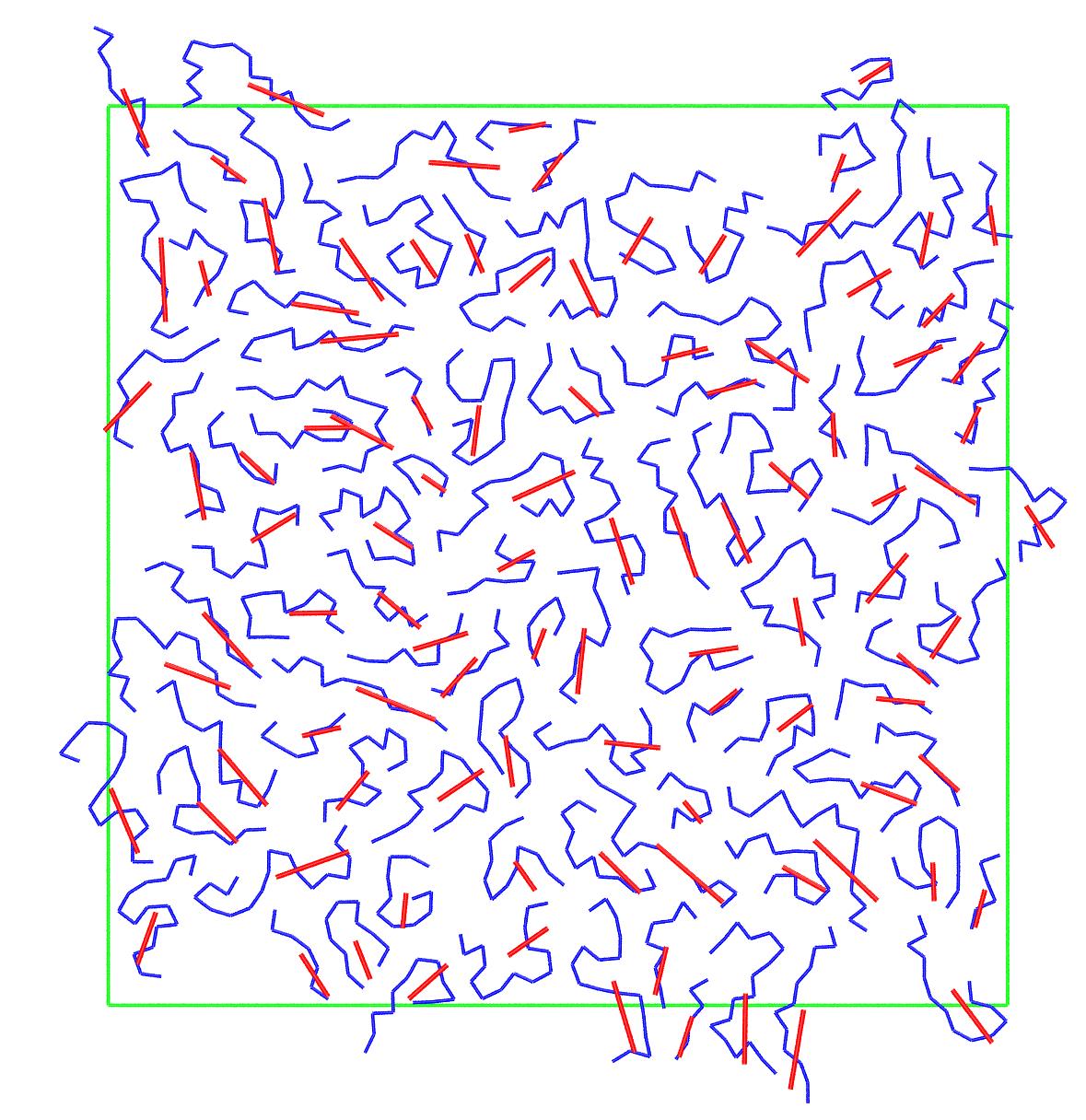}
\includegraphics[scale=0.40]{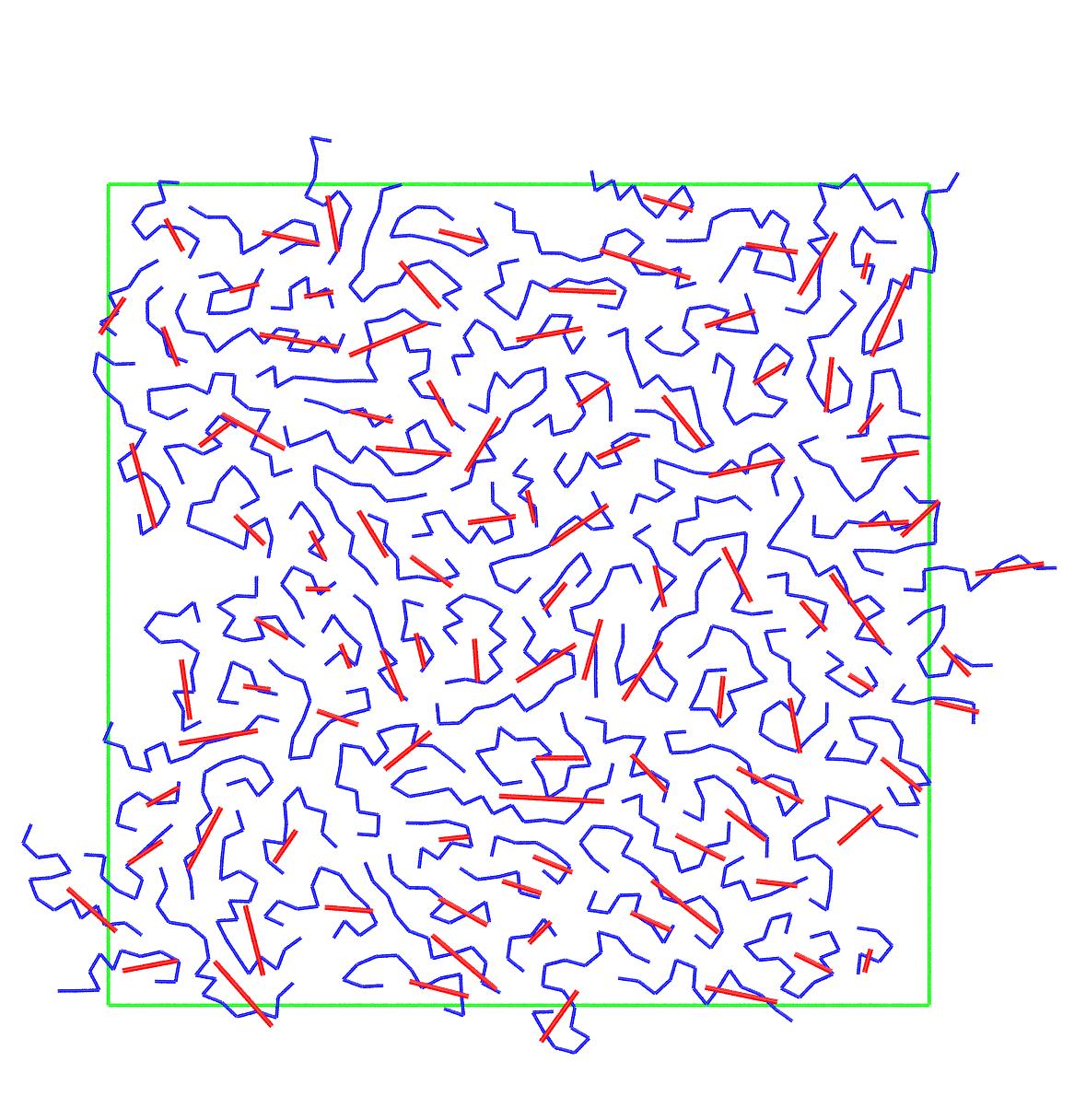}
\includegraphics[scale=0.40]{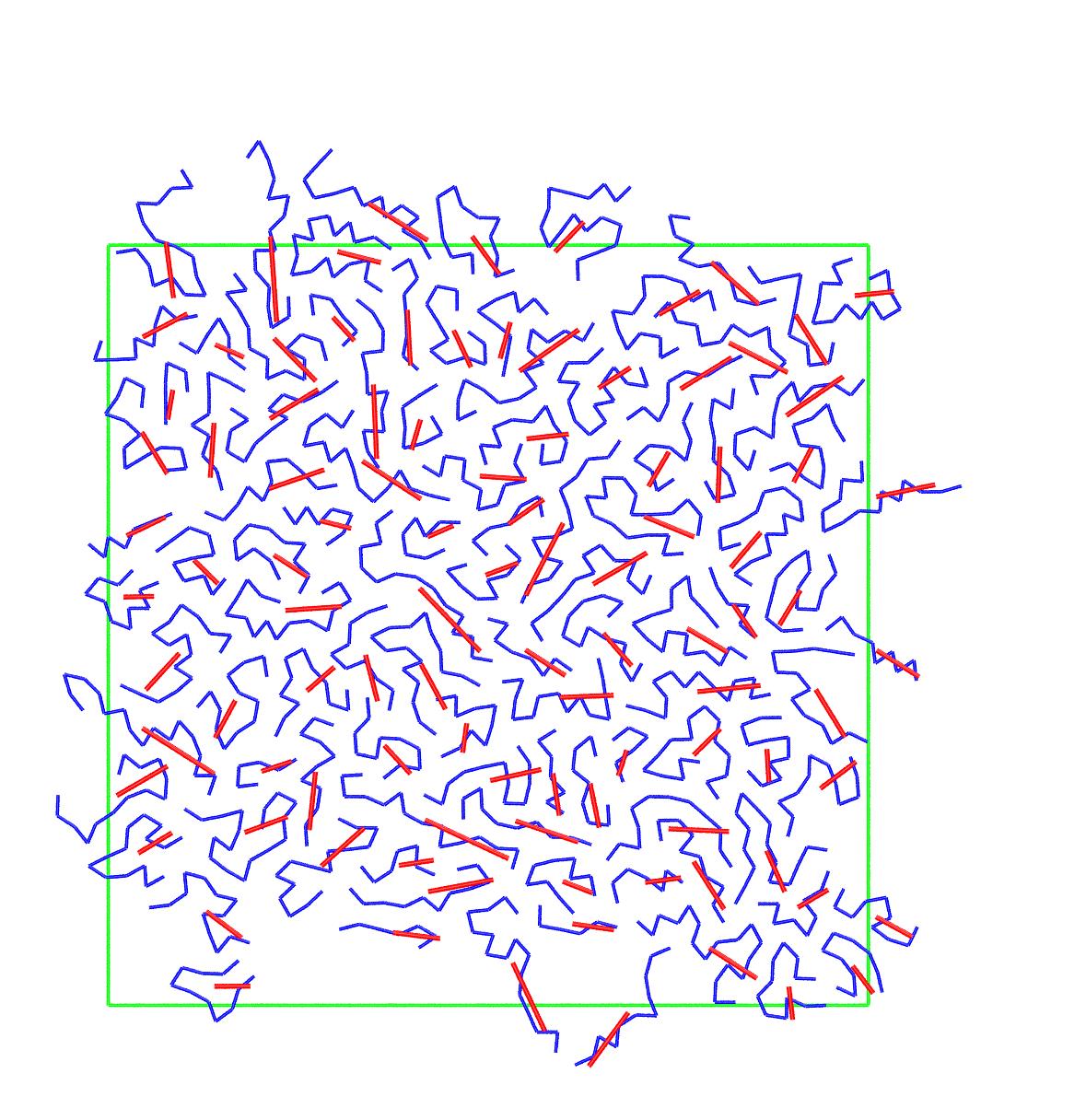}
\includegraphics[scale=0.40]{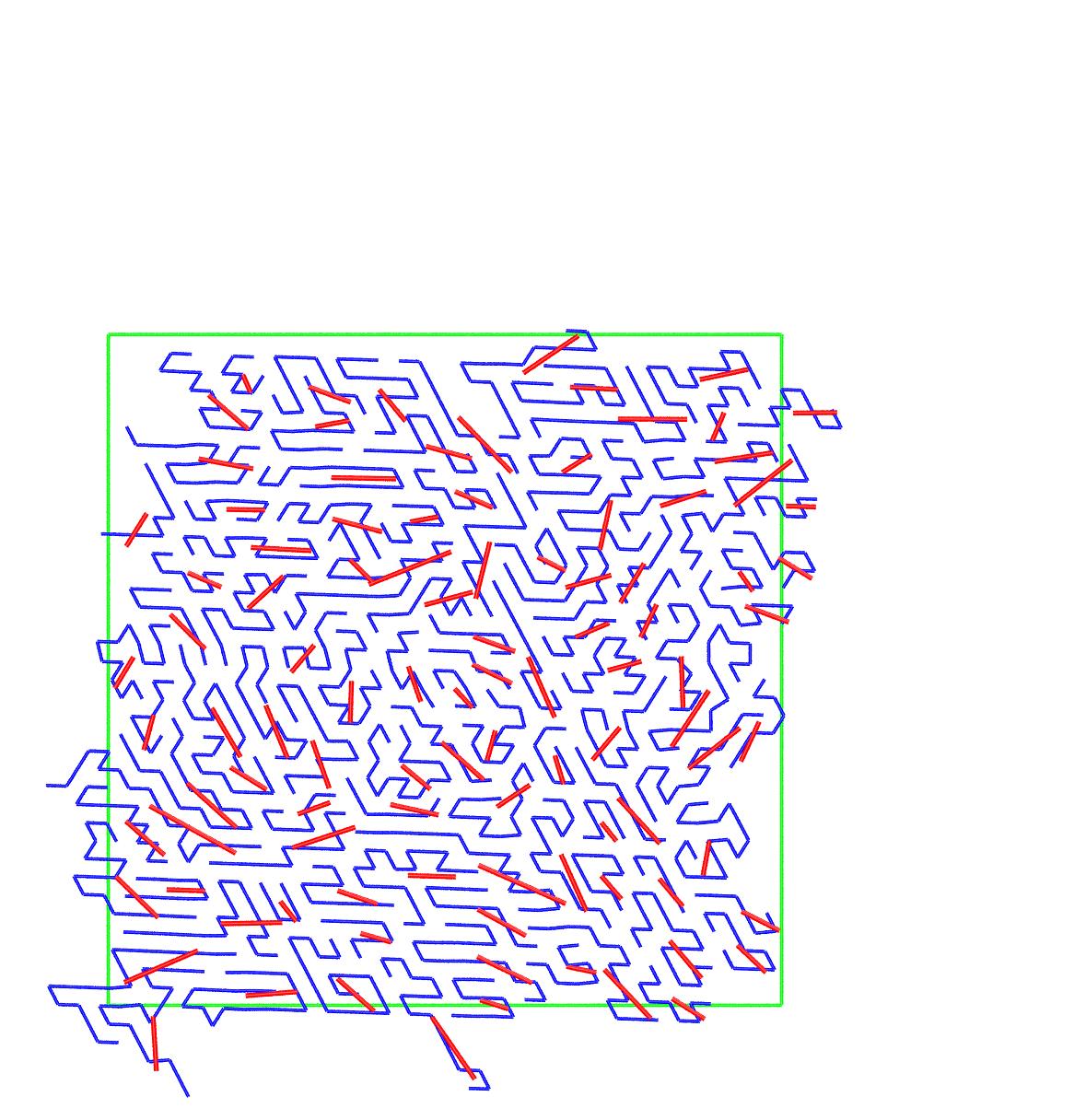}
\caption{System snapshots at the end of the MC simulation for freely-jointed (FJ) polymers where chains are represented by lines and are colored in blue. Also shown in red is the vector of the largest (semi)axis of the inertia ellipsoid, with a length proportional to the corresponding length of the axis. From left to right: $\varphi^* = 0.50$, 0.60, 0.70 and $\varphi^{*,RCP}_{2D}(FJ)$. Image created with the VMD software \cite{RN250}.} 
\label{LRO_finalFJ}
\end{figure*}

Figs. \ref{LRO_finalsnapshots} and \ref{LRO_finalFJ} show snapshots at the end of the MC simulations on semi- and fully-flexible polymers, respectively, where chains, shown in blue, are represented by lines. Also shown in red is the vector of the largest (semi)axis of the inertia ellipsoid, as calculated from the eigenvectors of the mass-moment of inertia tensor \cite{RN27}, with a length proportional to the length of the (semi)axis. Visual inspection of the final system configurations reveals a wealth of distinct global behavior ranging from isotropic state to the establishment of nematic and tetratic phases of varied level of perfection.

\par Interesting trends can be established from the evolution of the nematic ($q_2$) and tetratic ($q_4$) order parameters as a function of equilibrium bending angle and surface coverage. Starting with the most trivial behavior, the right-angle chains ($\theta_0 = 90^{\circ}$) remain isotropic at all packing densities, since $q_2, q_4 \rightarrow 0$ for all frames. We should note here that for all isotropic systems in two dimensions and due to finite system size the long-range order parameters, be $q_2$ or $q_4$, adopt positive values, which are expected to vanish in the limit of an infinitely-large system as originally discussed in Ref. \cite{RN2056}.
\par
\begin{figure*}[ht]
\centering
\includegraphics[scale=0.50]{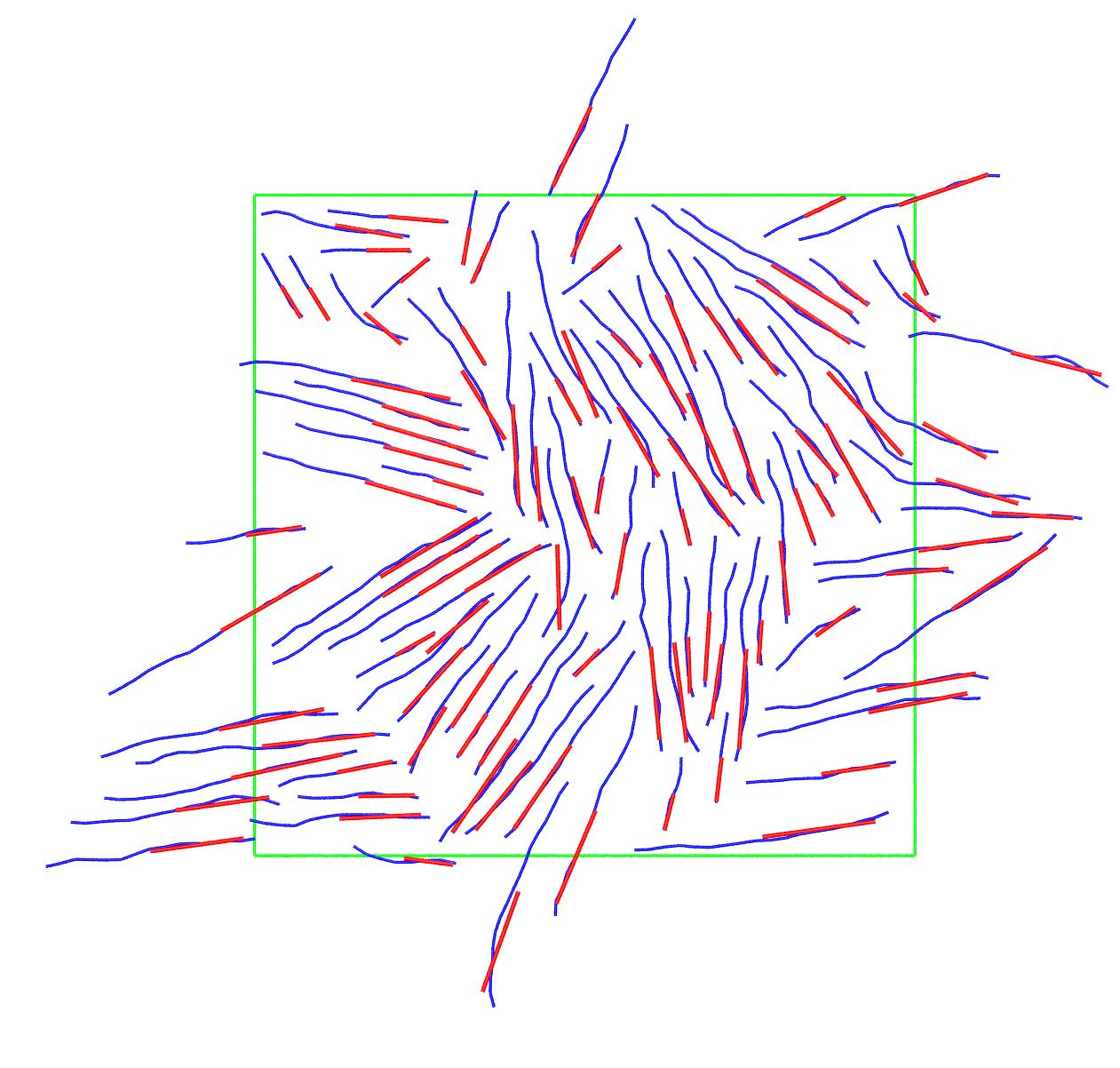}
\includegraphics[scale=0.25]{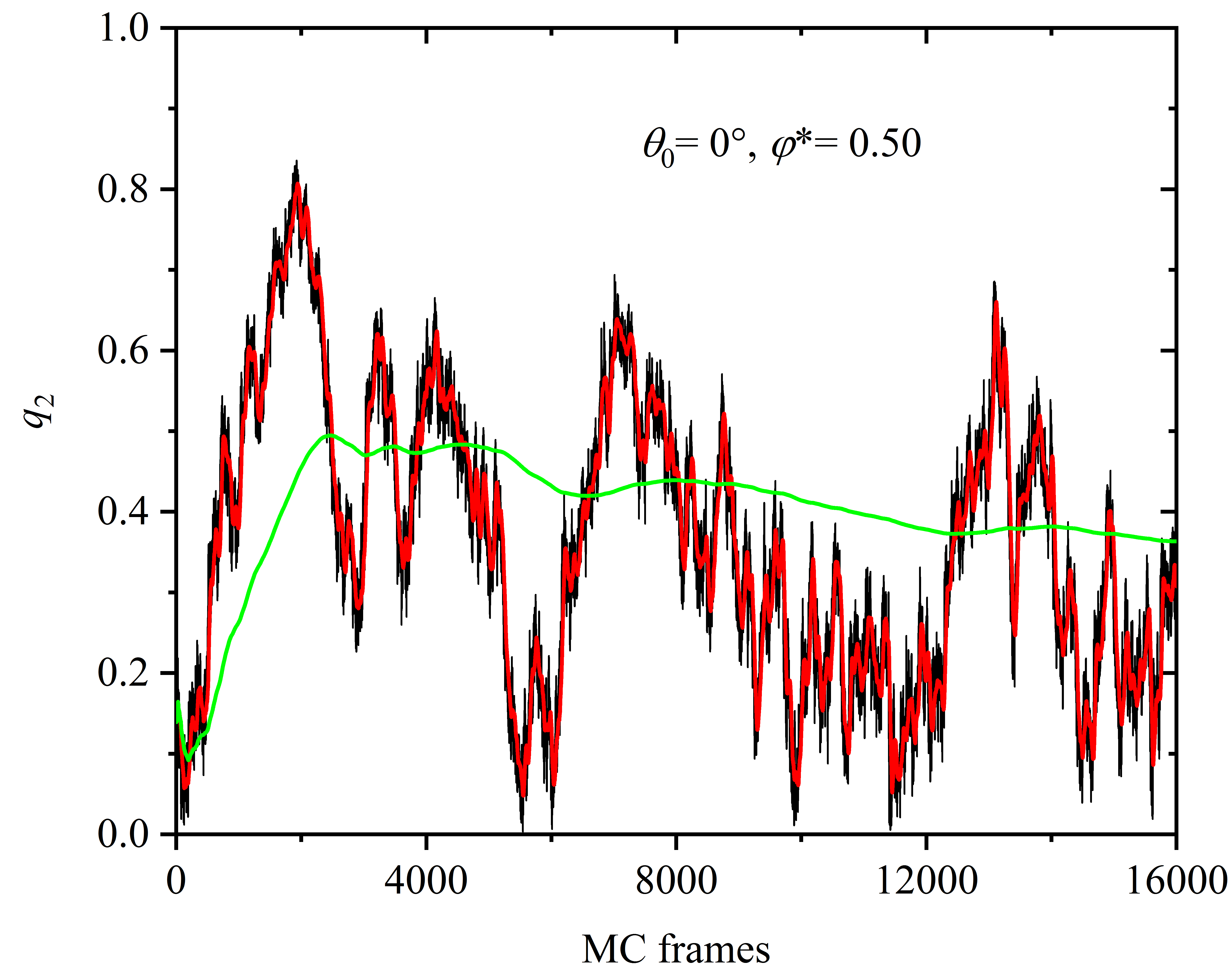}
\includegraphics[scale=0.50]{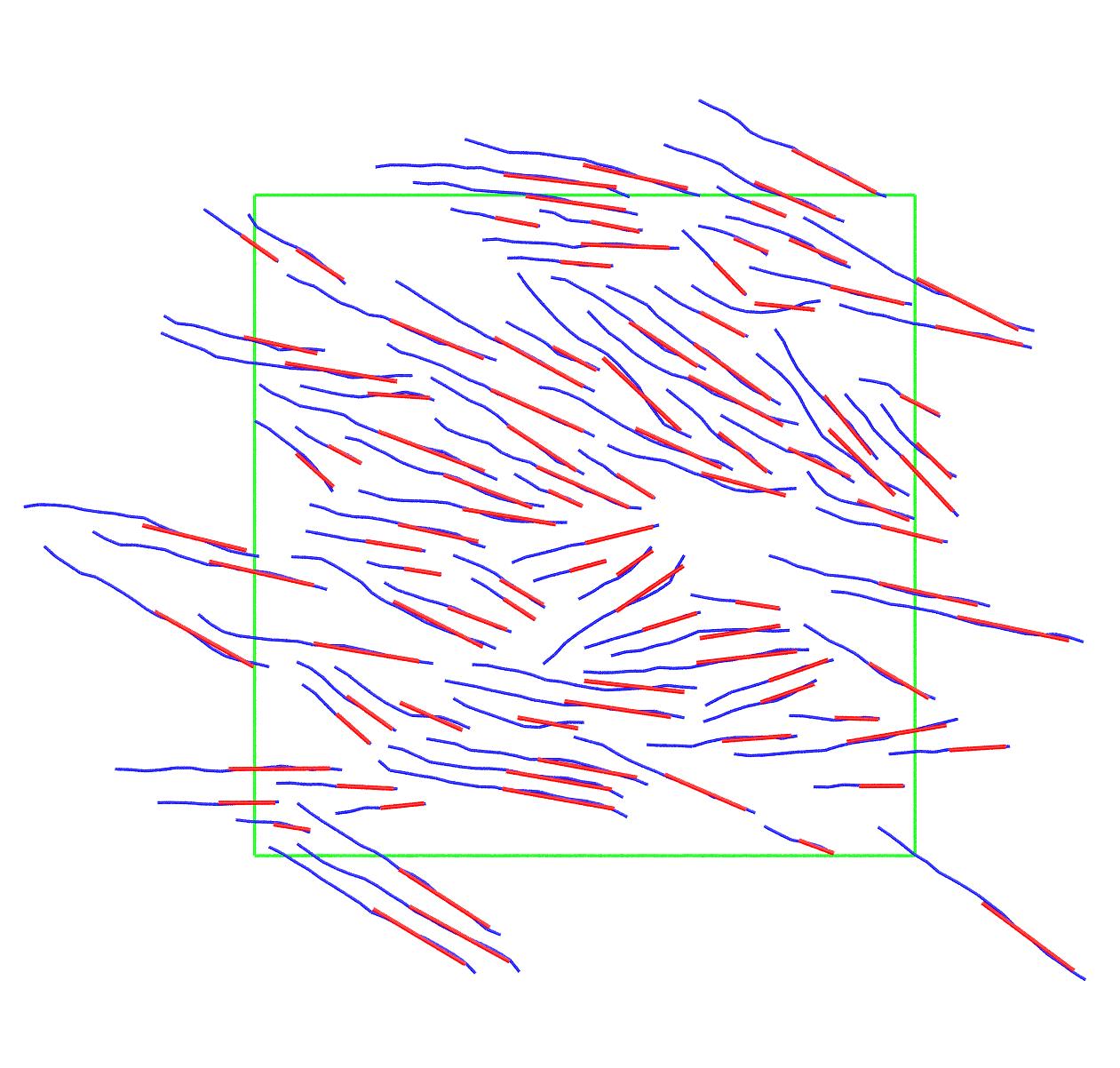}
\caption{Central panel: Nematic order parameter, $q_2$, as a function of MC frames for $\theta_0 = 0^{\circ}$ at $\varphi^* = 0.50$. Black, red and green curves correspond to instantaneous, running average (with a period of 50 frames) and cumulative running average values. 1 frame corresponds to $10^7$ MC steps. Left panel: Snapshot corresponding to an isotropic state characterized by the minimum registered $q_2$ value (= 0.004, frame \#5544); Right panel: Snapshot corresponding to an imperfect nematic state characterized by the maximum recorded $q_2$ (= 0.836, frame \#1927). Chains are shown in unwrapped coordinates, are represented by lines and are colored in blue. Also shown in red is the vector of the largest (semi)axis of the inertia ellipsoid, with a length proportional to the corresponding length of the axis.} 
\label{q2_vs_MCframes_and_snapshots_theta000_s0.50}
\end{figure*}

 The rest of polymer systems demonstrate, with minor deviations, a unified dependence on surface coverage. First, at intermediate volume fractions ($\varphi^* = 0.50$ and 0.60) the rod-like chains show a clear isotropic $\rightarrow$ nematic transition, as indicated by the sharp increase in $q_2$ in the central panel of Fig. \ref{q2_vs_MCframes_and_snapshots_theta000_s0.50}, where the system starts from an isotropic state ($q_2 \rightarrow 0$) and very rapidly becomes nematic ($q_2 \approx 0.82$). The resulting nematic phase is, however, unstable as evidenced by the very large fluctuations of the nematic order parameter as a function of MC frames. This instability is mainly attributed to the small chain lengths studied here leading to low aspect ratios for the polymers. As explained earlier, around $20\%$ of the chain population have an aspect ratio which can be considered below the stability threshold of $L / D > 7$ for the nematic state in two dimensions as gauged in \cite{RN2061} for regular rods, where $L$ is the length (fluctuating $R$ here) and $D$ the diameter (fixed $\sigma$ here). Visualizations of the systems snapshots corresponding to the lowest and highest $q_2$ values can be seen in the left and right panels of Fig. \ref{q2_vs_MCframes_and_snapshots_theta000_s0.50} further confirming the instability of the nematic phase for rod-like chains.
The formation of the unstable nematic state is evident also at higher volume fractions of $\varphi^* = 0.60$ and 0.70, with the latter configuration being more defect-ridden (lower $\langle q_2 \rangle$) compared to the former. However, at RCP the physical picture changes completely as rod-like chains adopt a highly imperfect tetratic phase since systematically $q_4 > q_2$ and $q_2 \rightarrow 0$ and as can be visually confirmed by the corresponding snapshot in the top-right panel of Fig. \ref{LRO_finalsnapshots}. 

\begin{figure}[ht]
\centering
\includegraphics[scale=0.30]{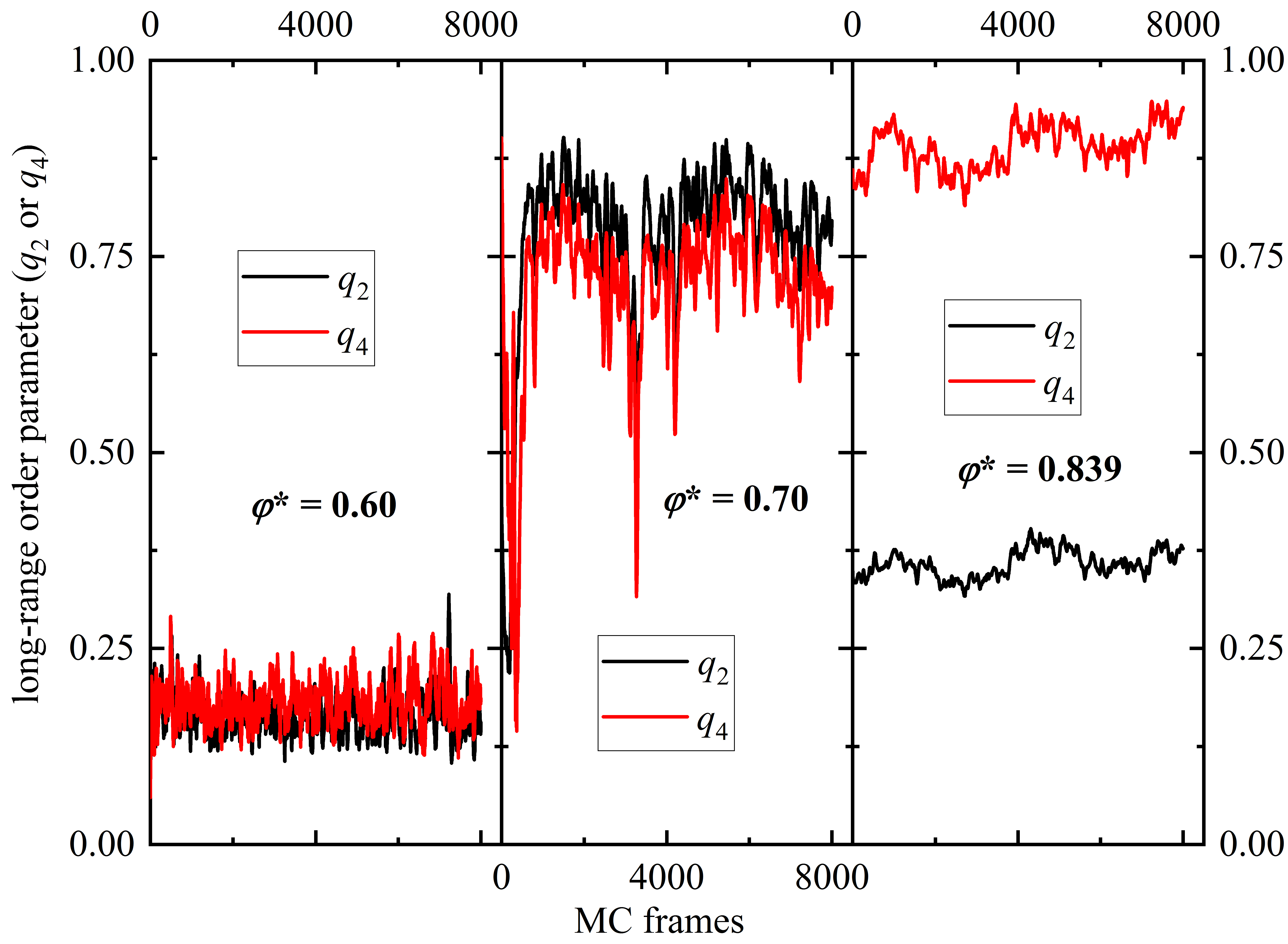}
\caption{Running average (with a period of 50 frames) of the nematic ($q_2$, black color), and tetratic ($q_4$, red color) order parameters, as a function of MC frames for $\theta_0 = 60^{\circ}$ at $\varphi^* = 0.60$ (left panel) 0.70 (middle panel) and RCP($60^{\circ}) = 0.839$ (right panel). 1 frame corresponds to $10^7$ MC steps.} 
\label{q2_q4_vs_MCframes_theta060}
\end{figure}

The extended ($\theta_0 = 60^{\circ}$) and compact ($\theta_0 = 120^{\circ}$) zig-zag chains exhibit an almost identical behavior: at $\varphi^* = 0.50$ both systems remain isotropic, while at higher concentrations ($\varphi^* = 0.60$ and 0.70) they transit to a stable nematic phase, being characterized by simultaneously high values of $q_2$ and $q_4$. Finally, at the RCP limit the tetratic state is established, being identified by high values for $q_4$ (tetratic order parameter) and significantly lower ones for $q_2$ (nematic order parameter). As a typical paradigm, the evolution of $q_2$ and $q_4$ versus MC frames for the $\theta_0 = 60^{\circ}$ at $\varphi^* = 0.60$, 0.70 and 0.839 (= $\varphi^{*,RCP}_{2D}(60^{\circ})$), corresponding to isotropic, nematic and tetratic states, can be found in the panels of Fig. \ref{q2_q4_vs_MCframes_theta060}. These trends are confirmed by the system configurations at the end of the simulations shown in Fig. \ref{LRO_finalsnapshots} where parallel and perpendicular inter-chain arrangements are visible.

\begin{figure}[ht]
\centering
\includegraphics[scale=0.40]{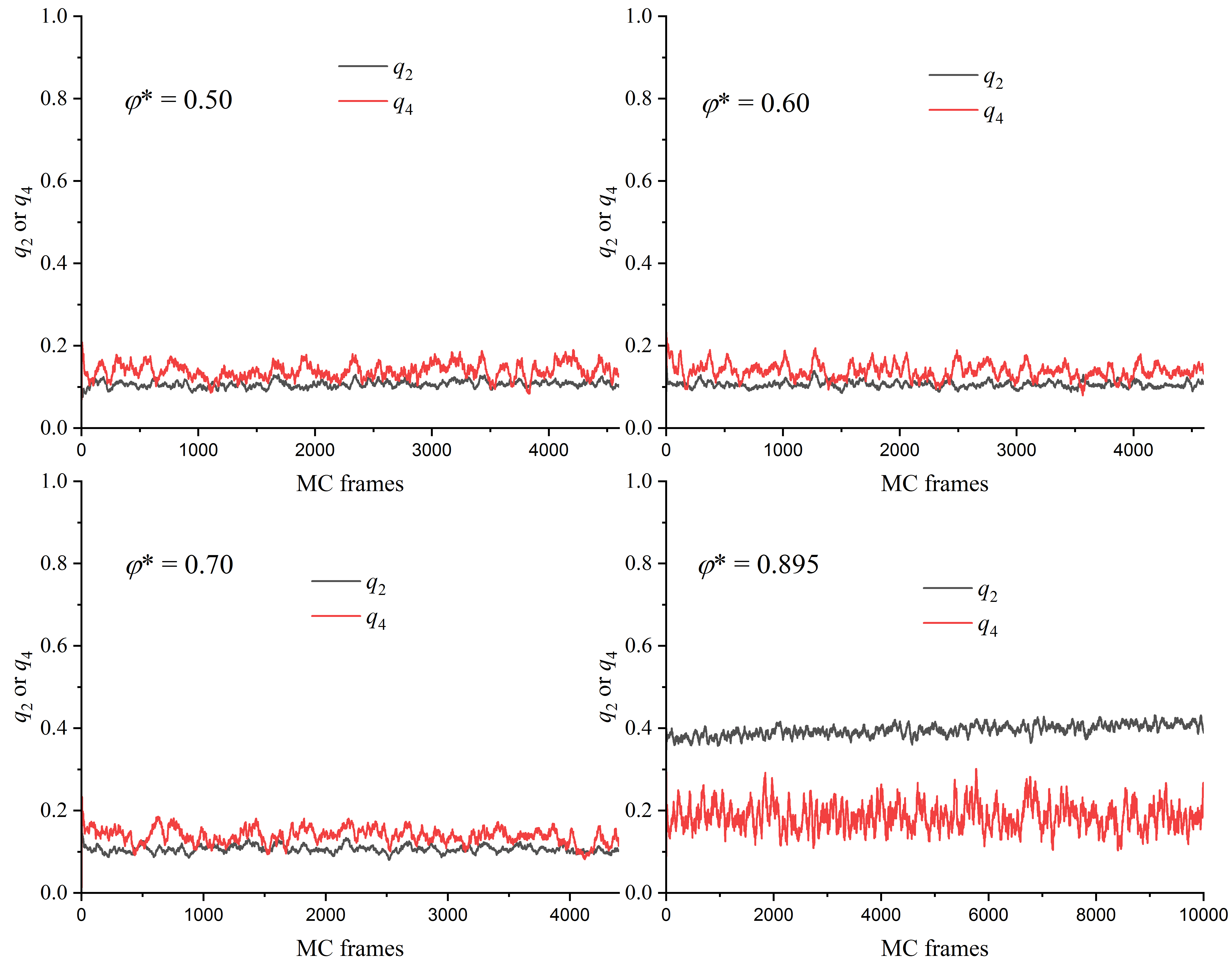}
\caption{Running average (with a period of 50 frames) of the nematic ($q_2$, black color), and tetratic ($q_4$, red color) order parameters, as a function of MC frames for freely-jointed (FJ) polymers at $\varphi^* = 0.50$ (top-left panel), 0.60 (top-right panel), 0.70 (bottom-left panel) and RCP(FJ) = 0.895 (bottom-right panel). 1 frame corresponds to $10^7$ MC steps.} 
\label{q2_q4_vs_MCframes_FJ}
\end{figure}

In comparison, as seen in Fig. \ref{q2_q4_vs_MCframes_FJ}, the fully flexible chains show isotropic behavior over the whole concentration range with a small increase in the nematic order occuring at RCP(FJ). This ordering is a rather localized trend, occuring in a small sub-group of chains (rightmost panel of Fig. \ref{LRO_finalFJ}.) 

\par
\begin{figure*}[ht]
\centering
\includegraphics[scale=0.40]{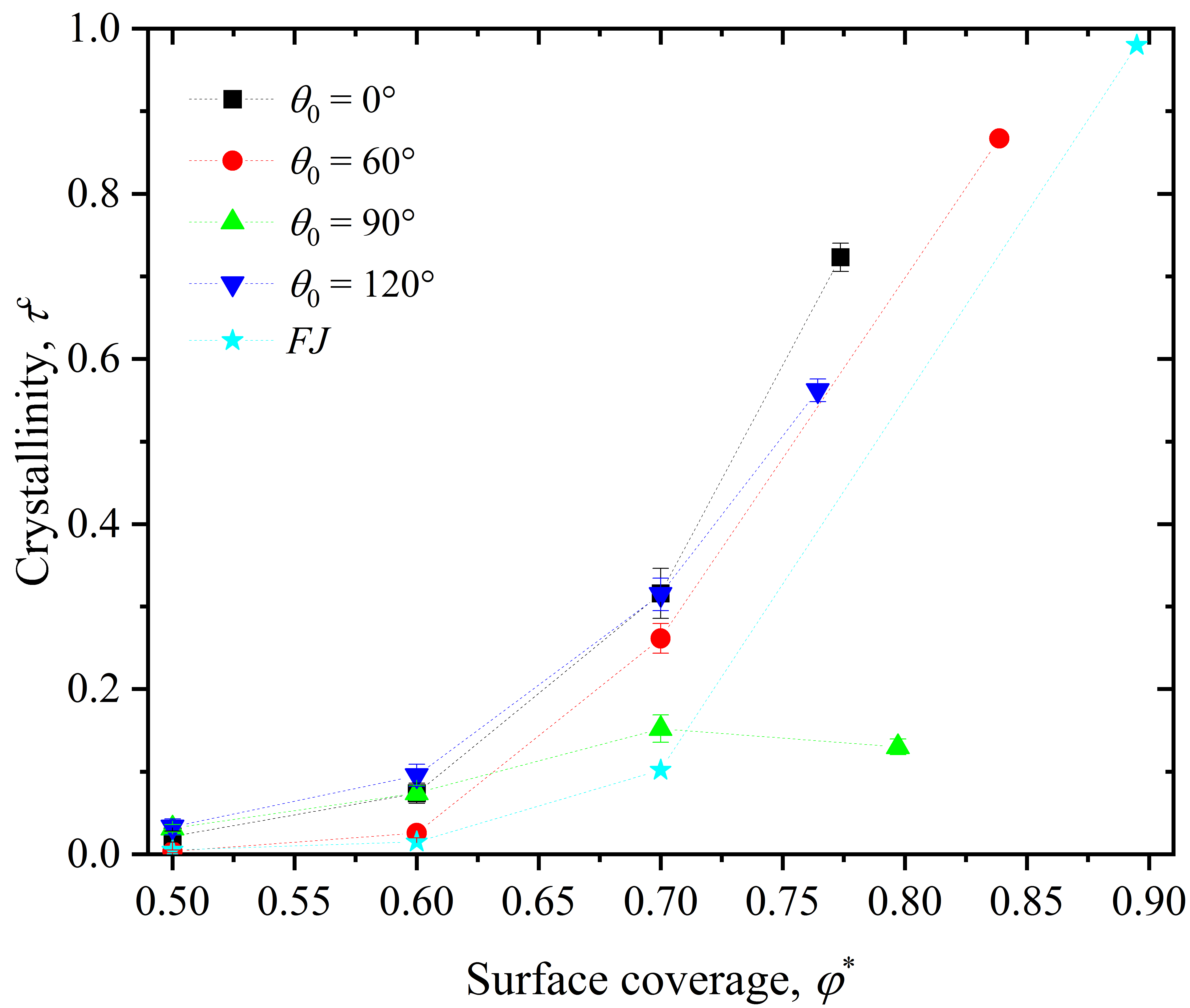}
\includegraphics[scale=0.40]{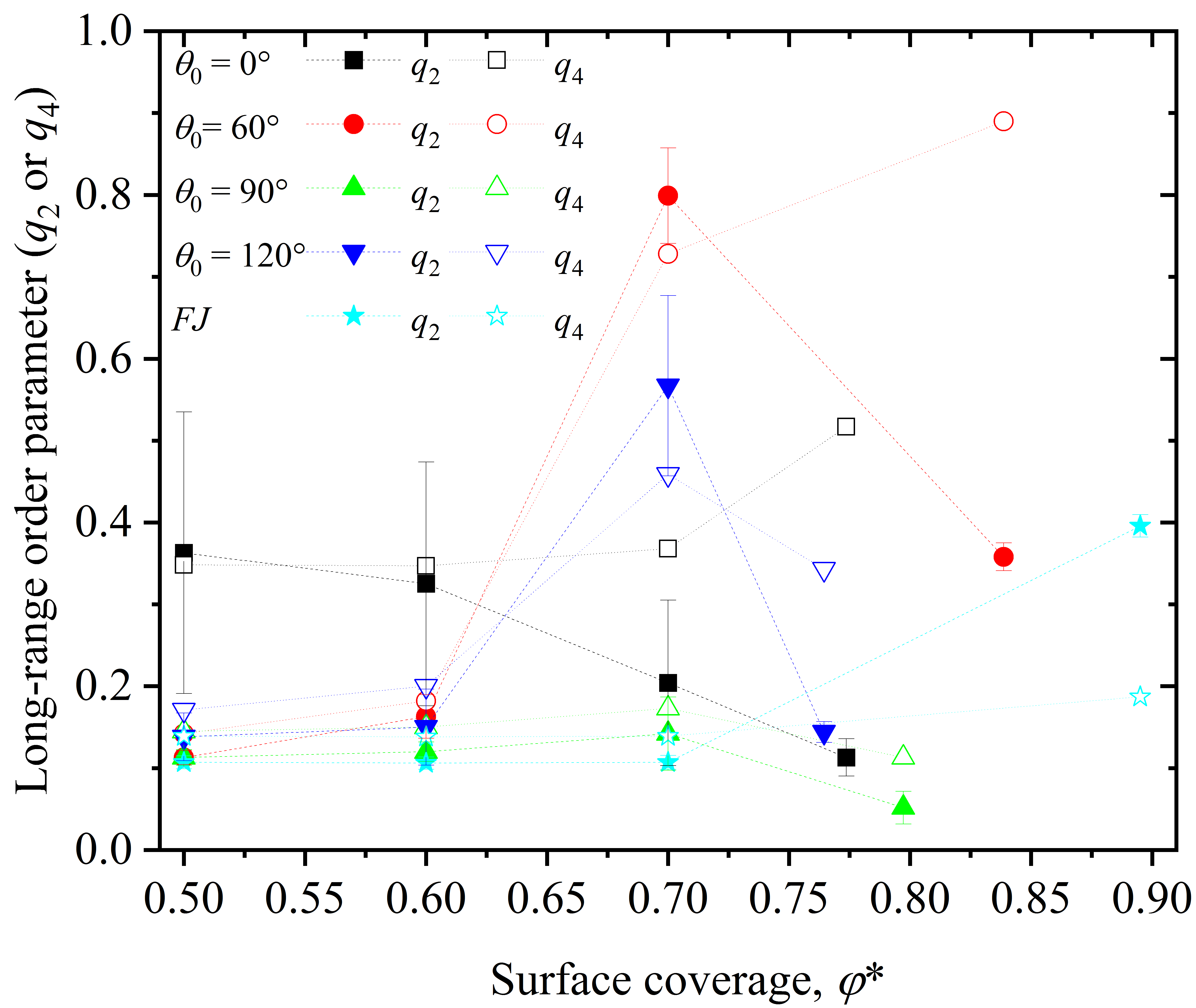}
\caption{(Left panel) Degree of crystallinity, $\tau_c$, and (right panel) nematic, $q_2$ (filled symbols), and tetratic, $q_4$ (open symbols), order parameters, as a function of surface coverage, $\varphi^*$, for all semi-flexible polymer systems studied here, including the freely-jointed (FJ) ones. Dashed or dotted lines connecting the scattered data serve only as guide for the eye. For visual clarity the error bars of $q_4$ are not shown.} 
\label{tc_q2_q4_vs_s}
\end{figure*}

 The results on the local and global order, as the RCP limit in two dimensions is approached, are summarized in Fig. \ref{tc_q2_q4_vs_s} showing the dependence of (left panel) the degree of crystallinity, $\tau^c$, and (right panel) of $q_2$ (filled symbols) and $q_4$ (open symbols) on surface coverage for all equilibrium bending angles studied here, which include the data for fully flexible (FJ) chains. 

The right-angle chains obviously constitute  a singular case,  due to $\theta_0 = 90^{\circ}$ being incompatible with the site geometry of the TRI crystal. Consequently, both their local and global order parameters are characterized by very low values over the whole concentration range, corresponding to a locally disordered and globally isotropic system. For the rest of the systems local order increases significantly with packing density and in all cases the RCP limit is characterized by the highest observed degree of crystallinity. However, the same is not true for global order: semi-flexible chains show the following long-range structural transition as RCP is approached: isotropic $\rightarrow$ nematic $\rightarrow$ tetratic. The nematic order of rod-like chains with average length $N = 12$ is highly unstable as demonstrated also by the very large error bars seen in Fig. \ref{tc_q2_q4_vs_s}, while the tetratic state seems to be an identifying characteristic of the RCP limit. 

\section{Conclusions}

\par Through extensive simulations we have studied the packing ability, local and global structure of semi-flexible athermal polymers extremely confined in monolayers, practically corresponding to systems in two dimensions. \ojoo{The combination of extreme confinement and very high surface coverage leads to a rich structural behavior as a function of equilibrium bending angle.}

As a first result we identify the trends on the densest configurations according to which: $\varphi^{*,RCP}_{2D}(FJ) > \varphi^{*,RCP}_{2D}(60^{\circ}) > \varphi^{*,RCP}_{2D}(90^{\circ}) > \varphi^{*,RCP}_{2D}(0^{\circ}) > \varphi^{*,RCP}_{2D}(120^{\circ})$. Simple calculations show that in three dimensions the RCP limit is approximately 13.6\% less dilute compared to the maximum achievable density of the HCP and FCC ideal crystals. In two dimensions and based on the present findings, the RCP limits range from the highest one of the freely-jointed (FJ) chains, which is just 1.3\% lower than the maximum achievable surface coverage, corresponding to the TRI crystal, to the least dense one of the compact, zig-zag chain configurations of $120^{\circ}$ (approximately 15.8\% less concentrated than $\varphi^{*,max}_{2D}$).     

The packing ability is intimately related to the bond geometry of the chains and their size, as imposed primarily by the equilibrium bending angle and secondarily by the surface coverage. A critical component in the identification of the RCP limit is the quantification of the inherent degree of order. Thus, order is here the combined contribution of local order at the level of monomers (crystallinity) and of global order at the level of chains. 
\par A wealth of distinct behaviors but also some universal trends can be recognized. Packings of right-angle ($\theta_0 = 90^{\circ}$), chains show a singular behavior by remaining amorphous (disordered) at the local level and isotropic at the level of chains. Interestingly enough, the RCP limit for right-angle chains  exceeds the maximum surface coverage of an ideal square crystal, i.e. $\varphi^{*,RCP}(90^{\circ}) > \varphi^*(SQU) = 0.785$. In parallel, the equilibrium bending angle of $\theta_0 = 90^{\circ}$ is incompatible with the geometric elements of the TRI crystal and thus this combination explains the disordered/isotropic state of the right-angle chain packing. Universal behavior is observed for the rest of semi-flexible systems: first, crystallinity increases monotonically as concentration increases and reaches its maximum at the RCP limit. This is in sharp contrast with the corresponding trends in three dimensions where disorder prevails at the RCP limit. Factors that lead to this discrepancy can be attributed to the protocol dependent nature of the RCP limit (which we should note is the same for all chain systems simulated here) in 2D and 3D, but also to the difficulty of crystal formation in 3D compared to 2D due to the increased coordination number at sufficiently high densities. 

\par Perhaps the most interesting trend is observed with respect to the global order. For rod-like polymers, and given the short chain lengths studied here, the nematic phase is highly unstable, fluctuating between configurations of high nematic order and ones of high tetratic order. Zig-zag chains of acute or obtuse supplementary angles show the following transition as surface coverages increases: isotropic $\rightarrow$ nematic $\rightarrow$ tetratic order, with the latter being the prevailing state at the RCP limit. Thus, based on the descriptors of order utilized here we can claim that it is the tetratic order that is intimately related to the establishment of random close packing in two dimensions for polymers. 
\par Based on the results presented here athermal polymers in two dimensions possess a richness of multi-scale structural behavior which is not encountered in their monomeric analogs and which appears even more complex than the one exhibited by the 3D analogs under bulk conditions.

\begin{acknowledgments}
\par Authors acknowledge support through project "PID2021-127533NB-I00" of MICINN/FEDER (Ministerio de Ciencia e Innovación, Fondo Europeo de Desarrollo Regional). D.M.F. acknowledges financial support through “Programa Propio UPM Santander” of Universidad Politécnica de Madrid (UPM) and Santander Bank. The authors gratefully acknowledge UPM for providing computing resources on the Magerit supercomputer through projects “r727”, “s341”, “t736” and “u242”.
\end{acknowledgments}

\section*{Author Declarations}

\subsection*{Conflict of Interest}
The authors have no conflicts to disclose.

\subsection*{Author Contributions}
\textbf{Daniel Martínez-Fernández:} Data curation (lead); Formal analysis (equal); Investigation (equal); Software (equal); Visualization (equal); Writing – original draft (equal).
\par \textbf{Clara Perdosa:} Data curation (equal); Formal analysis (equal); Investigation (equal); Software (supporting); Visualization (supporting); Writing – review \& editing (equal).
\par \textbf{Miguel Herranz:} Data curation (supporting); Formal analysis (supporting); Investigation (supporting); Software (equal); Writing – review \& editing (equal).
\par \textbf{Katerina Foteinopoulou:} Formal analysis (equal); Methodology (equal); Validation (lead); Writing – review \& editing (equal); Funding Acquisition (equal).
\par \textbf{Nikos Ch. Karayiannis:} Conceptualization (lead); Funding acquisition (equal); Investigation (equal); Methodology (equal); Writing – original draft (equal).
\par \textbf{Manuel Laso:} Methodology (equal); Funding acquisition (equal); Writing – review \& editing (equal); Funding Acquisition (equal).

\section*{Data Availability Statement}
The data that support the findings of this study are available from the corresponding author upon reasonable request.

\bibliography{Main_bib_LaTeX.bib}

\end{document}